# Integrated Photonic Quantum Computing: From Silicon to Lithium Niobate

Hui Zhang[1,2,3,4#*], Yiming Ma[5,6#], Yuancheng Zhan[7], Yuzhi Shi[1,2,3,4], Zhanshan Wang[1,2,3,4], Leong Chuan Kwek[7,8*], Anthony Laing[9*], Ai Qun Liu[10*], and Xinbin Cheng[1,2,3,4*]

[1] Institute of Precision Optical Engineering, School of Physics Science and Engineering, Tongji University, Shanghai, China

[2] MOE Key Laboratory of Advanced Micro-Structured Materials, Shanghai, China

[3] Shanghai Institute of Intelligent Science and Technology, Tongji University, Shanghai, China

[4] Shanghai Frontiers Science Centre of Digital Optics, Shanghai, China

[5] School of Microelectronics, Shanghai University, Shanghai, China

[6] Shanghai Collaborative Innovation Centre of Intelligent Sensing Chip Technology, Shanghai University, Shanghai, China

[7] Centre for Quantum Technologies, National University of Singapore, Singapore, Singapore

[8] National Institute of Education, Nanyang Technological University, Singapore, Singapore

[9] Quantum Engineering Technology Labs, H. H. Wills Physics Laboratory and Department of Electrical and Electronic Engineering, University of Bristol, Bristol, UK

[10] Institute of Quantum Technologies (IQT), The Hong Kong Polytechnic University, Hong Kong SAR, Hong Kong

[#]These authors contribute equally.

*Corresponding authors: jovie_huizhang@tongji.edu.cn (H.Z.); cqtlck@nus.edu.sg (L.C.K.); anthony.laing@bristol.ac.uk (A.L.); aiqun.liu@polyu.edu.hk (A.Q.L.); chengxb@tongji.edu.cn (X.C.);

**ABSTRACT**

Quantum technologies have surpassed classical systems by leveraging the unique properties of superposition and entanglement in photons and matter. Recent advancements in integrated quantum photonics, especially in silicon-based and lithium niobate platforms, are pushing the technology toward greater scalability and functionality. Silicon circuits have progressed from centimeter-scale, dual-photon systems to millimeter-scale, high-density devices that integrate thousands of components, enabling sophisticated programmable manipulation of multi-photon states. Meanwhile, lithium niobate, thanks to its wide optical transmission window, outstanding nonlinear and electro-optic coefficients, and chemical stability, has emerged as an optimal substrate for fully integrated photonic quantum chips. Devices made from this material exhibit high efficiency in generating, manipulating, converting, storing, and detecting photon states, thereby establishing a basis for deterministic multi-photon generation and single-photon quantum interactions, as well as comprehensive frequency-state control. This review explores the development of integrated photonic quantum technologies based on both silicon and lithium niobate, highlighting invaluable insights gained from silicon-based systems that can assist the scaling of lithium niobate technologies. It examines the functional integration mechanisms of lithium niobate in electro-optic tuning and nonlinear energy conversion, showcasing its transformative impact throughout the photonic quantum computing process. Looking ahead, we speculate on the developmental pathways for lithium niobate platforms and their potential to revolutionize areas such as quantum communication, complex system simulation, quantum sampling, and optical quantum computing paradigms.

# 1 INTRODUCTION

## 1.1 Silicon and lithium niobate: a comparative drive for advances in quantum photonics

The quest for quantum computing has created a new frontier in computational capabilities, holding the potential to tackle complex problems that are currently unsolvable by classical computers. Central to this revolution is the development of integrated photonic quantum computing, which leverages the manipulation of photons to perform quantum operations with unprecedented speed and precision. Integrated photonic quantum computing utilizes single photons as carriers of quantum information, combined with wafer-scale manufacturing processes, providing an attractive platform for the future of quantum technology. The low noise characteristics of photons and their weak coupling with the environment make photonic systems less susceptible to decoherence than solid-state material systems (e.g., superconducting circuits or trapped ions), thereby enabling operation without extreme cryogenic or ultra-high vacuum requirements. Pioneering research in quantum science, including tests of quantum entanglement[1], the generation of squeezed light[2], demonstrations of quantum teleportation[3], and loophole-free tests of Bell nonlocality[4], has proven the exceptional capability of photons in transmitting quantum information. These research outcomes have laid the basis for the development of fields such as quantum communication[5], quantum computing[6], and quantum simulation[7].

When deploying photonic quantum computing protocols on a large scale, it is desirable to integrate hundreds of different functional optical components on a single chip. Silicon photonics[8], thanks to its mature manufacturing processes, has been one of the leaders in integrated optical quantum technology. The extensive toolbox of photonic quantum components on the silicon-on-insulator (SOI) platform has enabled a variety of functionalities, including gate-based, arbitrary unitary operator-based, and measurement-based quantum computing, showcasing its versatility and immense potential in quantum technology applications[9]. However, from an optical property perspective, silicon is not the most ideal material platform[10,11]. It suffers from two-photon absorption loss and lacks second-order nonlinear effects, hindering the realization of non-classical light sources, nonlinear operations, and fast and low-loss modulation[12]. Other material platforms, such as indium phosphide (InP)[13], silicon nitride ($SiN_x$)[14], gallium arsenide (GaAs)[15], aluminium nitride (AlN)[16,17], and silicon carbide (SiC)[18], also fail to simultaneously provide ultra-low propagation loss, fast and low-loss optical modulation capabilities, and efficient nonlinear effects, which are crucial for integrated quantum photonic applications.

Lithium niobate ($LiNbO_3$, abbreviated as LN) is a promising candidate for photonics due to its superior electro-, nonlinear-, and acousto-optic properties, as well as the broad transparency window and relatively high refractive index. For example, periodically poled lithium niobate (PPLN) is crucial for wavelength conversion and photon pair generation. Despite its great potential for quantum photonics, LN has long lagged behind other integrated photonic platforms due to the difficulty in fabrication. LN waveguides are traditionally formed by titanium (Ti) diffusion or proton exchange in bulk LN, which suffer from low refractive index contrast and thus weak mode confinement, resulting in large device footprints and reduced nonlinear efficiency. In recent years, breakthroughs in ion-cutting and wafer bonding technologies have made the commercialization of high-quality thin-film LN wafers a reality[19,20], enabling the realization of ultra-low-loss, high refractive index contrast nanophotonic LN waveguides[21–31]. On the thin-film LN-on-insulator (LNOI) platform, a suite of discrete photonic devices has been developed, including ring resonators[32,33], electro-optic (EO) modulators[34–40], nonlinear wavelength converters[41–43], broadband frequency comb sources[44–46], non-classical light sources[47–49], detectors[50,51], and microwave-to-optical transducers[52,53]. These technological advancements have equipped thin-film LN with a rich component toolkit, making it prepared to become a multifunctional, high-performance integrated photonic quantum platform.

Despite its material advantages, the development of integrated photonic quantum circuits on LNOI platforms lags behind SOI systems. This disparity necessitates systematic knowledge transfer from mature SOI architectures while strategically leveraging LNOI's unique quantum capabilities. To bridge this gap, this review aims to establish a pioneering cross-platform analysis: comprehensively tracing the evolution on SOI from discrete components to integrated processors to extract scalable design principles for LNOI; concurrently identifying quantum domains—particularly where the superior electro-optic efficiency and nonlinearity of LNOI outperform SOI—to redefine competitive advantages in next-generation photonic quantum computing.

This article employs a hierarchical framework (**Fig. 1**) to review SOI and LNOI platforms for photonic quantum computing. The analysis progresses through three architectural tiers: the construction of fundamental quantum photonic toolboxes (sources, state manipulators, detectors); the integration of these elements into functional circuits; and the implementation of quantum algorithms and applications[54–56]. This sequential structure mirrors the physical implementation pathway of photonic quantum processors. Specifically, capitalizing on high-performance LNOI devices, we introduce three high-potential research vectors for advancing integrated photonic quantum circuits on LNOI: (1) High-fidelity squeezed-light sources for measurement-based quantum computing via cluster states—a promising avenue towards universal quantum computing; (2) Ultra-low-loss delay lines integrated with high-speed modulators enabling multiplexed single-photon sources essential to scalable quantum processing; (3) Microwave-photonic hybrid interfaces critical for quantum internet development. Through this review, we hope to characterize LNOI-specific component capabilities, demonstrate circuit-level synergies, and establish actionable guidance for large-scale integration of lithium niobate quantum computing circuits.

This review is organized as follows: Section 2 details the silicon photonic quantum computing toolbox; Section 3 surveys state-of-the-art SOI-based photonic quantum computing circuits; Section 4 introduces lithium niobate fundamentals; Section 5 examines progress in the lithium niobate photonic quantum toolbox; Section 6 introduces integrated LNOI circuit advancements and potential advantageous research vectors; and Section 7 discusses future technological outlooks.

### 1.2 Theoretical Framework for Integrated Quantum Photonics

Photonic quantum computing harnesses the unique properties of photons, i.e., their inherent speed, low decoherence, and robust manipulability via optical elements, to process quantum information. Its theoretical foundation is built upon a coherent pipeline that transduces abstract quantum algorithms into physical operations on light. This pipeline can be systematically deconstructed into a hierarchical framework comprising four interdependent layers: the physical encoding of quantum information, its controlled manipulation through quantum gates, final readout via quantum measurement, and the overarching computational architecture that defines the model of computation (**Fig. 2**). The following sections delineate this framework, articulating how each layer is grounded in the principles of quantum mechanics and integrated through photonic technology.

**1) Qubit encoding: defining the computational Hilbert space.** The framework begins by defining the computational substrate. Photonic qubits are physically encoded in the degrees of freedom of light, primarily falling into two paradigms:

- Discrete Variables (DV): Information is encoded in finite-dimensional spaces, such as a photon's path, polarization, or time-bin existence. A single qubit state is a superposition in a two-dimensional Hilbert space, $|\psi\rangle = \alpha|0\rangle + \beta|1\rangle$, with $|\alpha|^2 + |\beta|^2 = 1$. The required quantum states are typically generated via nonlinear processes like spontaneous parametric down-conversion (SPDC) or spontaneous four-wave mixing (SFWM).
- Continuous Variables (CV): Information is encoded in the infinite-dimensional space of the optical field's quadrature operators, $\hat{X}$ and $\hat{P}$. The computational basis states are non-classical field states like squeezed or coherent states, enabling a complementary approach to quantum computation.

**2) Quantum gate operations: engineering unitary evolution.**

- Linear optical operators: Universal gate sets can be constructed using interferometers composed of beam splitters, phase shifters, and modulators. A typical device is the Mach-Zehnder interferometer, where a tunable phase shifter enables a continuous family of unitary operations on path-encoded qubits. This approach provides high-fidelity control over single-qubit gates and specific two-qubit operations, albeit often with a probabilistic success rate that necessitates heralding or feed-forward correction.
- Nonlinear optical effects: Second- and third-order nonlinearities are essential for generating key quantum resources, such as heralded single photons and squeezed states. Beyond state generation, nonlinear processes such as sum-frequency generation and difference-frequency generation can enable quantum frequency conversion, which is critical for interfacing systems at different wavelengths. Engineering techniques like phase matching and dispersion engineering are critical for optimizing nonlinear efficiency.

**3) Quantum measurement: Projecting onto classical information.** The final step is the projection of the quantum state onto a classical outcome, a process fundamentally dictated by the encoding choice.

- DV Measurement: Relies on single-photon detectors, where the measurement statistics are compiled from coincidence counts. These statistics are the essential data for performing quantum state/process tomography, which fully characterizes an unknown quantum state/process.
- CV Measurement: Utilizes homodyne detection to measure the quadrature observables of the optical field. The key observable in a balanced homodyne detector is the differential photocurrent, $\hat{I}_{-}(t) \propto \beta_{LO}\hat{X}_{\theta}(t)$, where $\beta_{LO}$ is the amplitude of the strong local oscillator field, and $\hat{X}_{\theta}(t)$ is the quadrature operator of the signal field relative to the phase of the local oscillator.

**4) Computational Architectures: Defining the Model of Computation**. The specific arrangement and utilization of the layers above define the computational model. Three typical architectures in integrated photonic quantum computing are:

- Quantum circuit model: Computation proceeds via a sequence of single- and two-qubit gates, applying a unitary transformation $U = U_L U_{L-1} \cdots U_1$ to an initial state. Each gate $U_k = e^{-i\theta_k H_k}$ corresponds to a unitary matrix, and the overall effect of the circuit is the product of these gate matrices.
- Measurement-based quantum computation: All qubits are initialized in the $|+\rangle$ state and create a highly entangled cluster state through controlled-Z gate operations according to a specific entanglement graph $E$, yielding $|\psi\rangle = \prod_{(i,j)\in E} \mathrm{CZ}_{i,j} |+\rangle^{\otimes \mathrm{n}}$. Computation is driven by subsequent single-qubit measurements on this cluster state, rather than by a pre-determined gate sequence.
- Specialized quantum simulation models. These models (e.g., boson sampling, quantum walk) exploit the inherent complexity of linear optical networks, specifically, the hardness of calculating matrix permanents, to demonstrate a quantum computational advantage for specific tasks. Hamiltonian simulation studies quantum systems by replicating their evolution, as $|\psi\rangle_t = e^{-iHt/\hbar}|\psi\rangle_0$, enabling the analysis of material and molecular properties on a quantum computer.

The following sections will review the silicon and lithium niobate platforms using this defined photonic quantum computing pipeline, with a comparative analysis centred on the following performance metrics:

- **Photon pair generation rate (PGR):** The rate at which photon pairs are generated, scaling as $R \propto \left|\chi^{(n)}\right|^2 P_p L^2$ for a waveguide of length $L$ and pump power $P_p$. PGR quantifies the brightness of the source.
- **Second-order correlation function $g_H^{(2)}(0)$:** Measures the conditional probability of detecting a second photon after a herald event. $g_H^{(2)}(0) < 0.5$ marks the threshold for non-classicality, whereas a high-purity single-photon source requires a value much closer to zero. The value is degraded by multi-pair generation events and noise.
- **Squeezing level:** For CV states, this is the measured noise reduction below the shot noise level, expressed in decibels. A higher squeezing level (more negative dB value) enables higher fidelity quantum computation.
- **Modulation efficiency $V_\pi L$:** The product of the half-wave voltage and the modulator length. A lower $V_\pi L$ enables the implementation of low-power and compact phase shifters, which are essential for dynamic circuit reconfiguration.
- **Bandwidth $\Delta f$:** The operational frequency range of a modulator or the phase-matching bandwidth of a nonlinear process. A high bandwidth is critical for high-speed operation and spectral multiplexing.
- **Propagation Loss $\alpha$:** The attenuation coefficient, quoted in dB/m. Low loss is essential for scaling to large circuit sizes and maintaining high photon flux.
- **Extinction ratio:** A key metric for evaluating modulators or switches, defined as the ratio of optical power in the "on" state to that in the "off" state. A high ER is a prerequisite for high-fidelity quantum operations.

# 2 SILICON-BASED PHOTONIC QUANTUM DEVICE TOOLBOX

Silicon exhibits a strong third-order nonlinear effect and offers compact mode confinement. Leveraging semiconductor micro-nano processing technology, it is possible to achieve high-density integration of basic photonic quantum chip devices on-chip, including waveguides, beam splitters, couplers, and modulators. The key to harnessing these fundamental devices for large-scale quantum computing, quantum simulation, and other applications is the efficient generation, manipulation, and detection of photons on the silicon-based photonic quantum chip, which will be discussed in this section. To offer a more thorough review, we will also include similar materials with third-order nonlinearity, such as $SiO_2$ and SiN, in the silicon segment.

## 2.1 Quantum light sources

Quantum light sources are essential resources for a wide range of quantum-enhanced technologies and foundational tests of quantum mechanics. At the moment, high-quality photons are fabricated into two main categories: (i) probabilistic photon pair generation based on nonlinear parametric processes, including third-order ($\chi^{(3)}$) nonlinear processes known as spontaneous four-wave mixing (SFWM) and second-order ($\chi^{(2)}$) nonlinear processes known as spontaneous parametric down-conversion (SPDC), and (ii) deterministic single photon emission based on quantum dot (QD). SFWM/SPDC sources can maintain high indistinguishability between photons. However, these photons are generated probabilistically, and there is a trade-off between generation efficiency and multi-photon purity. QD sources, while deterministic, face challenges in maintaining the indistinguishability of photons from separate QDs.

### 2.1.1 SFWM photon pair sources in $\chi^{(3)}$ silicon

#### 2.1.1.a Characterization of SFWM photon pair sources

Silicon waveguides offer inherent advantages for integrated photonic quantum circuits through established foundry infrastructure. Their centrosymmetric crystalline structure precludes $\chi^{(2)}$ nonlinearity but enables relatively large $\chi^{(3)}$ effects via tight modal confinement, enabling efficient nonlinear photonic interaction of modes therein at moderate pump power. Thereby, in silicon, **SFWM sources** are widely adopted as heralded single-photon sources, entangled photon pairs, and squeezed light sources, utilizing the intrinsic $\chi^{(3)}$ nonlinearity of the material itself. During the SFWM process, two pump photons are annihilated through nonlinear interaction, producing a pair of signal and idler photons with related frequencies (**Fig. 3a**). Depending on the relationship between the signal and idler photons, the SFWM process can be further divided into two subcategories: the non-degenerate SFWM, where the two photons produced have different wavelengths, and the degenerate SFWM, where the two photons have identical wavelengths. The two processes can be expressed respectively as $2w_p = w_s + w_i$ (non-degenerate) and $w_{p1} + w_{p2} = 2w_{s/i}$ (degenerate). The distinction between these two types of SFWM is crucial for applications in quantum information processing, as it affects the properties of the generated photons, such as their spectral distinguishability and the feasibility of multiplexing. ***SPDC sources*** generate photons following $w_p = 2w_{s/i}$ (degenerate) and $w_p = w_s + w_i$ (non-degenerate). SPDC sources are challenging to generate in silicon as the material itself does not exhibit second-order nonlinearity, unless by employing $\chi^{(2)}$ waveguides like GaAs[57], AlN[58], and LN[59], which will be detailed in Section 5.

Using parametric sources, single photons can be generated in heralding: the detection of one photon signals the presence of its correlated partner. The produced state can be expressed in the photon number basis as

$$|\Psi\rangle = \sum_{n=0}^{\infty} \xi_n |n\rangle_s |n\rangle_i \quad (1)$$

where $|n\rangle$ is photon number state per mode, with $s$ and $i$ denoting the signal and idler modes, respectively. The probability of detecting $n$ photons at mode $s$ or mode $i$ are the same, which is expressed by the photonic number distribution $P_n = |\xi_n|^2$, which depends on the material nonlinearity and pump intensity. In practice, there is a certain probability of producing vacuum state and multiphoton states. To suppress multi-photon noise, the pump intensity should be sufficiently low to ensure that only the vacuum and single pair state are generated, while the vacuum state can be eliminated by the heralding process. This creates a trade-off between brightness and photon number impurity.

High spectral purity is also essential for interference-based photonic quantum circuits. Photon pairs generated via nonlinear parametric processes are typically correlated, projecting the heralded photon into a mixed state upon detection of its partner. It is known that the state describing a composition system is decomposed as

$$|\Psi\rangle = \sum_{i=1}^{n} \sqrt{\lambda_i} |u_i\rangle \otimes |v_i\rangle \quad (2)$$

where $\{|u_i\rangle\}$ and $\{|v_i\rangle\}$ are orthogonal basis states called Schmidt modes. The Schmidt coefficients $\lambda_i$ are the weights of each subsystem satisfying $\sum_i \lambda_i = 1$. The degree of correlation can be quantified by Schmidt number $K$ which is defined as $K = \frac{1}{\sum_{i=1}^{n} \lambda_I^2} \in [1, n]$. The photon purity $P$ of this states is defined as $P = \frac{1}{K} \in \left[\frac{1}{n}, 1\right]$, where $P = 1$ represents the Schmidt number $K = 1$, corresponding to a factorizable pure two-photon state[60]. Achieving this often requires spectral filtering to post-select specific Schmidt modes, and hence reducing brightness. If $P < 1$ is measured, it means the state also contains other degrees of entangled pairs. A maximally entangled state where $n \to \infty$, $P \to 0$, indicates that the state has almost no purity (maximally mixed) and is not suitable for a heralded single photon source. In the weak pump regime, the multi-photon probability is relatively low, and the photonic purity $P$ can be estimated as $P = g^{(2)}(0) - 1$, where $g^{(2)}(0)$ is the second-order correlation, which is experimentally measurable by

$$g^{(2)}(\Delta t) = \frac{P_{ss}(\Delta t)}{P_s P_s} \quad (3)$$

where $P_{ss}(\Delta t)$ are the probability of measuring coincidence counts at the delay time of $\Delta t$ and $P_s$ is the probability of measuring signal photon at the detector. With the heralded photon measured, the remaining photon state can be used as a single photon state and the purity of this heralded single photon is $g_H^{(2)}(\Delta t)$, which describes the quantity of single photon against multi-photon emission[60]

$$g_H^{(2)}(\Delta t) = \frac{P_{ssi}(\Delta t)}{P_{s1i}(\Delta t) P_{s2i}(\Delta t)} P_i \quad (4)$$

where $P_{s1i}(\Delta t)$ and $P_{s2i}(\Delta t)$ are the probaboilities of measuring coincidence count at the delay time and $P_i$ is the probability of measuring signal photon. The testing setup and measured results for $g^{(2)}$ and $g_H^{(2)}$ is shown in **Fig. 3b** and **Fig. 3c**, respectively. While $g_H^{(2)}(0) < 0.5$ is the benchmark for non-classical behaviour, a high-purity single-photon source strives for a value closer to 0. Another metric assessing the quality of heralded single photon source is the coincidence to accidental ratio (CAR, **Fig. 3d**), which estimates the noise of the measured photon counts. Coincidence counts between signal and idler photons from the same pair of photon generation (i.e., $R_{si}$) are desired counts while the spurious coincidence between time uncorrelated different pairs or other noises are called the accidental coincidence (i.e., $R_{ac}$). The CAR is defined as

$$CAR = \frac{R_{si} - R_{ac}}{R_{ac}} \tag{5}$$

The CAR value is strongly dependent on the pump power. At low pump power, the probability of multi-pair generation and noise is low, leading to a high CAR value. As the pump power increases, multi-pair generation events and various noise sources increase, decreasing the CAR value.

#### 2.1.1.b SFWM photon pair sources in silicon

SFWM sources have been widely demonstrated in Si[61–63], $SiO_2$[64], and $Si_3N_4$[65] waveguides. Common structures adopted include long straight waveguide structures arranged in spiral shapes[66] and microring resonator structures[61]. The spiral waveguides produce a broad photon spectrum, while the microring resonators, despite requiring more precise fabrication, take up less on-chip space and yield a narrower photon spectrum. For the potential in large-scale integrated photonic quantum computing, sources of single photons are expected to be highly indistinguishable and pure, and either near-deterministic or heralded with high efficiency. Multiple efforts have been reported to enhance the desired characteristics: In **Fig. 3e**, to suppress the strong spectral anticorrelations between emitted photons in silicon, a dual-mode pump-delayed excitation scheme has been demonstrated and achieves a near-ideal spontaneous photon source, by engineering the emission of spectrally pure photon pairs through inter-modal SFWM in low-loss multi-mode waveguides[66]. As one increases the pump power to enhance the generation rate, parasitic nonlinear effects in silicon, such as two-photon absorption (TPA) and free carrier absorption (FCA), may potentially damage the pair generation properties[67,68]. To address this inherent limitation, in the silicon microring resonator depicted in **Fig. 3f**, a reverse-biased p-i-n structure is proposed to mitigate the FCA effects, thereby achieving a doubled generation rate without compromising the high purity of the photon pairs[61].

The spectral purity of conventional MRR used for generating heralded single photons is theoretically bounded by 0.93, presenting an inherent trade-off between brightness and heralding efficiency[69]. At critical coupling, the heralding efficiency is capped at 50%. Strong over-coupling or under-coupling regimes can achieve near-unity efficiency, but at the cost of negligible photon rates[69]. To circumvent the trade-off, an advanced resonator source embedded in asymmetric MZIs is proposed (**Fig. 3g**), which enables the independent engineering of the linewidths of the pump and signal (idler) resonances, thus achieves a preparation heralding efficiency of 52.4% and spectral purity of 0.95 at relatively low pump power[70]. To generate spectrally unentangled photon pairs in MRRs, an alternative approach is to employ a phase shift between two identical, temporally delayed Gaussian pulses, which experimentally has achieved a spectral purity of 0.98[71]. Generally, since SFWM/SPDC sources produce photons non-deterministically, typically with a 5–10% probability, time-domain[72,73], wavelength[74], or spatial multiplexing techniques have been proposed to improve the heralded photon rate[75]. A bulk-optics multiplexed source achieves state-of-the-art performance: 66.7% fiber-coupled single-photon probability with ~90% indistinguishability[73]. SFWM photon sources are promising in generating on-chip heralded identical photons and are made into arrays, like the array of 18 SFWM photon sources in $SiO_2$ waveguides[64] (**Fig. 3h**) and 4 SFWM photon sources in Si microring resonators[76] (**Fig. 3i**).

Considering the physical limitations of TPA on silicon-based single-photon sources in the telecommunications band, alternative material platforms such as $Si_3N_4$[77–79] and high-index doped silica (Hydex)[80,81] exhibit certain advantages: Although their nonlinear indexes are smaller than that in silicon, their zero TPA characteristics support high-power pumping operations, while also offering lower propagation losses and an ultra-wide spectral transparency ranging from the ultraviolet to the mid-infrared. Resonators in $Si_3N_4$ and Hydex demonstrates proper integrated quantum light generation when the pump power is kept below the threshold of parametric oscillation. The $Si_3N_4$ platform realizes spectral translation from the visible to the telecommunications band through dispersion engineering, which can be used to construct interfaces for interconnecting multiple quantum systems[65]; Hydex, on the other hand, generates quantum frequency combs containing multi-photon entangled qubits, which can be directly applied in quantum communication and computation[82].

### 2.1.2 QD single photon sources

**QD sources** operate primarily on the emission of semiconductor materials. A pair of carriers, termed the exciton, is excited by the injected laser pulse in the QD. The decay of exciton emits a single photon via the spontaneous emission process. It is deterministic in nature, since each laser pulse should theoretically generate only one photon. However, when extracting photons from a deterministic source, there is often a nonnegligible loss. As the extraction loss increases, the deterministic source can, in practice, degenerate into a probabilistic one. Self-assembled InGaAs/GaAs QDs demonstrated record-high performance for single photon sources[83]. A near-optimal single-photon emission from a single QD was demonstrated using a resonant excitation technique[84]. A 99.1% (99.7%) single-photon purity, 66% (65%) extraction efficiency, and 98.5% (99.6%) photon indistinguishability have been achieved in InGaAs (GaAs) QDs[85,86]. However, these QDs require a critical working environment of ultra-low temperatures and high vacuum chambers. As illustrated in **Fig. 4a**, the QD is inserted in a cryostat typically operating between 4 and 30 K, and the pillar microcavity used to collect single photons, presenting ease of fibre coupling but difficulty in waveguide integration[83]. Unlike these micropillar-based QDs that emit photons out-of-plane (**Fig. 4a**), QDs in photonic-crystal waveguides allow near-unity preferential emission into the waveguide[87] (**Fig. 4b**). Maintaining the indistinguishability of photons emitted from separate QDs is challenging, and researchers employ active demultiplexing technologies to address this issue, aiming to treat a multiphoton source as if it were a single-photon source[88] (**Fig. 4c**).

While QDs can generate single photons, they are currently unable to produce other non-classical quantum states, such as squeezed states. Squeezed states, which have reduced uncertainty in one of the quadratures of the electromagnetic field at the expense of increased uncertainty in the other, are also a fundamental resource for quantum computing. They play a crucial role in continuous-variable one-way quantum computing and quantum error correction. Four-wave mixing or parametric down-conversion processes, however, are versatile in generating not only heralded single-photon sources and entangled photons but also squeezed quantum light (**Fig. 4d**). For example, in $\chi^{(3)}$ materials, microring optical parametric oscillator on silicon nitride (**Fig. 4e**)[89] and microring-based Kerr frequency combs on silica (**Fig. 4f**)[90] have been explored for squeezing generation, achieving 1.7 and 1.6 dB of squeezing, respectively. $\chi^{(2)}$ material can generate squeezed state more efficiently. In section 5, we will delve into the specifics of squeezed states generated on $\chi^{(2)}$ materials, accompanied by a comparison of their state-of-the-art advancements.

## 2.2 Quantum state manipulation

### 2.2.1 Basic components

A complex photonic quantum chip is composed of various building blocks, including waveguides, couplers, beam splitters, and modulators (**Fig. 5a**). Waveguides are the most fundamental components that are used to connect other components. The performance of waveguides is crucial for the quality of the entire chip. The propagation loss in silicon waveguides has been reduced to 2.7 dB/m by minimizing sidewall scattering[91]. Waveguides typically involve strip waveguides and ridge waveguides. Strip waveguides, with their simple and compact design and small bending radius, are ideal for fabricating passive components. Ridge waveguides, with larger lateral dimensions, are suitable for low-loss end-fire coupling, and are often utilized in active components. Photons transmitted through waveguides are routed, split, and combined in energy, wavelength, and polarization through beam splitters. Commonly utilized beam splitters include directional coupler, multimode interferometer (MMI), and Y-branch. MMI is widely adopted for its broad bandwidth and stability. Additionally, wavelength division multiplexing devices[92] and polarization controllers[93] are investigated based on beam splitters, enabling the manipulation of wavelength and polarization degrees of freedom.

Couplers are used to connect the chip to external devices. End-fire couplers and grating couplers are the most commonly adopted types. End-fire couplers optimize the adiabatic taper at the edge of the chip. It has the feature of polarization insensitive and large bandwidth. Cheben et al. achieved an ultra-low coupling loss of 0.32 dB per facet in end-fire couplers, with a polarization-independent coupling bandwidth exceeding 100 nm[94]. Grating couplers couple optical waves into the chip at a certain vertical angle. It offers the features of simple design, flexible layout, and large alignment tolerance. Ding et al. designed a chirped photonic crystal grating that achieved an ultra-low loss of 1.74 dB per facet[95]. The applications of grating couplers are also expanding; for instance, two-dimensional grating couplers can convert between polarization and path degrees of freedom[96].

Modulation on silicon photonic chips typically involves electro-optic (EO) modulators and thermo-optic (TO) modulators. The EO modulators make use of carrier-depletion EO effects that leverage plasmonic dispersion effects, enabling GHz modulation at low power consumption but high optical loss[97]. TO modulators utilize the resistive heating overlay on the waveguides[98], offer slow modulation speed (~MHz) due to the slow heat diffusion, but are known for their simplicity, compact size, and low optical loss[99], thus are the most commonly adopted on silicon chips. The power consumption of TO modulators typically ranges from 10 to 100 mW and can be further reduced to several mW by deep trenching[100] and waveguide coiling[101]. Additionally, MEMS-based modulators, which rely on the physical deformation of waveguides, exhibit high tuning efficiency but require complex fabrication processes[102]. The combination of modulators with beam splitters leads to the creation of Mach-Zehnder interferometers (MZIs), which are essential for chip programming and configuration on silicon photonic quantum chips and are critical for computational accuracy. Wilkes et al. designed a cascaded MZI that effectively compensates for fabrication errors, achieving an extinction ratio greater than 60 dB[103].

### 2.2.2 Photonic quantum operators

Integrated platforms have consistently excelled in precise and accurate control of photons. Based on advanced CMOS technology, thousands of components can now be integrated within a few $mm^2$. Today, many large-scale quantum photonic experiments have been reported, utilizing circuits that are tailored for specific applications and reconfigurable for diverse functionalities through software. In a reconfigurable photonic circuit, photons are directed and manipulated by the waveguide structures, whose connectivity is the key determinant of the feasible operations and configurations. The manipulation of photons encoded in their path degree of freedom has been extensively studied. The dominant device for manipulation is the MZI, which can be interpreted as a programmable $2 \times 2$ circuit, consisting of waveguides, 50:50 beam splitters (either MMI or directional coupler), and phase shifters. Here, we illustrate the photonic state path manipulation operators that are implemented on-chip, showcasing the sophisticated control mechanisms that are at the heart of integrated photonic chips.

First, the single-photon operators are illustrated in **Fig. 5b**. A common encoding strategy on photonic quantum circuits is dual-rail encoding, where the qubit is encoded on a dual rail of two parallel waveguides, representing two logical states $|0\rangle$ and $|1\rangle$:

$$|0\rangle \equiv |1\rangle_0|0\rangle_1 = a_0^\dagger|\mathrm{vac}\rangle \tag{6a}$$

$$|1\rangle \equiv |0\rangle_0|1\rangle_1 = a_1^\dagger|\mathrm{vac}\rangle \tag{6b}$$

where $a_k^\dagger$ is the photon creation operator for mode $k$ (where the mode corresponds to a specific waveguide), $|\mathrm{vac}\rangle$ represents the vacuum state which means no photons present. Operating on single-qubit state, the Pauli-X operators $\hat{X}$ is realized by a cross structure, which swaps the two logical states. The phase shifter operator $\hat{P}$ is induced by changing the refractive index of one arm, by TO effect. The operator $\hat{O}_{MZI}$ can reliably perform classical and quantum interference[104], in which the relative phase is induced by changing the refractive index of one arm, by TO effects. The MZI is composed of two beam splitters ($\hat{O}_{BS}$) and a phase shifter $\hat{P}$, which can be expressed as

$$\hat{O}_{BS} = \frac{1}{\sqrt{2}}\begin{bmatrix} 1 & i \\ i & 1 \end{bmatrix}, \hat{P} = \begin{bmatrix} e^{i\theta} & 0 \\ 0 & 1 \end{bmatrix} \tag{7}$$

$$\hat{O}_{MZI} = \hat{O}_{BS}\hat{P}\hat{O}_{BS} = ie^{i\frac{\theta}{2}}\begin{bmatrix} \sin\left(\frac{\theta}{2}\right) & \cos\left(\frac{\theta}{2}\right) \\ \cos\left(\frac{\theta}{2}\right) & -\sin\left(\frac{\theta}{2}\right) \end{bmatrix}. \tag{8}$$

A combination of one $\hat{O}_{\mathrm{MZI}}$ and an additional $\hat{P}$ enables an arbitrary $SU(2)$ unitary transformation, which allows arbitrary preparation, operation, and analysis of a path-encoded single qubit. When all elements are lossless, the circuits can be described by

$$\hat{O}_{SU(2)} = ie^{i\frac{\theta}{2}}\begin{bmatrix} e^{i\phi}\sin\left(\frac{\theta}{2}\right) & \cos\left(\frac{\theta}{2}\right) \\ e^{i\phi}\cos\left(\frac{\theta}{2}\right) & -\sin\left(\frac{\theta}{2}\right) \end{bmatrix} \tag{9}$$

which transforms the logical basis states in the following way

$$\{|0\rangle, |1\rangle\} \rightarrow \{e^{i\phi}\sin\left(\frac{\theta}{2}\right)|0\rangle + \cos\left(\frac{\theta}{2}\right)|1\rangle, e^{i\phi}\cos\left(\frac{\theta}{2}\right)|0\rangle - \sin\left(\frac{\theta}{2}\right)|1\rangle\} \tag{10}$$

forming an arbitrary orthonormal basis set, spanning the two-dimensional space. Projective measurements should perform the adjoint transformation $\hat{O}_{SU(2)}^\dagger$, projecting arbitrary bases back onto the computational basis so that the pattern of photon-detection events in each mode yields information about the probability distribution associated with measurements projected into that specific basis, requiring

$$\hat{O}_{SU(2)}^{\dagger}\left\{e^{i\phi}\sin\left(\frac{\theta}{2}\right)|0\rangle+\cos\left(\frac{\theta}{2}\right)|1\rangle, e^{i\phi}\cos\left(\frac{\theta}{2}\right)|0\rangle-\sin\left(\frac{\theta}{2}\right)|1\rangle\right\}$$

$$=\hat{O}_{SU(2)}^{\dagger}\hat{O}_{SU(2)}\{|0\rangle,|1\rangle\}=\{|0\rangle,|1\rangle\} \quad (11)$$

Projecting back into the computational basis can be realized by reserving the operation order and reversing the sign of the phase applied to the $|1\rangle$ mode.

Two-photon quantum interference occurs when two indistinguishable photons meet at an $\hat{O}_{\mathrm{MZI}}$, and photons at outputs present a superposition of bunching state and antibunching state, dependent on the phase in the MZI[105]. This is referred to the Hong-Ou-Mandle (HOM)-like or reverse-HOM interference[106]. Photonic two-qubit entangling is enabled by quantum interference. The circuit diagrams in **Fig. 5c** illustrate four types of two-qubit operations for path-encoded qubits: $\hat{O}_{\mathrm{CNOT}}$ presents the Knill-Laflamme-Milburn CNOT operation, consisting of several beam splitters with different reflectivity, allows the execution of a CNOT gate; similarly, $\hat{O}_{\mathrm{CZ}}$ is the Knill-Laflamme-Milburn CZ operation, which allows the generation of cluster state; $\hat{O}_{\mathrm{Bell}}$ and $\hat{O}_{\mathrm{Fusion}}$ are two bell analysis[76] operators which allow the entangling of two independent qubits or projecting an entangled state onto one of the Bell states. The non-interaction nature of single photons leads to non-deterministic two-qubit operations; thus, all of these gates have a successful probability, which is 1/9 for $\hat{O}_{CNOT}$ and $\hat{O}_{CZ}$, and 1/2 for $\hat{O}_{\mathrm{Bell}}$ and $\hat{O}_{\mathrm{Fusion}}$, respectively.

Notably, the two-photon quantum interference that occurs at a single MZI represents the most basic scenario and can be further generalized to $n$-photon quantum interference in a $N$-mode unitary circuit, as depicted in **Fig. 5d**. The $N$-mode circuit is composed of multiple MZI units. The $n^{\mathrm{th}}$ MZI, acting on modes $i$ and $j$, is represented by a transformation matrix $M_n$ – which is an $N$-dimensional identity matrix with its elements at positions $(i,i)$, $(i,j)$, $(j,i)$, and $(j,j)$ replaced by the submatrix $\hat{O}_{SU(2)}$. The unitary matrix of this $N$-mode photonic circuit $U_m$ can be represented as the product of MZI transform matrices in the designed order as $\hat{O}_{SU(N)}=\prod_n M_n$. A well-known example utilizing this generalized $\hat{O}_{SU(N)}$ is the boson sampling problem[107], which is believed to be an intractable task for classical computers but efficiently executable in a specific quantum machine. There are many architectures for the general linear optical circuit, some of the possible configurations include the Reck scheme[108] and Clements scheme[109].

We will illustrate **a computational example** in which a Bell state measurement (BSM) circuit is designed and executed using above-mentioned photonic quantum operators. BSM is crucial to quantum teleportation, which is a cornerstone of quantum information processing. In quantum teleportation, Alice and Bob pre-share an entangled Bell pair $(|00\rangle + |11\rangle)/\sqrt{2}$. This entanglement provides the quantum correlation that will let Bob's qubit become the carrier of the state to be teleported. Alice performs a BSM on (a) the qubit that holds the unknown state $|\psi\rangle$ she wants to teleport and (b) her half of the entangled pair. The BSM projects these two qubits onto one of the four Bell states $\{|\Phi^+\rangle, |\Phi^-\rangle, |\Psi^+\rangle, |\Psi^-\rangle\}$.The measurement collapses both qubits into a specific Bell state and simultaneously leaves Bob's qubit in a state that is related to $|\psi\rangle$ by a simple Pauli correction (I, X, Z, or Y). Alice then sends two classical bits indicating which Bell state she obtained. Bob uses those two bits to apply the corresponding Pauli operator to his qubit, after which it is in the exact original state $|\psi\rangle$—even though it was never physically transmitted. Thereby, an efficient BSM circuit that can unambiguously determine each of the four Bell states is required to quantum teleportation.

As known, any two-qubit pure state can be rotated into the Bell basis by applying a Hadamard gate followed by a CNOT gate. Intuitively, reversing the sequence implements the inverse rotation, converting Bell states back into computational-basis states. However, in linear optics, full Bell-state discrimination remains probabilistic; nevertheless, schemes that reliably identify two of the four Bell states are experimentally feasible with certain success probability. Here we present the derivation of BSM implementation using $\hat{O}_{\mathrm{Bell}}$ (**Fig. 6a**) as a case study in photonic quantum-circuit design. Consider first the evolution of the states $|\Phi^{\pm}\rangle$. The initial state can be written in terms of the creation operators for the photons on the vacuum as

$$\frac{1}{\sqrt{2}}\left(\hat{S}^{\dagger}_{\uparrow,1}\hat{S}^{\dagger}_{\uparrow,2} \pm \hat{S}^{\dagger}_{\downarrow,1}\hat{S}^{\dagger}_{\downarrow,2}\right)|\mathrm{vac}\rangle \tag{12}$$

where the notation $\hat{S}^\dagger_{\uparrow,1}$ ($\hat{S}^\dagger_{\downarrow,2}$) describes a photon arising in the zero $|0\rangle$ (one $|1\rangle$) mode of the *i*-th qubit. Swapping the waveguide modes applies the transformation $\hat{S}^\dagger_{\downarrow,1} \leftrightarrow \hat{S}^\dagger_{\uparrow,2}$, thus produces the state

$$\frac{1}{\sqrt{2}}\left(\hat{S}^\dagger_{\uparrow,1}\hat{S}^\dagger_{\downarrow,1} \pm \hat{S}^\dagger_{\uparrow,2}\hat{S}^\dagger_{\downarrow,2}\right)|\text{vac}\rangle \tag{13}$$

Next, each qubit interferes on a Hadamard operator, which gives the state

$$\frac{1}{2\sqrt{2}}\left(\left(\hat{S}^\dagger_{\uparrow,1} + \hat{S}^\dagger_{\downarrow,1}\right)\left(\hat{S}^\dagger_{\uparrow,1} - \hat{S}^\dagger_{\downarrow,1}\right) \pm \left(\hat{S}^\dagger_{\uparrow,2} + \hat{S}^\dagger_{\downarrow,2}\right)\left(\hat{S}^\dagger_{\uparrow,2} - \hat{S}^\dagger_{\downarrow,2}\right)\right)|\text{vac}\rangle \tag{14}$$

which, upon expansion, becomes

$$\frac{1}{2\sqrt{2}}\left(\hat{S}^\dagger_{\uparrow,1}\hat{S}^\dagger_{\uparrow,1} - \hat{S}^\dagger_{\downarrow,1}\hat{S}^\dagger_{\downarrow,1} \pm \hat{S}^\dagger_{\uparrow,2}\hat{S}^\dagger_{\uparrow,2} - \hat{S}^\dagger_{\downarrow,2}\hat{S}^\dagger_{\downarrow,2}\right)|\text{vac}\rangle \tag{15}$$

Finally, another waveguide swapping is applied:

$$\frac{1}{2\sqrt{2}}\left(\hat{S}^\dagger_{\uparrow,1}\hat{S}^\dagger_{\uparrow,1} - \hat{S}^\dagger_{\uparrow,2}\hat{S}^\dagger_{\uparrow,2} \pm \hat{S}^\dagger_{\downarrow,1}\hat{S}^\dagger_{\downarrow,1} - \hat{S}^\dagger_{\downarrow,2}\hat{S}^\dagger_{\downarrow,2}\right)|\text{vac}\rangle \tag{16}$$

Crucially, since we post-select on four-fold coincidences and the photons are indistinguishable, the single photons bunch completely at the output and therefore never register as coincident events; they remain invisible to our detectors (**Fig. 6b**). When we consider the other two Bell states

$$|\Psi^\pm\rangle = \frac{1}{\sqrt{2}}\left(\hat{S}^\dagger_{\uparrow,1}\hat{S}^\dagger_{\downarrow,2} \pm \hat{S}^\dagger_{\downarrow,1}\hat{S}^\dagger_{\uparrow,2}\right)|\text{vac}\rangle$$

Under the same transformation the two states become

$$|\Psi^\pm\rangle = \frac{1}{\sqrt{2}}\left(\hat{S}^\dagger_{\uparrow,1}\hat{S}^\dagger_{\downarrow,2} \pm \hat{S}^\dagger_{\uparrow,2}\hat{S}^\dagger_{\downarrow,1}\right)|\text{vac}\rangle$$

$$\Rightarrow \frac{1}{2\sqrt{2}}\left(\left(\hat{S}^\dagger_{\uparrow,1} + \hat{S}^\dagger_{\downarrow,1}\right)\left(\hat{S}^\dagger_{\uparrow,2} - \hat{S}^\dagger_{\downarrow,2}\right) \pm \left(\hat{S}^\dagger_{\uparrow,2} + \hat{S}^\dagger_{\downarrow,2}\right)\left(\hat{S}^\dagger_{\uparrow,1} - \hat{S}^\dagger_{\downarrow,1}\right)\right)|\text{vac}\rangle$$

$$\Rightarrow \frac{1}{\sqrt{2}}\left(\hat{S}^\dagger_{\uparrow,1}\hat{S}^\dagger_{\uparrow,2} - \hat{S}^\dagger_{\downarrow,1}\hat{S}^\dagger_{\downarrow,2}\right)|\text{vac}\rangle \text{ for } |\Psi^+\rangle$$

$$\frac{1}{\sqrt{2}}\left(\hat{S}^\dagger_{\downarrow,1}\hat{S}^\dagger_{\uparrow,2} - \hat{S}^\dagger_{\uparrow,1}\hat{S}^\dagger_{\downarrow,2}\right)|\text{vac}\rangle \text{ for } |\Psi^-\rangle$$

$$\Rightarrow \frac{1}{\sqrt{2}}\left(\hat{S}^\dagger_{\uparrow,1}\hat{S}^\dagger_{\downarrow,1} - \hat{S}^\dagger_{\uparrow,2}\hat{S}^\dagger_{\downarrow,2}\right)|\text{vac}\rangle \text{ for } |\Psi^+\rangle$$

$$\frac{1}{\sqrt{2}}\left(\hat{S}^\dagger_{\downarrow,1}\hat{S}^\dagger_{\uparrow,2} - \hat{S}^\dagger_{\uparrow,1}\hat{S}^\dagger_{\downarrow,2}\right)|\text{vac}\rangle \text{ for } |\Psi^-\rangle \tag{17}$$

Hence, the symmetric input $|\Psi^+\rangle$ yields coincidences in the same qubit but opposite modes, whereas the input $|\Psi^-\rangle$ yields coincidences in opposite qubits and opposite modes (**Fig. 6b**). Since the state is guaranteed to be one of the four Bell states—as in teleportation or entanglement swapping—we identify which one simply by checking which pair of modes fires together.

### 2.3 Single photon detector

Detecting photonic quantum states is fundamental for measuring the outcome of photonic quantum operations. While detectors are required to possess unity detection efficiency, zero dark counts, dead-time-free operation, and perfect temporal resolution, but practical detectors can only approximate these characteristics due to inherent physical limitations. Most demonstrations to date have relied on off-chip detectors coupled via optical fibers, introducing significant losses. Chip-integrated detectors offer to minimize the coupling loss and device footprint, thus are essential for scalable quantum information readout. Single-photon detectors have multiple technology pathways, including photo-multiplier tubes, single-photon avalanche photodiodes (SPADs) and superconducting nanowire single-photon detectors (SNSPDs).

SPADs operate by biasing a semiconductor junction above its breakdown voltage in Geiger mode, where absorption of a single photon triggers an avalanche current pulse that is electronically quenched to detect the photon arrival. There is a known trade-off in silicon SPADs, i.e., thick-junction devices have decent detection efficiency but poor timing jitter, while thin-junction devices have good timing jitter but poor efficiency. A prior solution is to use a resonant cavity to enhance the efficiency without increasing the thickness, but it induces a small bandwidth of around a few nanometres[110]. Ma et al.[111] simulated a nanostructured silicon SPAD design that achieved high detection efficiency and low timing jitter across a broad spectral range, making use of light trapping enhancement. Subsequently, Zhang et al.[112] demonstrated a light-trapping, thin-junction Si SPAD that breaks the trade-off, by diffracting the incident photons into the horizontal waveguide mode, thus significantly increasing the absorption length (**Fig. 7a**). The photon detection efficiency has a 2.5-fold improvement in the near-infrared regime, while the timing jitter is 25 ps. The potential of on-chip SPAD arrays with high peak photon detection efficiency and low crosstalk probability is demonstrated by Lubin et al.[113], showcasing their utility in quantum correlation measurements (**Fig. 7b**). Recently, waveguide-coupled germanium-on-silicon SPADs were demonstrated with a detection efficiency of 5.27% at 80 K[114] (**Fig. 7c**) and 38% at 125 K[115] (**Fig. 7d**), both at 1310 nm, showing a great promise for future room-temperature chip-based quantum photonic applications.

SNSPDs operate by maintaining a superconducting nanowire well below its critical temperature while biasing its current near the critical value. The absorption of even a single photon deposits sufficient energy to trigger a transition from the superconducting to the normal resistive state, generating a measurable voltage pulse. This pulse undergoes amplification, filtering, and electronic discrimination to register a photon detection event. For integration, SNSPDs have been fabricated using several superconducting materials including niobium nitride (NbN), niobium titanium nitride (NbTiN), tungsten silicide (WSi), and molybdenum silicide (MoSi), and have been patterned on GaAs[116], Si[117], and $Si_3N_4$[118] waveguides. Sprengers et al.[116] proposed the first waveguide-integrated SNSPD in GaAs platform (**Fig. 7e**). Pernice et al.[117] patterned sub-wavelength absorbing NbN nanowire atop a silicon waveguide and achieved a breakthrough of 91% detection efficiency, 18 ps jitter, and 50 Hz dark count (**Fig. 7f**). Specifically, Najafi et al.[118] demonstrated integration of SNSPDs with a photonic circuit of MZIs, achieving an average system detection efficiency beyond 10%. Instead of direct deposition, SNSPDs fabricated on $Si_3N_4$ membrane can be flexibly transferred to other substrates, thus enabling scalable integration (**Fig. 7g**). Gyger et. al.[119] demonstrated low-power micro-electromechanical reconfigurable photonics integrated with SNSPDs. The same NbTiN layer was utilized to construct the MEMS actuators, electrical connections, contact pads, and single-photon detectors (**Fig. 7h**). It showcased high-extinction ratio routing and a high-dynamic range for single-photon detection. Munzberg et al.[120] presented an integrated SNSPD using a photonic crystal cavity on silicon, achieving a detection efficiency of 70% at telecom wavelengths, a recovery time of 480 ps, and a vanishingly low dark count rate (**Fig. 7i**).

Moreover, photon number resolving (PNR) detectors are required in certain quantum protocols, like imaging and heralding single-photon sources. PNR detectors based on transition-edge sensors (TESs) have been evanescently integrated on $SiO_2$[121], with demonstrations of up to five-photon resolution (**Fig. 7j**). A series of integrated SNSPDs atop a GaAs waveguide allow the on-chip PNR detection of up to four photons[122] (**Fig. 7k**). Till now, few detectors can provide high-fidelity photon number resolution at few-photon levels. Cheng et. al.[123] made a huge advancement with an innovative on-chip detector that is capable of resolving up to 100 photons, by spatiotemporally multiplexing an array of superconducting nanowires along a single waveguide (**Fig. 7l**). This detector offers unparalleled photon number resolution and high-speed response, enabling the unveiling of quantum photon statistics of true thermal sources. Both TESs and SNSPDs have to operate at cryogenic temperature.

## 3 SILICON-BASED PHOTONIC QUANTUM COMPUTING

The integrated photonic quantum circuits enable the robust and precise control of photonic quantum information by generating, manipulating, and detecting photons on a single chip. This miniaturizes quantum-optical experiments into chip-scale waveguide circuits, offering an ideal platform for generalized quantum-interference measurements (Section 3.1). Specifically, silicon-integrated circuits encompasses three major paradigms for processing quantum information: gate-based quantum computing (Section 3.2), which constructs algorithms through sequences of discrete quantum logic gates, measurement-based quantum computing (Section 3.3), where computation emerges from adaptive measurements on entangled cluster states, and arbitrary unitary operator-based quantum computing (Section 3.4), realized through programmable linear optical networks for tasks like quantum simulation and boson sampling. While gate-based and measurement-based models constitute universal quantum computing frameworks, unitary operator implementations—though operating as quantum simulators—share the same integrated photonic hardware architecture, thus are included here. The silicon photonics platform accommodates all three approaches by enabling high-fidelity reconfigurable components while leveraging CMOS compatibility for scalable quantum information processing.

### 3.1 Quantum interference

It is essential to distinguish between quantum interference from classical interference. A critical milestone for large-scale quantum technologies a demonstration that two individual photon sources within a waveguide circuit can exhibit high-visibility quantum interference between them. As shown in **Fig. 8a**, they integrated two SFWM sources in an interferometer with a reconfigurable phase shifter. The circuit generate and manipulate both non-degenerate and degenerate photon pairs. The input pump was split by the first MMI coupler between the two sources. By operating in the weak pump regime to ensure single-pair generation, simultaneous pumping yielded the path entangled N00N state. A thermal phase shifter dynamically controlled the relative phase in one arm, applying a shift $\phi$ to the pump and a $2\phi$ shift to the entangled biphoton state. **Figure 8b** showing the separation of signal, idler, and pump wavelength cannels using WDMs. For non-degenerate SFWM sources, waveguide modes are labelled A and B, with creation operators $a_s^\dagger$, $a_i^\dagger$, $b_s^\dagger$, $b_i^\dagger$. The intial state is:

$$|\psi\rangle = \left(e^{2i\phi} a_s^\dagger a_i^\dagger + b_s^\dagger b_i^\dagger\right)|0\rangle \tag{18}$$

After MZI transformation:

$$|\psi\rangle = \cos\phi|\psi\rangle_{bunch} + \sin\phi|\psi\rangle_{split} \tag{19}$$

where $|\psi\rangle_{bunch} = e^{i\phi}\left(a_s^\dagger a_i^\dagger + b_s^\dagger b_i^\dagger\right)|0\rangle$(both photons in one mode), $|\psi\rangle_{split} = ie^{i\phi}\left(a_s^\dagger b_i^\dagger - b_s^\dagger a_i^\dagger\right)|0\rangle$ (photons in different modes). Classical interference was measured (**Fig. 8c**) for modes A/B with the phase $\phi$ swept from $\pi/2$ to $5\pi/2$, showing $2\pi$-periodic power modulation.

For quantum interference (**Fig. 8d**), the two-photon coincidence count rates for $A_sA_i$ and $B_sB_i$ represents the probability of measuring the $|\psi\rangle_{bunch}$ state, varying as $|\cos\phi|^2 = \frac{1}{2}(1 + \cos2\phi)$, yielding $\pi$-periodic fringes. The measured coincidence rates exhibit different maximum values between fringes. The asymmetry arises from spurious photons generated near the bright pump before entering the sources. These photons produce $2\pi$-periodic classical interference; their peak contribution increases coincidences in peaks without reducing valleys. Due to the short input waveguide relative to spiral sources, peak differences remain modest. **Figure 8e** shows coincidences for $A_sB_i$ and $A_iB_s$, varying as $|\sin\phi|^2 = \frac{1}{2}(1 - \cos2\phi)$, which is $\pi$-periodic. Spurious photons do not affect count rates because at $\phi = \pi$ and $\phi = 2\pi$ points, they are all interfered into a single port, preventing coincidences. For degenerate SFWM sources (**Fig. 8f**), signal and idler wavelengths are identical and indistinguishable by WDMs. The evolution process of non-degenerated source can be directly applied here by mixing the signal and idler creation operator, thus only the $|\psi\rangle_{split}$ state can be observable. In short, interference between the pump and biphoton states at the second MMI coupler produced

Mach-Zehnder interference fringes in the pump transmission (period $2\pi$) and half-period fringes (period $\pi$) in the biphoton coincidence statistics.

**Figure 8g** depicts the off-chip measurement setup for photon pairs in the $|\psi\rangle_{split}$ state. After exiting the waveguide, one photon undergoes a controlled path delay, while the other is polarization matched. The pair is then interfered at a beam splitter. The phase-stable, two-color $|\psi\rangle_{split}$ in non-degenerate scheme yielded Hong-Ou-Mandel (HOM) interference fringes, exhibiting spectral beating between the non-degenerate signal and idler wavelengths (**Fig. 8h**). Due to color entanglement in $|\psi\rangle_{split}$ state, the filtered biphoton spectrum has two lobes, one lob from each of the signal-and idler-channel filters. The coincidence probability after interference is given by

$$p_{HOM} = \frac{1}{2} - \frac{V}{2}\cos\left(2\pi x \frac{\delta}{\lambda_p^2}\right)\sin\left(2\pi x \frac{\omega}{\lambda_p^2}\right) \tag{20}$$

where $x$ is the dealy displacement, $\lambda_p$ is the pump wavelength, $\omega$ is the WDM channel width and $\delta$ is the singal-idler channel spacing. For degenerate sources, this corresponds to off-chip HOM interference (**Fig. 8i**), where δ = 0 eliminates oscillations per Eq. (19).

### 3.2 Gate-based quantum computing

Quantum circuit model (i.e., quantum gate-based model) is a fundamental framework that employs a sequence of quantum gates acting on qubits to describe quantum computing. In this model, a quantum circuit typically starts with an initialization phase where qubits are prepared in a known state, often the ground state. Following this, a series of logical quantum gates, such as Hadamard (H), CNOT, and various single-qubit rotations, are applied to perform computations. In this pathway, the initial states of the qubits are independent, and entanglement is generated through the application of quantum gates. On-chip encoding of single-photon quantum states can be achieved through superposition of waveguide paths, where a photon simultaneously occupies two or more paths to form a quNit ($N \geq 2$). As previously noted, high-fidelity single-qubit operations are realized via waveguides, beam splitters, and phase shifters on a single chip. Implementing two-qubit operations is more challenging, as they require conditional operations on one qubit based on the state of another. Such nonlinear optical interactions are inherently weak at the single-photon level. Usually, nonlinear operations are probabilistically simulated using linear optics and measurements, achieving effective quantum gates through appropriate post-selection or heralding signals.

Enhancing the computational potential of a gate-based quantum system is an ongoing pursuit. Given a system with $n$ photons, each with $m$ dimensions, the system capacity is $m^n$. Therefore, there are intuitively two ways to enhance the system capacity: increasing the number of photons or enlarging the dimensions. We review these two strategies to uncover their respective developments.

### 3.2.1 Multi-photon gate-based

Due to the difficulty of maintaining indistinguishability and chip integration of QD sources on silicon, the development of on-chip photon generation has primarily focused on SFWM sources. Initially, entangled photon pairs are generated in silicon waveguides and microring resonators with different degrees of freedom[124–127], and are used to demonstrate various photonic quantum applications[128–132]. The milestone in 2011, as referenced[133], showcased quantum interference with two heralded photons from independent silicon waveguides, achieving a visibility of 73%. This two-photon interference process was then demonstrated on one single chip, with a visibility of 72% measured from two independent microring resonator sources[134]. Following that, by integrating the microring photon pair sources, spectral demultiplexers, and reconfigurable optics, a path-entangled two-qubit state was generated and validated[127] (**Fig. 9a**). Subsequently, complex four-photon states[135] and frequency-degenerate entangled four-photon states[136] are also generated in silicon waveguides. The record for the number of photons on a single silicon chip has reached eight[137], and optimistically with continued reduction in losses, even more photons could be achieved.

These fully-integrated photon-pair/heralded single-photon sources have driven extensive research into multi-photon-based applications on integrated photonic chips, including the creation of complicated photon states and the implementation of quantum information processing algorithms. On the other hand, a gate-based quantum model requires a diverse set of quantum logical gates, which will be combined to create sophisticated algorithms. Translating this model into a physical chip involves establishing a correspondence between the required optical circuitry and each quantum gate. Early works have been dedicated to this area. Single-qubit gates are straightforward to map onto the rotations of programmable photonic MZIs. The MZIs can be adjusted to perform precise phase shifts on single photons, effectively rotating the quantum state within the Bloch sphere. As the representative two-qubit gate, the CNOT gate was first demonstrated in a $SiO_2$ chip, employing an unheralded CNOT scheme that does not include the ancillary photons required for scaling[138]. This kind of CNOT implementation showcased a rudimentary version of Shor's factoring algorithm using two CNOT gates, successfully factorizing the number 15[139]. The same type of CNOT gate was also demonstrated with three photonic qubits, enabling a quantum teleportation demonstration, achieving single-chip teleportation of single-qubit states with an average fidelity of 89 ± 3%[140] (**Fig. 9b**). Subsequently, universal linear optical circuits capable of implementing all possible photonic quantum gates have been proposed and experimentally realized in a single programmable photonic circuit[141] (**Fig. 9c**). The circuit was fabricated in a single $SiO_2$ chip, comprising a six-mode triangularly arranged MZI network. It demonstrated a variety of single-/two-qubit gates, including the integrated heralded CNOT gate and six-dimensional complex Hadamard operations. Following this, universal linear optical circuits have also been demonstrated in $Si/Si_3N_4$ chips[76,142].

Quantum state teleportation is a vital element of photonic quantum computing, designed to transmit a particle's quantum state to another particle, rather than the particle itself. This method of communication, harnessing quantum entanglement to transmit quantum information encoded in states, lays the groundwork for scalable quantum networks and distributed quantum computing. The first quantum teleportation employed silica slab waveguides[140]. Later, teleportation between two silicon chips was achieved, utilizing four microring resonators to produce high-quality entangled states[76] (**Fig. 10a**). In this work, photons embodying the quantum states were sent to another chip through a quantum photonic interconnect that converted path to polarization degree of freedom, followed by reconstruction of the states via tomography. Furthermore, the chip-to-chip teleportation has been advanced for the resource-efficient transmission of high-dimensional quantum states[143], which involves employing an unsupervised, trained quantum autoencoder to compress the input state, which is subsequently decompressed at the receiver chip before state reconstruction (**Fig. 10b**).

Translating circuit models into physical gates on photonic chips presents inherent bottleneck due to the probabilistic nature of photonic quantum gates. The success probability diminishes exponentially, and the propagation of unwanted states to subsequent gates renders direct cascading impractical. Nonetheless, the advent of re-programmable photonic chips facilitates the processing of a multitude of quantum tasks and algorithms on a single chip, especially the variational learning tasks. As depicted in **Fig. 10c**, a variational eigenvalue solver is implemented on a photonic quantum processor[144]. The processor, featuring a CNOT gate and reprogrammable single-qubit gates, efficiently computes the Hamiltonian's expectation value, offering a significant speed advantage over traditional exact diagonalization methods. Coupled with a classical optimization algorithm, the system experimentally determined the ground-state molecular energy of $He–H^+$, and markedly optimizes quantum resource utilization. Similarly, an experimental quantum Hamiltonian learning was demonstrated, where a silicon-photonics quantum simulator is interfaced with a diamond nitrogen-vacancy center's electron spin[130]. Employing Bayesian inference, the simulator accurately deduces the Hamiltonian, precisely pinpointing the Rabi frequency with an uncertainty of about $10^{-5}$. Being interactive, the protocol can effectively characterize the operations of the quantum photonic device.

### 3.2.2 Multi-dimensional gate-based

High-dimensional encoding is another feasible approach to a larger system capacity. Besides, the strategy also shows noise robustness in quantum communications and higher efficiency and flexibility in quantum computing. On integrated chips, path encoding is the prevalent technique for high-dimensional encoding (or qudit encoding) due to its straightforward implementation. Utilizing on-chip beam splitters and MZIs, photons can be routed and manipulated across various paths.

In the generic qudit-on-chip architecture, $d$ identical SFWM sources are coherently pumped in parallel, producing a photon pair in superposition across the entire source array. Because both photons originate from the same spontaneous process, the generated bipartite state reads $\sum_{k=0}^{d-1} c_k |1\rangle_{i,k} |1\rangle_{s,k}$, where $|1\rangle_{i,k}$ and $|1\rangle_{s,k}$ respectively indicate the Fock state of the idler and signal photon in the $k$th spatial mode, and $c_k$ represents the complex amplitude in each mode (with $\sum|c_k|^2 = 1$). We identify the spatial mode index $k$ with the logical basis: a photon occupying mode $k$ is taken to represent the qudit state $|k\rangle$ $(k = 0, \dots, d-1)$. The global state therefore maps to the maximally entangled $d$-level two-qudit state

$$|\psi\rangle_d = \sum_{k=0}^{d-1} c_k |k\rangle_i |k\rangle_s \tag{21}$$

whose coefficient can be chosen at will by shaping the pump distribution across the $d$ sources and by tunning the relative phase of each mode with on-chip MZI networks. Uniform pump and equal phases yield the maximally entangled state $|\psi_d^+\rangle = \frac{1}{\sqrt{d}} \sum_{k=0}^{d-1} |k\rangle_i |k\rangle_s$. Asymmetric MZI filters then deterministically separate the non-degenerate photons, and a network of waveguide crossings routes all signal photons to the top modes and the idler photons to the bottom modes, enabling local manipulation and measurement of each qudit. Any unitary transformation $U_d$ on a single $d$-level system can subsequently be implemented with linear-optical elements.

Following the theoretical framework, experimental demonstrations have been carried out for dimensions ranging from $d = 3$ to $d = 15$. **Figure 11a** shows the generation, manipulation, and analysis of 3D path-entangled qutrit states using a silicon photonic chip[145]. They achieved high-quality quantum interference with visibilities exceeding 96.5% and the creation of a maximally entangled-qutrit state with a fidelity of 95.5%. The entangled qutrits are used for graph-related quantum simulations and high-precision optical phase measurements. Additionally, Qiang et al. constructed two pre-entangled ququarts, which is a four-dimensional system, to implement arbitrary two-qubit quantum operations, including entangling operations[131] (**Fig. 11b**). They achieved this by shifting from a pure circuit model to realizing arbitrary two-qubit unitary operations through the linear combination of four unitaries. A total of 98 different two-qubit unitary operations were executed with a high fidelity of 93%, and a quantum approximate optimization algorithm is implemented for solving combinatorial problems and simulating quantum walks. Later, a significant leap is noticed in the development of a multidimensional integrated photonic quantum chip, capable of generating high-dimensional entanglement with up to 15 dimensions[146] (**Fig. 11c**). The chip integrates over 550 photonic components, including 16 identical photon-pair sources. It leverages the natural encoding and processing capabilities of photons in various path degrees of freedom to achieve high precision and controllability of multidimensional quantum systems. It showcased applications such as quantum randomness expansion and self-testing on multidimensional states. The group was subsequently devoted to the development of qudit-based processors, implementing arbitrary single-qudit operations and two-qudit multivalued controlled operations for the quantum Fourier transform algorithm[136] (**Fig. 11d**), and demonstrated a generalized multipath delayed-choice experiment that demonstrates wave-particle duality with up to 8-path[137].

Regardless of the path taken, exploiting additional degrees of freedom can also enhance system capacity, such as frequency[80,149–151], transverse mode[152], and time bins[153]. In frequency-bin encoding, quantum information is encoded by the photon being in a superposition of different frequency bands. Commonly, frequency bin is generated by SFWM in single ring resonator, and manipulated by phase modulators. Since the efficiency of SFWM scales quadratically with the free spectral range (FSR) of the resonator[154], there is a trade-off between the photon generation rate and the number of accessible frequency bins: higher FSR results in higher generation rate but fewer accessible frequency bins. Clementi et al.[150] overcome this limitation by direct, on-chip controlling the interference of biphoton amplitudes generated in multiple ring resonators that are coherently pumped (**Fig. 12a**). The frequency-bin spacing is no longer related to the ring radius, thus enabling very high-finesse resonators, while reaching megahertz generation rates. Further, to remove the reliance on external bulky excitation lasers, Mahmudlu et al. demonstrated a prototype chip that hybrid integrated electrically pumped InP gain section and $Si_3N_4$ microring filter system (**Fig. 12b**), resulting in a turnkey quantum light source, and emitting frequency-bin entangled qubit and qudit states[151]. Within polarization-bin and time-bin, a dual-polarization 10-channel mode (de)multiplexer has been implemented using cascaded dual-core adiabatic tapers on silicon, showcasing transverse-mode expanding[152] (**Fig. 12c**), and quantum interference from pulsed time-bin entanglement within a silicon ring resonator has been successfully demonstrated[153] (**Fig. 12d**).

The advancement of on-chip conversion devices is expected to broaden the application potential of these degrees of freedom significantly. For instance, two-dimensional (2D) gratings and polarization rotator splitters facilitate the conversion of path-encoded entanglement and transverse-mode encoded entanglement in chip-to-chip teleportation[76,143].

The connection of quantum information between high-dimensional chips has also been developed via a multi-core optical fiber and an efficient multi-core fiber coupler[155–157].

### 3.3 Measurement-based quantum computing

#### 3.3.1 Proof-of-principle demonstration

In the gate-based quantum computing model, interactions between photons are realized by incorporating ancillary circuitry and photons. The partial state collapse, which results from detecting ancillary photons in ancillary modes, executes the necessary quantum logic. However, the overheads associated with this scheme are prohibitive at a practical level. Measurement-based quantum computing (MBQC), also referred to as one-way computing, presents a significantly more resource-efficient model for quantum computation[158–160]. In this approach, the challenge of requiring deterministic gates is transformed to constructing a generic entangled cluster state, upon which any quantum computation can be implemented through a sequence of measurements. Universal quantum computation is proven possible using MBQC protocol on sufficiently connected graph states, notably cluster states. **Figure 13** depicts the comparison of a gate-based quantum circuits and MBQC protocol.

The principle of MBQC consists of three key stages: preparation of an entangled resource state, execution of single-qubit projection measurements, and application of correction operations based on measurement outcomes. The resource state is a large-sized quantum state with a special entanglement structure and is generally constructed by superposing and entangling many single-qubit states. A common type of resource state has a structure consisting of vertices and edges just like graph and is therefore called a graph state. The corresponding mathematical representation is given by

$$|G\rangle = \prod_{(i,j)\in E} CZ_{ij} \, |+\rangle^{\otimes n}, \; |+\rangle = \frac{|0\rangle + |1\rangle}{\sqrt{2}} \tag{22}$$

where CZ denotes the controlled-Z operation. Two-qubit graph state constructed according to Eq. (7) is Bell state, and three-qubit graph state is Greenberger-Horne-Zeilinger (GHZ) state. After preparing the resource state, the next step is to perform projection measurement on individual qubit in the order required by the targeting algorithm. Each qubit is measured in a basis expressed by

$$M_k(\theta, \phi) = \cos\theta \cdot X_k + e^{i\phi} \sin\theta \cdot Y_k \quad (23)$$

where the measurement angle $(\theta, \phi)$ for qubit $k$ is dynamically adjusted based on prior outcomes. This measurement-driven process teleports quantum information through the graph. The final computational result is decoded by applying Pauli corrections $X^{C_x} Z^{C_z}$ to the output qubits, where $C_x$ and $C_z$ depend on the parity of all measurement outcomes, preserving determinism despite inherent quantum randomness.

There have been several demonstrations of MBQC on silicon chips. In 2005, Walther et al. experimentally demonstrated the feasibility of MBQC using bulk optics. They encoded four-qubit cluster states into the polarization state of four photons and then demonstrated MBQC through a universal set of one- and two-qubit operations[161]. Grover's search algorithm was implemented to illustrate that MBQC is ideally suited for such tasks. Then, progress has been achieved on chip. In **Fig. 14a**, graph states, the key entangled resources for MBQC, were reported[162]. The authors produced four on-chip-generated photons to construct the star and line type of four-photon graph states. The stabilizers of the star and line states were characterized, and the multipartite nonlocality was tested. Based on the graph-state, researchers demonstrated error-protected qubits by an integrated silicon photonic scheme that entangles multiple photons and encodes multiple physical qubits on individual photons[163] (**Fig. 14b**). They compared the running of the phase-estimation algorithm without and with error correction and observed an increase in the success rate from 62.5% to 95.8%. Since the entanglement structure and connectivity of entangled qubits are crucial for the execution of quantum algorithms in graph-state computing, hypergraph entanglement has been developed[164] (**Fig. 14c**). Unlike graph-type entanglement, where only the nearest qubits interact, hypergraph entanglement allows any subset of qubits to be arbitrarily entangled through hyperedges. A basic measurement-based protocol and efficient resource state verification using color-encoded stabilizers have been implemented with local Pauli measurements. **Figure 14d** shows an experiment on a silicon photonic chip that involves a five-qubit linear cluster state to implement the quantum error correction code, and demonstrates its capability of identifying and correcting a single-qubit error[165]. The encoded quantum information is reconstructed from the single-qubit error with an average fidelity of 86.3 ± 3.2 %. The scheme is extended for a fault tolerant MBQC on stabilizer formalism that allows the redo of qubit operation against the failure of the teleportation process.

Recent advancements also combined high-dimensional quantum states with MBQC, for example, using orthogonal, maximally entangled *d*-level two-partite states to construct cluster states through projection measurement, and then demonstrating proof-of-concept high-dimensional one-way quantum operations[81]. More recently, a large high-dimensional cluster state has been generated, encoding multiple qubits on each photon through high-dimensional spatial encoding, controlled via a spatial light modulator (**Fig. 14f**). It generates cluster states with over nine qubits at a rate of 100 Hz[166].

### 3.3.2 Discrete-variable MBQC vs. Continuous-variable MBQC

Most of the aforementioned MBQC demonstrations utilize entangled single-photon sources and are categorized as discrete-variable (DV) approaches. MBQC can also be implemented with continuous variables (CV), where cluster states are generated from squeezed states. There are some related works that contribute to a broader understanding of MBQC, which is considered a highly promising strategy for the foreseeable future, although most of these works have not yet been demonstrated on chips, including the generation of squeezed states[89,90], which have reduced quantum noise below the standard quantum limit, the generation of large-scale cluster states from the squeezed states[167,168], and the measurement strategy to execute quantum computing algorithms[159,169].

While DV-MBQC and CV-MBQC share the same core paradigm, i.e., leveraging highly entangled resource states (graph states or cluster states) and performing adaptive measurements on their quantum units to drive the computation, bypassing the execution of traditional gate sequences – they differ significantly in physical implementation, information encoding, resource state properties, and computational capabilities. **Table 1** provides a detailed comparison of the operational principles underlying continuous-variable and discrete-variable MBQC. The most fundamental distinction lies in the information carrier and the dimension of the Hilbert space. DV-MBQC uses quantum bits (qubits) as information carriers, whose states reside in a two-dimensional Hilbert space (e.g., photon polarization states $|H\rangle$, $|V\rangle$ or electron spin states $|\uparrow\rangle$, $|\downarrow\rangle$). Conversely, CV-MBQC uses quantum modes (qumodes) as carriers. Information is encoded in the orthogonal quadrature of the optical-field, i.e., position (Q) and momentum (P), which are operators of continuous observables. Its Hilbert space is infinite-dimensional.

The nature and preparation of the resource states also differ. DV-MBQC relies on discrete cluster states/graph states, typically highly entangled multi-qubit states formed on a lattice. These states are inherently non-Gaussian (referring to the complexity of their entanglement structure, even though local measurement bases are discrete). Preparing such discrete cluster states requires precise qubit control and high-fidelity two-qubit entangling gates. CV-MBQC, however, employs continuous-variable cluster states, typically prepared by feeding multiple squeezed-state light fields into a designed linear optical network. Ideally, such cluster states, produced from Gaussian sources and linear operations, are Gaussian states. Their Wigner function is a Gaussian distribution, and entanglement is generated solely by quadratic Hamiltonians. Optical techniques enable the efficient theoretical generation of CV cluster states with specific topologies, like linear clusters and square lattices.

The adaptive measurement, which is the core operation driving the computation, also differs significantly. In DV-MBQC, computation is driven by discrete projective measurements performed on individual qubits in chosen bases (typically the Pauli bases {X,Y,Z} or superpositions thereof), yielding discrete outcomes. The sequency of measurements and the choice of basis (adaptively adjusted based on prior measurement results) collectively encode the logic of the simulated quantum circuit. In CV-MBQC, the core operation is homodyne detection: by adjusting the phase $\theta$ of a local oscillator, a specific quadrature value of a single qumode (i.e., $Q_\theta = Q\cos\theta + P\sin\theta$) is measured, yielding a continuous real number outcome. Computation is similarly driven by sequentially performed homodyne measurements, and the choice of the measurement angle $\theta$ also requires complex adaptive adjustment based on prior results. Processing and feeding back these continuous measurement results is inherently more complex than dealing with discrete outcomes.

A crucial distinction concerns computational universality. DV-MBQC is proven to be a universal model of quantum computation in theory when combined with adaptive measurements on a sufficiently large cluster state, it can simulate any quantum algorithm. This universality stems from the discrete cluster state itself, which inherently embodies the non-classical resources required for universal quantum computing. However, for CV-MBQC, if restricted solely to Gaussian operations, i.e., using Gaussian resource states, Gaussian measurements, and adaptive feedback processes, it can only simulate Gaussian quantum circuits which composed of linear optical operators and some quadratic gates. While CV-MBQC is efficient for specific tasks, it cannot achieve universal quantum computation without the non-Gaussian operations. Universal CV computing must introduce non-Gaussian gates like cat states into the resource state, create non-Gaussian entanglement during the resource state preparation, and perform difficult non-Gaussian measurement like photon counting. Utilizing quantum error correction based on non-Gaussian states (e.g., Gottesman-Kitaev-Preskill, GKP code states) is also desired. Generating, manipulating, and maintaining high-quality non-Gaussian states remains the most significant experimental challenge in CV quantum computing.

Fault-tolerant strategies also diverge considerably. DV-MBQC can relatively directly apply well-developed discrete quantum error correction (QEC) codes (e.g., surface codes, color codes) to its cluster state framework. Error models like bit-flip and phase-flip are clearly defined. In CV-MBQC, continuous-variable quantum error correction is far more challenging. The issue lies in quantum information being distributed across a continuous phase space, with errors also being continuous and correlated. A core strategy involves using the GKP encoding scheme: encoding a quasi-discrete logical bit of information into the periodic wavefunction of a physical qumode, requiring infinite squeezing ("infinitely sharp peaks"). However, achieving high-fidelity preparation and maintenance of such states experimentally is extremely difficult. Consequently, while fault-tolerance theory for DV is relatively mature, the fault-tolerant path for CV, although theoretically viable, faces immense experimental hurdles.

The advantage of DV-MBQC lies in the discreteness of its information processing, which affords conceptual clarity, a complete proof of universality, and a relatively mature error correction theory, making it the primary pathway towards large-scale fault-tolerant quantum computing. CV-MBQC, while demonstrating potential scalability advantages in preparing Gaussian resource states on optical platforms, faces the fundamental obstacle of lacking efficient and controllable non-Gaussian operations, which presents a key barrier to achieving universal quantum computation with continuous variables.

### 3.3.3 Pathway to universal quantum computing

Million-qubit universal quantum computing is desired. However, the core bottleneck presents: although photonic quantum computing offers advantages of room-temperature operation, its quantum state is vulnerable to optical loss (single-photon fiber loss ~0.2 dB/km). Traditional linear optical schemes rely on probabilistic operations, causing computational success rates to decay exponentially with scale. Early photonic architectures of measurement-based quantum computing require pre-generated large-scale entangled states, but error accumulation during this process proves difficult to suppress, while single-photon encodings (e.g., dual-rail) lack intrinsic error correction capabilities. PsiQuantum and Xanadu both aims to address the challenge of achieving scalable, fault-tolerant quantum computation on photonic platforms. Their shared goal is to build a "quantum transistor" for photonics – transforming fragile physical qubits into stable logical qubits through fault-tolerant mechanisms while leveraging semiconductor manufacturing for scalability. The potential breakthrough will enable quantum computing to transition from laboratory demonstrations to solving practical problems (e.g., materials design, cryptanalysis), unleashing quantum advantage.

The two companies adopt distinct technical approaches: PsiQuantum proposes the Fusion-based quantum computation (FBQC) model, which features a DV-MBQC architecture[167,168]. This design utilizes entangled resource states (i.e., 6-qubit ring states) and fusion measurements to achieve fault tolerance. The architecture comprises three core units: (1) Resource state generators produce constant-size entangled states; (2) Fusion routers connect qubits via fixed routing; (3) Fusion measurement units perform projective entangling measurements. This design dramatically reduces operational depth – qubits are measured immediately after generation, significantly minimizing optical loss accumulation. Fault tolerance relies on syndrome graph redundancy: Fusion measurement outcomes simultaneously construct entanglement networks and error detection frameworks. Through XX/ZZ measurement combinations, a cubit lattice syndrome graph is formed (**Fig. 15a**). When combined with minimum-weight perfect matching decoders, this achieves an 11.98% fusion erasure threshold – representing a 73% improvement over traditional 4-star schemes (**Fig. 15b**). Based on the integration of fixed routing and (2,2)-Shor encoding, the photon loss tolerance is increased to 10.4% (**Fig. 15c**).

PsiQuantum has demonstrated fully integrated quantum modules (**Fig. 15d**) using 300-mm silicon photonic chips (GlobalFoundries process)[170], achieving 99.98% single-qubit SPAM fidelity and 99.22% two-qubit fusion fidelity. The scalability bottleneck of PsiQuantum at this stage is that classical error correction codes (e.g., surface code) demand physical error rates below thresholds (typically $10^{-2}$-$10^{-3}$), but current photonic components fall short: PNRD efficiency must exceed 99% (99.89%), and optical path transmission must exceed 99.5%.

Xanadu adopts a CV-MBQC path, which converses gaussian state to logical qubits and centers around GKP encoding[171]. The architecture is divided into three stages (**Fig. 15e**): (1) Gaussian boson sapling sources generate squeezed states; (2) Refineries enhance state quality through multiplexing and adaptive interference; (3) Quantum processing units (QPUs) synthesize GKP Bell pairs into cluster states. They recently achieve a core breakthrough in GKP logical qubit synthesis: using four-mode GBS chips combined with ultra-high-efficiency photon-number-resolving detectors (PNRDs, 99.89% efficiency), Wigner-negative GKP states are prepared on integrated photonic circuits (**Fig. 15f**), demonstrating 3 × 3 lattice structures and 9.75 dB symmetric effective squeezing (**Fig. 15g**), marking the first optical-platform validation of logical qubit core features. Xanadu's system integration milestone is enabled by the customized silicon nitride chips[172]. The resulted Aurora system demonstrates GKP state generation, with four-peak position/momentum distributions and stabilizer expectation values of 0.273, surpassing the 0.208 Gaussian limit. Integration of 35 photonic chips (84 squeezers and 36 PNRDs) achieves synthesis of cluster states spanning 12 modes × 8.64 billion temporal modes, setting a scale record for photonic quantum systems. Despite the achievements, challenges still persist: Current optical path transmission rates of 78-82% must exceed 99.5% to meet fault-tolerance requirements (**Fig. 15h**), making low loss waveguides (loss 0.5 dB/m) a critical breakthrough point.

We summarized the differences between the two technical approaches of PsiQuantum and Xanadu in the Table inside **Fig. 15**. The essential distinction lies in: PsiQuantum prioritizes operational redundancy: Fusion measurements directly couple entanglement generation with error correction. Hardware fault-tolerance thresholds (>10% photon loss tolerance) better suit early-stage scaling but require improved single-photon source efficiency. Xanadu prioritizes encoding efficiency: GKP logical qubits inherently resist Gaussian noise and support deterministic Clifford gates but demand extreme optical path transmission fidelity (<0.5 dB loss). Common challenges centre on optical loss control: PsiQuantum must overcome fusion success rate bottlenecks, while Xanadu needs to achieve 99.5% component transmission, which also poses challenges for LNOI platforms. Both must resolve million-chip manufacturing and packaging challenges, with the ultimate goal being the realization of practical fault-tolerant quantum computers.

### 3.4 Arbitrary unitary operator-based quantum computing

Besides logic gate-based approaches, significant research in silicon photonic quantum computing focuses on implementing arbitrary unitary operators. Arbitrary unitary operator-based quantum computing constitutes, by strict definition, quantum simulation. The core concept involves mapping computationally hard real-world problems onto quantum systems, enabling their solution through quantum state measurement. Within this framework, several principal paradigms emerge: Quantum walks address the quantization of discrete stochastic processes, Hamiltonian simulation enables dynamical reconstruction of physical systems, and Boson sampling provides a pathway to demonstrate quantum computational advantage by tackling #P-hard problems through optical interference.

#### 3.4.1 Quantum walk

Quantum walk represent a distinct category of quantum simulation that similarly employs output photon distribution sampling but exhibits lower computational complexity than boson sampling. As quantum analogy of classical random walks, they describe particle dynamics in discrete or continuous space governed by unitary evolution rather than classical probabilistic rules. Crucially, the quantum walker exists in a superposition of position states, enabling simultaneous traversal of multiple paths. This evolution is dictated by a unitary operator, which deterministically evolves the wavefunction[173]. Most quantum-walk models fall into two classes: discrete-time quantum walks (DTQW) and continuous-time quantum walks (CTQW)—with the latter being the one most frequently realized on integrated photonic chips.

DTQW are characterized by evolution proceeding in discrete time steps and typically require an additional quantum coin to determine the direction of each step. The DTQW occurs within an extended Hilbert space, formed by the tensor product of the coin space ($H_c$) and the position space ($H_p$). The $H_c$ describes the internal degree of freedom of the walker. Its basis states typically represent directions of movement. For instance, on a one-dimensional line, the basis states are $\{|\uparrow\rangle, |\downarrow\rangle\}$, denoting left and right, respectively. The $H_p$ describes the position of the walker on a graph or lattice. Its basis states are $\{|x\rangle\}$, where $x$ represents the lattice sites. The quantum state of the system can be written as

$$|\psi\rangle = \sum_x \sum_m \alpha_{x,m} |x\rangle \otimes |m\rangle \tag{24}$$

where $\alpha_{x,m}$ are complex probability amplitudes, and $|\alpha_{x,m}|^2$ represents the probability of finding the walker at position xwith coin state $m$. Each steap of the DTQW evolution is governed by a unitary operator $U$, which is the composition of two operations:

$$U = S \cdot (C \otimes I_p) \tag{25}$$

where $S$ is the conditional shift operator, acting the entire system, $C$ is the coin operator, acting on the coin space, and $I_p$ is the identify operator on the position space. Common choice of $C$ includes Hadamard coin and Grover coin, which is defined by

$$C_H = \frac{\sqrt{2}}{2}\begin{bmatrix} 1 & 1 \\ 1 & -1 \end{bmatrix},\ C_G = \begin{bmatrix} 0 & 1 \\ 1 & 0 \end{bmatrix} \tag{26}$$

The typical $S$ operator for a one-dimensional line is typically defined as:

$$S = |\uparrow\rangle|\uparrow\rangle \otimes \sum_i |i-1\rangle\langle i| + |\downarrow\rangle|\downarrow\rangle \otimes \sum_i |i+1\rangle\langle i| \tag{27}$$

This operation means: if the coin state is $|\uparrow\rangle$, move one position to the left; if it is $|\downarrow\rangle$, move one position to the right. After $n$ steps of evolution, the initial state $|\phi_0\rangle$ becomes $|\phi_n\rangle = U^n|\phi_0\rangle$.

The evolution of CTQW is described by the Schrödinger equation, and its Hamiltonian H is typically related to the adjacency matrix or Laplacian matrix of a graph. For instance, for a CTQW on a simple graph, the Hamiltonian is taken as $H = \gamma A$, where $A$ is the graph's adjacency matrix and $\gamma$ is the transition rate. The evolution of the system state $|\psi(t)\rangle$ is governed by

$$i\hbar \frac{d}{dt}|\psi(t)\rangle = H|\psi(t)\rangle \tag{28}$$

Therefore, the state at time $t$ is given by $|\psi(t)\rangle = e^{-iHt}|\psi(0)\rangle$, where $e^{-iHt}$ is the time evolution operator.

Over the past decade, research in quantum walk has undergone a significant transition from single-particle simulations to multiparticle quantum correlation control. Early work implemented two-dimensional single-photon quantum walks via waveguide networks but was limited by the inability of classical light sources to exhibit quantum characteristics[174] (**Fig. 16a**). Using photons as quantum walkers and integrated waveguide arrays as the evolution network for quantum random walks, integrated photonic chips have demonstrated large-scale multi-particle quantum walks on complex graphs. In laser-written chips, each waveguide is coupled to several neighbours, enabling architectures that directly map onto highly connected graphs with tunable coupling strengths between nearest and non-nearest neighbours. In this way, the waveguide lattice embodies the adjacency matrix that governs CTQW evolution. A critical breakthrough emerged in 2010 when Peruzzo et al. first realized dual-photon quantum walks on $SiO_xN_y$ waveguide chips, observing nonclassical correlation phenomena in a 21-node 1D array-experimental data violated classical limits by 76 standard deviations, confirming that multiparticle interference exponentially expands state space[173] (**Fig. 16b**). More recently, quantum walk experiments have expanded into multidimensional implementations. Using femtosecond laser direct-writing technology, researchers now enable three-photon walks in triangular lattices, mapping to a $19^3$ state space. Quantum correlations in these systems are validated through statistical eigenvalue analysis combined with machine learning techniques[175] (**Fig. 16d**). Besides CTQW simulations, Crespi et al. investigated the DTQW, where the interplay between the Anderson localization mechanism and the bosonic/fermionic symmetry of the wavefunction[176] (**Fig. 16e**). By tuning the entangled state phase and amplitude, a single device can cover continuous transitions from Bose to Fermi statistics, providing a paradigm for complex quantum simulations.

Meanwhile, reconfigurable on-chip optical networks composed of MZIs and phase shifters can be programmed to implement arbitrary unitary transformations, thereby realizing programmable quantum-walk dynamics. A millstone arrived in 2021 with the silicon photonic quantum walk processor. This chip integrates dual reconfigurable five-mode optical networks and on-chip entanglement sources, achieving full parametric control of dual-particle walk Hamiltonians, particle statistical properties and indistinguishability[132] (**Fig. 16c**). They extended the detection scale to 15-node graphs, achieving polynomial-time discrimination via hierarchical certificates[132]. Quantum search algorithms verified $O(\sqrt{N})$ acceleration advantages in 5-node complete graphs, with dual-boson walks further expanding search space to 15 nodes. In addition to the graph-related problems, quantum walk processors has also been used to simulate the environment-assisted quantum transport[177] (**Fig. 16f**), by controlling and mapping static and dynamic disorder in a re-programmable silicon chip, which is composed of 88 MZIs and 176 free parameters. Current research still faces challenges including photon loss limitations (Ref.[178] shows similarity dropping to 91.4% after three steps) and low-probability multi-particle events (Ref.[175] reports only 1.4% three-photon bunching probability). Future directions focus on nonlinear interaction simulations, topological quantum walk realization, and bridging programmable processors with universal quantum algorithms. Photonics platforms—through integrated waveguides, synthetic dimensions, and chip-scale control—continue advancing quantum walks from fundamental research toward practical quantum technologies.

### 3.4.2 Hamiltonian simulation

Another category of unitary simulation is to use the reconfigurable linear optical circuit for simulating quantum dynamic behaviours by calculating the Hamiltonian evolution. For example, using the Hamiltonian as a bridge, researchers exploit a mapping between vibrations in molecules and photons in waveguides, which then allows the simulation of vibrational dynamics of the atoms within molecules[179] (**Fig. 17a**). The process can be expressed as

$$|\Psi\rangle_t = e^{-iHt/h}|\Psi\rangle_0 = U_L^{\dagger} e^{-i\Sigma t/\hbar} U_L |\Psi\rangle_0 \tag{29a}$$

$$\hat{H} = U_L^{\dagger} \Sigma U_L \tag{29b}$$

The Hamiltonian $\hat{H}$ is decided by the target molecule and can be decomposed to the unitary matrix $U_L$ and diagonal term $\Sigma$ via the singular value decomposition. Thus, the time evolution can be treated by two stationary unitary matrices $U_L^{\dagger}$, $U_L$ and a diagonal phase term $e^{-i\Sigma t/\hbar}$ evolution with time $t$. The input state $|\Psi\rangle_0$ decides which dynamic process is studied and the sampling results can indicate the probability of a mode at the given evolution time. Using this scheme, several four-atom molecules with up to six vibrational modes are simulated by a $SiO_2$ chip. In the molecular vibration simulation, the system is governed by a Hermitian Hamiltonian ($\hat{H} = \sum_i \hbar\omega_i a_i^{\dagger} a_i$), which describes a closed, conservative quantum system. This Hamiltonian ensures unitary evolution that inherently conserves probability across all vibrational modes. The photonic chip directly implements the unitary operator for sequential time steps, mapping multi-photon Fock states to vibrational excitations. The key results, like the vibrational relaxation and anharmonic evolution in $H_2O$, as shown in **Fig. 17b**, demonstrate the simulated probability evolution for both single-excitation and two-excitation states. These data shows smooth energy transfer between local modes without probability leakage.

For non-Hermitian quantum systems, significant challenges arise. For example, PT-symmetric Hamiltonians

$$H_N(\gamma) = -\sum_{k=1}^{N-1}(|k\rangle\langle k+1| + |k+1\rangle\langle k|) + i\gamma(|1\rangle\langle 1| - |N\rangle\langle N|) \tag{30}$$

are typically non-Hermitian, which violate $H = H^{\dagger}$ but preserve combined parity-time symmetry. Here, the imaginary terms $\pm i\gamma$ explicitly model open-system dynamics, introducing gain and loss at boundary modes. Such Hamiltonians exhibit exotic phenomena like exceptional points, where eigenstates coalesce and the system undergoes phase transitions from real-valued spectra (unbroken phase, oscillatory dynamics) to complex-conjugate pairs (broken phase, exponential decay/growth). Crucially, the non-unitary evolution operator $G_N(\gamma, t) = e^{-iH_N(\gamma)t}$ cannot be natively implemented on quantum hardware. To overcome this, as shown in **Fig. 17c**, Ref.[180] employs unitary dilation, embedding $G_N(\gamma, t)$ into an expanded $2N$-mode unitary matrix $U_{2N}(\gamma, t)$:

$$U_{2N}(\gamma, t) = \begin{bmatrix} \tilde{G}_N(\gamma, t) & iD_N(\tilde{G}_N) \\ iD_N(\tilde{G}_N^{\dagger}) & \tilde{G}_N^{\dagger}(\gamma, t) \end{bmatrix} \tag{31}$$

where $\tilde{G}_N$ is rescaled to unit norm and $D_N = \sqrt{I - \tilde{G}_N \tilde{G}_N^{\dagger}}$ is the defect operator. This technique doubles the optical modes and demands precise singular-value decomposition to construct $D_N$. Experimental data in **Fig. 17d** validates the non-Hermitian features: vacuum-state probability emerges due to excitation tunneling into the time-reversed subsystem, while phase-dependent behaviors—oscillatory (unbroken), polynomial decay (EP), and exponential saturation (broken)—underscore the system's sensitivity to $\gamma$.

Research on random controllable unitary matrices has spurred the flourishing development of another field, the optical neural networks. In this context, random unitary matrices $U$ serve as the weight matrices within neural network architectures. This approach has led to the proposal of various innovative architectures and algorithms across a broad range of applications[108,109,181,182], including complex-valued neural networks[183] and sophisticated computer vision networks[184].

### 3.4.3 Boson sampling

Most of other quantum simulation works originate from and are centred around the concept of boson sampling and rely on the sampling of photons. Within an intermediate-scale device, boson sampling is designed as such a task that offers the possibility of demonstrating a quantum advantage over classical devices[7]. A boson sampling experiment is defined as injecting $n$ bosons in different modes of an $m$-mode linear interferometer, followed by the probabilistic sampling of output photon distributions resulting from the bosonic interference[107]. If performed with indistinguishable bosons, this experiment results in an output distribution that is hard to sample, even approximately, on classical computers. The calculation of the probability associated with each observed boson sampler event requires the estimation of permanent, a well-known #P-hard problem. Using classical computers, there have been benchmarks showing that the computation of one permanent for a $48 \times 48$ matrix takes a supercomputer over an hour[185,186], and the largest sampling problem simulated is 50-photon Gaussian boson sampling performed by Sunway TaihuLight Supercomputer[187]. Using photonic quantum systems, early five-photon experiments were claimed to surpass early electronic computers[188], and recent milestones include 18-photon measurements on integrated chips[189] and a major 76-photon experiment with bulk optics[162].

To illustrate the principle of boson sampling, we consider a set of $m$ modes with associated creation operators $a_i^\dagger$, for $i = 1, \ldots, m$, satisfying bosonic commutation relations (**Fig. 18a**). A Fock state of $n$ photons in these modes can be written as

$$|S\rangle = |s_1 s_2 \ldots s_m\rangle = \prod_{i=1}^{m} \frac{\left(a_i^\dagger\right)^{s_i}}{s_i!} |0\rangle \tag{32}$$

where $s_i$ are the non-negative integers that count the number of photons in each mode and $\sum s_i = n$. When $s_i \leq 1$ for all $i$, the state is a no-collision state, meaning that there is no individual mode containing two or more photons. Injecting the photons into an $m$-mode linear-optical transformation described by $U \in SU(m)$. Its action is fully determined by the evolution of the creation operators:

$$a_i^\dagger \rightarrow \sum_{j=1}^{m} U_{ij} a_j^\dagger \tag{33}$$

The transition probability between an input state $|S\rangle = |s_1 s_2 \ldots s_m\rangle$ and $|T\rangle = |t_1 t_2 \ldots t_m\rangle$ can be written as

$$\Pr[S \rightarrow T] = \frac{|\mathrm{Per}(U_{S,T})|^2}{s_1! \ldots s_m! t_1! \ldots t_m!} \tag{34}$$

where $U_{S,T}$ is an $n \times n$ submatrix of $U$ constructed by taking $t_i$ copies of the $i'$th row of $U$ and $s_j$ copies of its $j'$th column, and

$$Per(B) = \sum_{\sigma \in S_n} \prod_{i=1}^{n} b_{i,\sigma(i)} \tag{35}$$

is the permanent of the matrix $B$, and $S_n$ is the symmetric group.

The implementation of boson sampling on a chip initially use probabilistic source based on SFWM or SPDC. One of the two generated photons is detected, heralding the other photon. Early experimental demonstrations of boson sampling were reported with three to five photons (**Fig. 18b**), mainly in integrated photonics[190–193,193–198]. The scheme suffers from the inherent probabilistic nature of the photon generation process, and the requirement for low pump intensities to suppress multi-pair generation per pulse. Therefore, scaling boson sampling systems presents exponential challenges due to the stringent requirement for strictly synchronized, spectrally pure, and high-brightness single-photon sources. To overcome limitations of probabilistic SPDC/SFWM photon sources, a generalized approach termed scattershot boson sampling (SBS) was proposed (**Fig. 18c**). In SBS, $k$ distinct sources ($k > n$) simultaneously feed photon pairs into different input ports of a linear interferometer. This configuration enhances the expected photon number per pulse by a combinatorial factor of $\binom{k}{n}$, yielding an exponential improvement in the generation rate for $k \gg n$ compared to conventional boson sampling with fixed inputs.

Another variation is Gaussian boson sampling (GBS) which uses squeezed light sources as input and calculates Hafnian instead of Permanent (**Fig. 18d**). Paesani et al.[137] demonstrated a silicon chip with eight-mode $U$ and up to eight photons, switching between different optical pumping regimes and implementing the scattershot, Gaussian, and standard boson sampling protocols on the same chip (**Fig. 18e**). Recent milestones include 18-photon measurements on an integrated full-stack GBS chip. This chip provides up to eight modes of squeezed vacuum initialized as two-mode squeezed states in single temporal modes, a fully programmable four-mode interferometer, and photon number-resolving readout on all outputs[189] (**Fig. 18f**). It enables remote users to execute quantum algorithms, including those for molecular vibronic spectra and graph similarity. Subsequently, they extend the scheme to 216 squeezed modes entangled with 3D connectivity, using a time-multiplexed architecture[199] (**Fig. 18g**), which offers quantum computational advantage well beyond the current state of the art supercomputers. This work, though, is not integrated on chip. A list of related experiments reported in integrated chips is summarized in **Table 2**, where the main features of the photon sources, linear interferometers and detector are detailed. A large-scale arbitrary unitary simulator, the graph processor Boya, was reported recently[200], which integrates 2500 components to form a synthetic lattice of nonlinear photon-pair sources and linear optical circuits, all fabricated on a silicon chip. It can be reconfigured to process complex-weighted graphs of various topologies and perform tasks related to graph properties (**Fig. 18h**).

Moving beyond proof-of-concept demonstrations of the variations of boson sampling, identifying tractable problems solvable by these approaches is fundamental to their utility. Boson sampling has found a famous application in the simulation of molecular vibronic spectra[201–204] (**Fig. 19a**). The spectra depend on the Frank-Condon profile, which is given by

$$FCP(\omega_{vib}) = \sum_{m}^{\infty} \left|\langle m|\hat{U}_{Dok}|0\rangle\right|^2 \delta(\omega_{vib} - \sum_{k}^{N} \omega'_k m_k) \tag{36}$$

This equation is the sum of the production of two terms to form a full spectrum. The wave number of each peak is decided by the second term $\delta(\omega_{vib} - \sum_{k}^{N} \omega'_k m_k)$. The intensity of the peak is decided by the first term $\left|\langle m|\hat{U}_{Dok}|0\rangle\right|^2$, which is called the Frank-Condon Factor. The main part of the calculation is the vacuum state $|0\rangle$ evolution to $|m\rangle$ under the operator $\hat{U}_{Dok}$. This part is computationally intractable for a supercomputer but yet it can be mapped to a quantum boson sampling process given by

$$\hat{U}_{Dok}|0\rangle = \hat{R}_{C_L} \hat{S}_{\Sigma}^{\dagger} \hat{R}_{C_R}^{\dagger} \hat{D}_{\frac{1}{\sqrt{2}} J^{-1} \delta} |0\rangle \tag{37}$$

The boson sampling process starts with a squeezed coherent state and evolves in a linear optical circuit $\hat{R}_{C_L}$. The output photon combination corresponds to the state $|m\rangle$ and its intensity can be mapped to the peak intensity of the spectra. All the parameters like $C_L$, $C_R$, $\sum$, $J$ are extracted from the target molecular structure itself.

Boson sampling is inherently a mathematical process with few direct practical uses, GBS bridges this gap since the computation of Hafnian provides a pathway to graph-theoretic applications, like finding the dense subgraphs[205,206], as shown in **Fig. 19b**. The dense graph is related to the calculation of maximum perfect matching of a graph, which is mathematically related to calculating Hafnian as

$$Pr(\bar{n}) = \frac{1}{n!\sqrt{\det \sigma_Q}} haf A_s \tag{38}$$

where $\bar{n}$ is the selected combinations of subgraph, $A_s$ is related to the structure of graph connections. The connection between the graph and GBS, further enables its application to molecular docking and enhancing image recognition. Ref.[207] proposes to transform molecular docking problems into maximum weighted clique searches through constructing a labelled distance graph that captures the binding site and ligand pharmacophore interactions (**Fig. 19c**). GBS intrinsically identifies dense subgraphs by leveraging photon interference patterns, where the Hafnian term inherently favours fully connected subgraphs (cliques). In GBS-enhanced image recognition[208], the GBS generates high-dimensional feature representations through frequency distributions over selected computational bases, effectively replacing traditional neural network hidden layers (**Fig. 19d**). This nonlinear transformation surpasses classical SVM and physical extreme learning machine implementations. The quantum advantage stems from GBS generating complex feature correlations beyond linear separability, with optimal performance attained using just 5 million samples and 3,000 computational bases, demonstrating resource efficiency without overfitting.

Scalability of boson sampling is essential to establish quantum computational advantage and enable practical applications such as large-scale graph property verification. Besides the probabilistic nature of the photon generation process, the scaling of boson sampling also depends critically on controlling imperfections, such as the presence of losses, partial distinguishability, and generic experimental errors[209]. Limited photon loss preserves quantum advantage, but losses scaling logarithmically with photon count or leaving only a fractional remnant enable efficient classical simulation. Similarly, fully distinguishable photons trivialize the problem, while partial distinguishability enables classical approximation. Crucially, compounding imperfections, like loss with dark counts, accelerate classical simulatability even at constant error rates. Fabrication flaws and measurement noise further degrade quantum advantage. Critically, persistent noise of any type fundamentally limits scalability, as constant noise enables efficient classical simulation via tensor networks[210]. Thus, quantum advantage requires suppressing individual errors while ensuring their collective impact stays below critical thresholds.

### 3.5 The Transition to Thin-Film Lithium Niobate: A Paradigm Shift

Silicon photonics has established a significant position in the field of quantum computing, primarily owing to its seamless compatibility with mature Complementary Metal-Oxide-Semiconductor (CMOS) processes. This compatibility enables the realization of unprecedented levels of integration density and the fabrication of complex photonic quantum circuits. This platform has demonstrated exceptional performance in specific computational tasks such as Boson Sampling and quantum walks, proving its capability to address particular quantum problems and providing a viable physical path toward demonstrating quantum advantage.

However, the inherent physical properties of silicon impose fundamental constraints on its potential for building universal, scalable quantum computing systems. Firstly, its crystal structure possesses inversion symmetry, resulting in the absence of a linear electro-optic (Pockels) effect. Consequently, silicon-based modulators must rely on slow thermo-optic efforts or the nonlinear carrier dispersion effect, which introduces optical loss. This makes achieving high-speed, low-power dynamic circuit reconfiguration and real-time feedback control exceptionally challenging, yet these capabilities are crucial for many quantum algorithms and error-correction protocols. Secondly, although silicon possesses a strong third-order nonlinearity, which can be used to generate photon pairs via four-wave mixing, this process is often accompanied by two-photon absorption and free carrier absorption, which severely limit the usable pump power, thereby constraining the efficiency of nonlinear conversion and the brightness of photon pairs. Finally, the silicon platform is relatively limited in functionality; it is inherently unsuitable for key operations requiring strong second-order nonlinearity, such as efficient quantum frequency conversion and entanglement photon pair generation based on parametric down conversion, which limits its interoperability with other quantum systems.

To construct practical and scalable quantum computing systems, it is necessary to overcome the bottlenecks of the silicon platform. This has spurred the exploration of new material platforms. Lithium niobate, particularly in its thin-film form, is a historically significant optical material that has experienced a revival in recent years. Advances in micro-nanofabrication technologies, such as ion slicing, have propelled TFLN into becoming a highly competitive emerging integrated photonics platform. In contrast to silicon, lithium niobate possesses a strong linear electro-optic effect, with an electro-optic coefficient of 31 pm/V. This enables TFLN-based modulators to support high-speed, low-power pure phase modulation at rates up to hundreds of GHz, enabling rapid feedback and circuit reconfiguration – key features for realizing programmable photonic quantum processors. Furthermore, while silicon's third-order nonlinearity grapples with TPA issues, lithium niobate concurrently has second-order nonlinear coefficients ($d_{33} \approx -25.2$ pm/V). This not only allows for more efficient generation of entangled photon pairs via spontaneous parametric down conversion but also enables the direct generation of highly squeezed states, paving the way for continuous-variable one-way quantum computing. Leveraging its strong second-order nonlinearity, TFLN also facilitates efficient quantum frequency conversion, acting as a wavelength bridge to connect different quantum systems.

Moreover, through advanced nanofabrication techniques, the TFLN has successfully demonstrated ultra-low propagation losses (<0.027 dB/cm) comparable to those of silicon nitride, alongside a broad transparency window ranging from 350 nm to 5 um, which provides a foundation for multi-functional photonic integration from the visible to the mid-infrared spectrum. The material properties of silicon and lithium niobate, along with their derived key performance metrics for heralded single-photon sources, squeezed states, electro-optic modulation, and frequency conversion-are summarized in the **Table 3**.

# 4 FUNDAMENTALS OF LITHIUM NIOBATE

Lithium niobate (LN) is a promising material in photonics, owing a significant position thanks to its exceptional nonlinear and EO properties. LN crystal boasts a transparency window that spans from the ultraviolet to the mid-infrared, exhibiting a high refractive index and a significant EO coefficient, making it an ideal choice for modulators, frequency converters, and sensors. Traditional LN has existed in bulk form, performing well in discrete components for decades. However, integrating these functionalities into compact and low-loss integrated photonic circuits has been challenging due to the difficulty in processing them into thin films with high crystalline quality.

With the development of fabrication technologies, thin-film lithium niobate (TFLN) has become an advent, and transferring TFLN onto an insulating substrate, such as $SiO_2$, produces lithium niobate-on-insulator (LNOI). This thin-film variant, produced through techniques such as ion slicing and wafer bonding, retains the excellent optical properties of bulk LN (e.g., EO, acousto-optic (AO), and nonlinear optical properties) while offering the advantages of integrated photonics (e.g., high stability and integrability). LNOI has enabled the fabrication of high-quality, low-loss waveguides and resonators, making it an important material for high-performance photonic integrated circuits.

TFLN creates nanophotonic components with tight mode confinement, which is essential for high-density integration and enhanced optical nonlinearity. It also led to breakthroughs in the performance of EO modulators, nonlinear frequency converters, and photonic quantum circuits. These integrated devices offer improved performance in terms of speed, power consumption, and footprint, and open up new possibilities for on-chip photonic applications in communications, sensing, and quantum information processing.

Within TFLN, specifically, X-cut and Z-cut TFLN play unique roles in integrated photonics due to their different crystallographic orientations. X-cut TFLN, where the crystal's X-axis is perpendicular to the waveguide surface, is commonly used to leverage its largest EO coefficient $d_{33}$, which is particularly important in the fabrication of EO modulators. In contrast, Z-cut TFLN, with its Z-axis perpendicular to the waveguide surface, is suitable for piezoelectric-based applications, like acousto-optic modulators and sensors. These two cutting orientations of TFLN offer flexibility in designing photonic devices with specific performance requirements.

## 4.1 Nonlinear optical effects

LN exhibits both second- and third-order nonlinearities and possesses a broad transparency window extending from ultraviolet to mid-infrared. Its low linear propagation loss (across visible and near-infrared wavelengths) and absence of multiphoton absorption above 800 nm make it suitable for ultra-broadband nonlinear processes, even at low pump powers. In this section, we review the applications of second-order, third-order nonlinearity and the engineering of nonlinear optical interactions through phase matching and dispersion engineering.

**Second-order nonlinearity (Fig. 20a)** – LN is renowned for its strong second-order nonlinear optical properties, which stem from its non-centrosymmetric crystal structure. The magnitude of the second-order nonlinearity in LN is substantial, with the largest tensor component $d_{33}$ being 27 pm/V, making it an attractive material for various nonlinear optical applications. The second-order nonlinearity in LN allows for the conversion of frequencies through processes such as second-harmonic generation (SHG), sum-frequency generation (SFG), difference-frequency generation (DFG), and optical parametric oscillation (OPO). LN is also used for generating nonclassical light via SPDC. These implementations primarily occur in bulk crystals or microscale waveguides, facilitated largely by periodic poling. One of the key advantages of utilizing LN's second-order nonlinearity in integrated systems is the ability to achieve quasi-phase matching (QPM) through periodic poling. This technique involves the periodic inversion of the crystal's ferroelectric domains, which allows for the efficient conversion of optical frequencies over a broad range of wavelengths. The QPM approach has been instrumental in realizing devices such as periodically poled lithium niobate (PPLN) waveguides and resonators, which exhibit high conversion efficiencies and low thresholds for nonlinear optical processes.

**Third-order nonlinearity (Fig. 20b) -** Besides the inherent second-order nonlinearity, for third-order processes, LN utilizes its nonlinear refractive index of $1.8 \times 10^{-19}$ m$^2$/W, comparable to that of $Si_3N_4$ ($2.5 \times 10^{-19}$ m$^2$/W at 1.55 μm). Crucially, efficient third-order phenomena in LN manifest not only through direct $\chi^{(3)}$ effects but also via cascaded $\chi^{(2)}$ processes. Third-harmonic generation has been demonstrated through sequential SHG and SFG cascades[211,212], while effective four-wave mixing arises from cascaded SHG and DFG[213]. This nonlinearity mechanism enables Kerr frequency comb generation, which leverages the interaction between $\chi^{(3)}$ nonlinearity and high intracavity power to produce coherent comb spectra via four-wave mixing. By phase-locking to Kerr solitons—dispersive waves dynamically balancing nonlinearity, dispersion, and cavity gain/loss—LN devices achieve octave-spanning combs essential for *f*-2*f* self-referencing. Such capabilities underpin critical applications in optical clocks, terabit communications, and molecular fingerprinting. TFLN further advances this field through monolithic photonic circuits that integrate electro-optic comb tuning, soliton injection, and on-chip self-referencing[46]. Unlike $Si_3N_4$ or silica platforms requiring external modulators and detectors, TFLN uniquely provides simultaneous $\chi^{(2)}$ nonlinearity and Pockels effect on a single chip, enabling full comb-on-chip functionality with unprecedented compactness.

### 4.2 Engineering of nonlinearity

On the TFLN photonic platform, phase matching and dispersion engineering are the two most significant techniques in nonlinear science and applications (**Fig. 20c-d**). In $\chi^{(2)}$-mediated three-wave mixing, material and waveguide dispersion prevent momentum conservation at interacting frequencies, creating phase mismatch that requires engineered phase matching solutions. Common phase matching techniques include birefringent phase matching (BPM), modal phase matching (MPM), and quasi-phase matching (QPM) (compared in **Table 4**). **BPM** leverages refractive index differences between polarization states in birefringent crystals by precisely orienting light propagation directions. In practice, crystals are typically cut at specific angles to achieve phase-matched beam interactions at predetermined incidence angles, to achieve the effect of phase matching. However, BPM is rarely implemented in LNOI waveguides, and perfect BPM synchronization is rarely achieved in the telecommunication band. **MPM** resolves phase matching straightforwardly through waveguide geometry optimization, enabling dispersion engineering to match fundamental and higher-order modes. For example, this approach can work at arbitrary azimuthal angles in Z-cut LN films that are isotropic in the device plane. However, MPM suffers from reduced mode overlap penalties between fundamental and higher-order spatial modes[43], which results in degradation in nonlinear conversion efficiency.

Optimal $\chi^{(2)}$ frequency conversion requires simultaneous phase matching and identical polarization/spatial modes for all three interacting waves. As phase matching ensures that at least one of the interacting frequencies occupies a different spatial and/or polarization mode, one solution is to employ **QPM** rather than true phase matching. By periodically reversing the sign of the nonlinear coefficient of a crystal or waveguide, i.e., the polarization direction, an artificially periodic phase factor is created. This periodically resets the accumulated phase mismatch, enabling the converted light to restart coherent growth upon each reversal. Thus, QPM allows for phase matching in a crystal with the maximum effective nonlinear coefficient, free from the restriction of the intrinsic birefringence in that direction. QPM fabrication requires periodic inversion of the sign of $\chi^{(2)}$ nonlinear coefficient. This is possible specifically in ferroelectric materials, where high-voltage pulses induce ferroelectric domain polarity inversion – a process termed periodic poling. LN fortunately offers the ferroelectric property. The fabrication techniques of PPLN on TFLN have become mature in recent years. Specifically, domain inversion occurs parallel to the film plane in X-cut lithium niobate, restricting periodic poling to linear gratings. Conversely, domain inversion along the film normal in Z-cut films enables arbitrary poling geometries like radial gratings.

Dispersion engineering critically enhances nonlinear processes in QPM-enabled systems[214]. As QPM accommodates nearly arbitrary waveguide geometries, geometric dispersion becomes a direct design parameter. This freedom establishes new design paradigms where multiple dispersion orders, including the group velocities and group-velocity dispersion of interacting waves, can be co-engineered to achieve broadband wavelength operation[215]. Fejer et al. formalized these design rules for dispersion-engineered QPM devices, focusing particularly on bandwidth engineering for nonlinear interactions[216], being applied to the deisgn of nonlinear components that generate and manipulate quantum light.

### 4.3 Fabrication technologies

The "Smart-Cut" technology, widely employed for the fabrication of SOI wafers, has been refined and has emerged as the go-to method for producing high-quality thin-film LNOI wafers[217]. The Smart-Cut fabrication process for LNOI (**Fig. 21a**) starts from the high-dose implantation of $He^+$ (or $H^+$) ions into a high-quality LN substrate to define a cleavage plane at the desired film thickness. In parallel, the handle wafer, usually LN or Si with an oxide layer on top, is prepared. Then the LN substrate is bonded to the handle wafer, through either adhesive (e.g., using benzocyclobutene, BCB)[32] or direct wafer bonding[218]. After the bonding, thermal annealing is employed to enhance the bonding strength and split the LN substrate along the cleavage plane, thereby defining the thin film. Finally, an additional annealing process is conducted to diminish crystal defects caused by ion implantation, followed by a chemical mechanical polishing (CMP) process to improve surface smoothness. Direct bonding is preferred over adhesive bonding for many applications, as it allows higher annealing temperatures for better recovery of the nonlinear and EO properties of the LN thin film sliced by ion implantation[217]. It is noted that most LNOI fabrication is performed using congruent instead of stoichiometric LN, implying that the crystals are lithium deficient. A congruent LN crystal has a very low intrinsic absorption of light.

### 4.3.1 LN Waveguide Fabrication

**Diffused Waveguides in Bulk LN.** LN waveguides are conventionally fabricated using titanium (Ti) in-diffusion or proton exchange[219]. Ti-diffused waveguides are formed by depositing Ti strips on bulk LN substrates and then thermal annealing. Ti diffuses into LN crystals with a typical diffusion length of a few micrometres, resulting in an increase in the refractive index of LN (**Fig. 21b**). Proton-exchange waveguides are constructed by immersing LN into a liquid source of hydrogen, where protons ($H^+$) replace lithium ions ($Li^+$) and cause a refractive index change. The refractive index perturbation can also be realized by ion implantation[220] and femtosecond laser writing[221]. A major limitation of these index perturbation-based waveguides is the low index contrast, resulting in weak mode confinement and large bending radius, thus hindering dense integration.

**Monolithic Waveguides on TFLN:** The availability of thin-film LNOI facilitates the fabrication of monolithic ridge/rib waveguides with large index contrast and strong mode confinement. LN waveguides can be fabricated by mechanical methods, such as ultra-precision cutting[222,223] (**Fig. 21c(i)**), optical grade dicing[224,225], and diamond polishing[226,227]. Simple structures, such as straight waveguides and curved waveguides with large curvature, can generally be fabricated individually using these methods. The wet etching of bulk LN by hydrofluoric acid (HF) or potassium hydroxide (KOH) has been a well-established practice for many years (**Fig. 21c(ii)**). LNOI waveguides have also been demonstrated using Ar ion implantation and subsequent KOH wet etching[228,229].

Compared to mechanical dicing/polishing and wet etching, dry etching is much preferred for integrated photonics as it is more controllable and more anisotropic. However, dry etching of LN had been notoriously challenging. Unlike other common photonic materials like Si and SiN, a suitable reactive ion etching (RIE) recipe was not available for LN. Fluorine-based RIE can etch LN effectively[230], while inducing the formation and redeposition of non-volatile lithium fluoride (LiF) on the surface, which increases sidewall roughness and scattering loss[231]. Introducing argon (Ar) gas into the fluorine atmosphere can mitigate LiF redeposition by physically bombarding the target, thereby improving the sidewall smoothness[232] (**Fig. 21c(iii)**).

Nowadays, a more popular dry etching process for LNOI is Ar ion milling, which is a pure physical etching process free from LiF redeposition. Nonetheless, pure physical etching has low selectivity with respect to resists, limiting the achievable etch depth. Rib waveguides[233] and rib microring resonators[32] have been fabricated by Ar ion milling. Hard masks like amorphous Si[33] and chromium (Cr)[27] are commonly used to increase the etch depth. The other issue with Ar ion milling is that it forms nonvertical sidewalls due to LN redeposition. Unlike LiF redeposition, LN redeposition can be smooth and does not induce large scattering loss. Additionally, it can be effectively removed through post-etch wet chemical treatment, which further reduces optical losses. The sidewall angle, typically ranging from 40° to 80°, ultimately limits the minimum feature size and spacing between adjacent structures. The combination of chemical etching by fluorine and physical sputtering by Ar improves not only the sidewall smoothness but also the sidewall angle[232].

Other than Ar ion milling, focused ion beam (FIB) milling has also been utilized to fabricate LNOI waveguides[234]. Due to the small focal spot size of an ion beam, FIB possesses very high fabrication resolution and is suitable for defining fine patterns such as photonic crystals[235]. However, FIB is expensive and time-consuming, thus is not suitable for large-scale fabrication.

The emerging photolithography-assisted chemo-mechanical etching (PLACE) approach opens an avenue for wafer-scale ultralow-loss photonic devices[236]. First, a Cr mask deposited on LNOI is patterned by femtosecond laser ablation. Then, the pattern is transferred to the LN thin film by removing uncovered LN through CMP. Finally, the Cr mask is removed by chemical wet etching. The PLACE approach yields ridge waveguides with sub-nanometer surface roughness and propagation loss as low as 0.027 dB/cm[30]. Microring and microdisk resonators with quality factors above $10^7$ have also been demonstrated[30,237]. An issue that remains unresolved is the low aspect ratio of nanostructures fabricated by CMP, which is below 1.5 due to the limited selectivity of the metal mask and LN. Fabricating coupled devices with gaps of less than 2 μm is difficult. Given the nascent stage of the PLACE technique, significant efforts are anticipated to enhance the aspect ratio in the coming period.

**Rib-Loaded Waveguides (Fig. 21d):** An alternative category of waveguides in TFLN is the rib-loaded waveguide, which circumvents the need for LN etching. In this category, another material is deposited or bonded on TFLN and etched into a strip. The patched material typically possesses a refractive index similar to or higher than that of LN but is easier to etch. Consequently, an effective rib waveguide is formed, with a portion of the optical mode confined within the LN layer, thereby leveraging its distinctive material properties. Various loading materials have been explored, including Si[238–242], SiN[38,25,243,244], chalcogenide glass (ChG)[245], titanium dioxide ($TiO_2$)[246,247], and tantalum pentoxide ($Ta_2O_5$)[248,249]. Recently, it has been demonstrated that the loading ridge can also be low-index material (e.g., photo- or electron-beam resist), based on the concept of bound states in the continuum (BIC)[250,251]. A disadvantage of the rib-loading approach is that only a portion of the optical mode resides in the LN, worsening the performance of the LN photonic devices.

### 4.3.2 PPLN Fabrication

In Z-cut LNOI, two approaches can be utilized for PPLN fabrication: (1) exfoliation of the LN thin film from an already poled bulk crystal; (2) directly poling on the LN thin film. In the first case, bulk LN is poled with standard bulk PPLN fabrication techniques, such as electrode poling[252], electron beam writing[253], femtosecond laser irradiation[254], Czochralski method[255], and so on. Then the bulk PPLN is bonded onto a handle wafer and sliced into thin film using the aforementioned "Smart-Cut" technology. This method was employed in one of the earliest demonstrations of thin-film PPLN[217] and several subsequent demonstrations[256–258]. Direct poling on TFLN has also been demonstrated recently. One approach involves utilizing an atomic force microscope (AFM) tip to generate a high localized electric field for domain inversion, which is capable of achieving feature size down to 100 nm[259,260]. An alternative approach is to pattern electrodes onto the thin film (with some buffer dielectric material), and pole through the LN/oxide/handle wafer stack at an elevated temperature[43].

In X/Y-cut LNOI, PPLN is typically fabricated by depositing surface electrodes in the form of coplanar gratings. The poling period, i.e., the gap between the adjacent electrodes in each pair is often larger than 600 nm, thereby requiring a poling voltage on the level of several hundred volts[261]. It is worth noting that X/Y-cut PPLN only allows the poling grating to be along one direction, as the domain inversion needs to be along the Z axis, while Z-cut PPLN allows arbitrary directions in the *xy* plane (e.g., radial gratings for poled ring).

# 5 LITHIUM NIOBATE-BASED PHOTONIC QUANTUM DEVICE TOOLBOX

## 5.1 Quantum light sources

LN emerges as a preeminent platform for integrated quantum photonics, leveraging its strong second-order nonlinearity ($\chi^{(2)}$) to generate nonclassical light states essential for quantum information processing. This section comprehensively examines quantum light sources on LN with two primary aspects: SPDC-based photon-pair generation and continuous-variable squeezed states. Critically, LN's electro-optic tunability and mature fabrication techniques facilitate monolithic integration of these sources with programmable circuits, establishing a scalable pathway toward fully chip-based quantum systems.

### 5.1.1 SPDC photon pair sources in $\chi^{(2)}$ LN

As detailed in Section 2.1, SPDC constitutes the predominant method for quantum light generation via second-order nonlinearity in LN. In this process (**Fig. 22a**), a short-wavelength pump photon spontaneously splits into longer-wavelength signal and idler photon pairs under strict energy conservation, requiring vanishing phase mismatch for efficient nonlinear interaction. Compared to silicon/silicon nitride-based SFWM[61,65,77,133], SPDC leverages significantly larger nonlinear coefficients in LN, enabling high-efficiency photon-pair generation at lower pump powers and shorter waveguide lengths. However, the substantial wavelength differences among interacting waves in SPDC induce significant phase mismatch, necessitating QPM mitigation in LN platforms. QPM implementation requires periodic inversion of the material's $\chi^{(2)}$ sign (**Fig. 22b**), exploiting LN's ferroelectric nature where high-voltage pulses invert ferroelectric domain orientations during periodic poling.

In a standard SPDC process within a single-mode waveguide, a high-frequency pump is incident into the LN waveguides, that will split into lower-frequency signal photon ($\omega_s$) and idler photon ($\omega_i$). This process must satisfy both energy conservation and momentum conservation (also known as the phase-matching condition). In a single-mode waveguide where spatial modes are strictly confined and under low pump power conditions where higher-order terms leading to multi-pair generation are negligible, and after using heralding technology to eliminate the vacuum component (i.e., heralding the existence of the signal photon by detecting the idler photon), the resulting effective two-photon state can be approximated as:

$$|\psi\rangle = \iint dw_s dw_i f(w_s, w_i)|1_s, 1_i\rangle \tag{39}$$

Here, $|1_s, 1_i\rangle$ represents the Fock state with one photon at the signal frequency and the other at the idler frequency. The function $f(w_s, w_s)$ represents the joint spectral amplitude (JSA), which describes the probability amplitude for the signal and idler photons to be generated at different frequency combinations. Integration over all possible frequencies satisfies the normalization condition. The JSA can typically be decomposed into the product of two main factors:

$$f(w_s, w_i) \propto \alpha(w_s, w_i) \times \phi(w_s, w_i) \tag{40}$$

where $\alpha(w_s, w_i)$ is the pump envelope function, determined by the spectral shape of the pump light; $\phi(w_s, w_i)$ is the phase matching function, originates from the conservation condition that the SPDC must satisfy. By optimizing the phase matching conditions, the shape of JSA can be altered. If the shape of the JSA is distributed along the diagonal line $w_s + w_i = C$ (where $C$ is a constant), it indicates strong frequency correlation between the signal and idler photons. Through control methods such as using a narrowband pump or filtering the generated photons, the JSA can be made to approach a symmetric circular or elliptical off-diagonal distribution in frequency space. This indicates that the frequency correlation between the signal and idler photons is weakened. For the heralded single photon, its state is described by a reduced density matrix. When the JSA can be factored into the product of two independent functions $f(w_s, w_i) = f_s(w_s) \cdot f_i(w_i)$, it implies no frequency entanglement between the signal and idler photons. In this case, the heralded single photon itself has high frequency purity. In quantum interference, a high-purity single-photon state is the key to achieving interference visibility. Frequency correlation can lead to "which-path" information, thereby reducing interference visibility. After weakening spectral correlations, the single photon frequency purity increases, and consequently, the interference visibility is enhanced.

SPDC photon-pair emission in LNOI has been realized in microdisks[262], PPLN waveguides[47–49,59,263,264], and PPLN rings[265]. It was first demonstrated in a microdisk resonator coupled by a fiber taper[262]. Then, in a 5 mm X-cut PPLN waveguide (**Fig. 22c**), photon pair generation reached 11 MHz at 250 μW pump power with coincidence-to-accidental ratios (CARs) exceeding 700[49]. Using waveguides with sub-micron mode size, high CARs of over 1000 were achieved in very short waveguides of only 300 μm length[48], and over broad spectral ranges[47,48]. Photon-pair generation has also been demonstrated in a fully integrated PPLN ring resonator[265], with photon-pair rates up to 36 MHz using just 13 μW pump power and CARs of up to 15000 (**Fig. 22d**). Reconfigurable LN waveguide circuits have also enabled on-chip entangled photon generation and manipulation. This architecture integrates two independent photon-pair sources within a Hong-Ou-Mandel interferometer featuring electro-optic phase control, generating deterministically separated identical photon pairs[59]. These integrated sources show promising performance for quantum photonics, with measured photon rates matching or exceeding optimized silicon counterparts[266], even without employing techniques like the slow light[267] for enhancement of generation efficiency. **Table 5** presents a comparison of the typical experimental results of photon pair sources (also known as heralded single photon sources) on $\chi^{(2)}$ and $\chi^{(3)}$ waveguides and microcavities. Compared to SFWM sources, SPDC sources enable efficient photon-pair generation with lower pump power and higher brightness. It is evident that PPLN photon pair sources exhibit exceptional efficiency, with a brightness several orders of magnitude greater than that of the previously demonstrated state-of-the-art photon pair sources. The unparalleled brightness observed in PPLN sources is attributed to the combination of giant single-photon nonlinearity, excellent single-mode conditions, and the purity of SPDC process.

Besides the photonic pair generation, pure single photons are needed. Similar to SFWM sources on silicon, pure single photons can be generated with photon-pair sources using heralding, where only one of the photons of a pair a used as a quantum resource, and the other heralds the presence of this resource. To achieve spectrally pure heralded single photons, narrow band filtering is needed at the cost of pair generation rate. This can be realized by dispersion engineering of LNOI waveguides without filtering[268]. The drawback of SPDC source is still the probabilistic nature of the generation process, which can be further improved by multiplexing techniques, as we will discuss later in Section 6.3. On the other hand, pure single photons can also be deterministically created using atomic emitters or QDs[269]. The material LN itself does not provide suitable color centers or similar atom-like defects suitable as single-photon sources. However, such centers could be created by selective doping of the LN host material. Doping of LN has been demonstrated with Tm[270], Er[271–274], and Yb ions[275,276], which were used to demonstrate photoluminescence, amplification, and lasing. Semiconductor QDs can be coupled to the LNOI platform by hybridization, using specifically designed tapers. Aghaeimeibodi et al.[277] directly integrate efficient single photon sources in the telecom band with LN photonic platform with coupling efficiency of 40.1% (**Fig. 22e**). Another promising deterministic technical pathway is QDs and color centers in 2D materials, for example, 2D $WSe_2$ flake was placed on a facet of Ti-diffused MZI in bulk LN for coupling of single photons to integrated photonic structures (**Fig. 22f**)[278], which is extendable to LNOI platform. Compared to QD emitters integrated on silicon platform, LN platform offers effective mechanism for cryogenic-compatible EO modulation, thus enabling a full reconfigurable integrated source on demand. Some quantum-photonic applications are not based on the number states of light like the single photons and photon pairs discussed above, but rather use squeezed light as an input, which will be discussed as follows.

### 5.1.2 Squeezed light sources

#### 5.1.2.a Principle of squeezing

The squeezed state is an important non-classical light field in quantum optics, whose core idea is to achieve suppression of noise in specific quadrature components of the light field through quantum manipulation. The quantization of the light field is the foundation for understanding squeezed states. The light field is described by the annihilation operator $\hat{a}$ and the creation operator $\hat{a}^\dagger$, which themselves are not Hermitian operators. Therefore, we define two observable Hermitian operators – the quadrature amplitude component and the quadrature phase component:

$$\hat{X} = \frac{1}{2}(\hat{a} + \hat{a}^\dagger), \hat{P} = \frac{1}{2i}(\hat{a} - \hat{a}^\dagger) \tag{41}$$

These two operators satisfy the commutation relation $[\hat{X}, \hat{P}] = i/2$, which leads to a fundamental limit on their measurement precision. According to the Heisenberg uncertainty principle, a pair of conjugate observables cannot be measured precisely simultaneously, and their variances satisfy:

$$V(\hat{X}) \cdot V(\hat{P}) \geq \frac{1}{16} \tag{42}$$

where $V(\hat{X}) = \langle \hat{X}^2 \rangle - \langle \hat{X} \rangle^2$ represents the noise fluctuation (variance) of the quadrature component. Both the vacuum state and the coherent state are minimum uncertainty states, satisfying:

$$V(\hat{X}) = V(\hat{P}) = \frac{1}{4} \tag{43}$$

This noise level is known as the standard quantum limit (SQL) or shot noise limit. Even the vacuum state, which has zero mean photon number, still exhibits unavoidable quantum fluctuations in its quadrature components. This constitutes the ultimate limit of classical precision measurement.

The defining feature of a squeezed state is that the noise of one quadrature component of the light field is suppressed below the SQL, while the noise of the conjugate component necessarily increases, yet the product of their uncertainties still satisfies the uncertainty relation. Mathematically, a squeezed state is generated by the action of the squeezing operator $\hat{S}(\zeta) = \exp[(\zeta\hat{a}^2 - \zeta^*\hat{a}^{\dagger 2})/2]$, where $\zeta = re^{i\theta}$ is the complex squeezing parameter: $r \geq 0$ is the squeezing factor, determining the degree of squeezing; $\theta$ is the squeezing angle, determining the direction of squeezing in phase space. The squeezed vacuum state is expressed as $|\zeta\rangle = \hat{S}(\zeta)|0\rangle$. When $\theta = 0$ is chosen, the variances of the quadrature components are:

$$V(\hat{X}) = \frac{1}{4}e^{-2r} < \frac{1}{4}, V(\hat{P}) = \frac{1}{4}e^{2r} > \frac{1}{4} \tag{44}$$

The degree of the squeezing is usually expressed in decibels (dB):

$$Squeezing\ (dB) = -10 \cdot log_{10}(e^{-2r}) \tag{45}$$

The Wigner functions of the vacuum state (**Fig. 23a**) and coherent state (**Fig. 23b**) are symmetric and circular, with isotropic noise. The Wigner function of the squeezed vacuum state is elliptical, with noise below the SQL along the minor axis and noise above the SQL along the major axis (**Fig. 23c**). The squeezed coherent state combines both displacement and squeezing characteristics (**Fig. 23d**).

In SPDC, the frequency, wave vector, and polarization of the generated photons are determined by the phase-matching conditions. If the two photons are degenerate, meaning they are indistinguishable in all degrees of freedom such as frequency, direction, and polarization, a single-mode squeezed state is produced. If the two photons are non-degenerate, meaning they are distinguishable in at least one of these degrees of freedom, a two-mode squeezed state is generated. The single-mode squeezed vacuum state contains only even photon number states:

$$|\zeta\rangle = \frac{1}{\cosh r}\sum_{m=0}^{\infty}(-\tanh r)^m \frac{\sqrt{(2m)!}}{2^m m!}|2m\rangle \quad (46)$$

This shows that the squeezed vacuum state has a mean photon number $\langle n\rangle = \sinh^2 r > 0$, even though it is derived from the vaccum state via squeezing. The two-mode squeezed state (EPR entangled state) is generated by the operator $\hat{S}(\zeta) = \exp[(\zeta \hat{a}_1 \hat{a}_2 - \zeta^* \hat{a}_1^\dagger \hat{a}_2^\dagger)/2]$. In two-mode squeezed state, each individual mode has an average photon number of $\langle n_1\rangle = \langle n_2\rangle = \sinh^2 r$. Although the photon number distribution within each mode itself may exhibit fluctuations, measurement reveal that the noise of the photon number differences between the two modes, $\langle (n_1 - n_2)^2\rangle$, is suppressed below the standard quantum limit – which is a direct manifestation of EPR entanglement.

The experimental measurement of squeezing relies on the balanced homodyne detection (BHD) technique. The core principle of BHD involves using a strong coherent light beam, the local oscillator (LO), as a reference. This LO interferes with the squeezed light under test on a 50:50 beam splitter.  The noise information is extracted by measuring the photocurrent difference from the two output ports. A piezoelectric ceramic (or phase shifter on chip) is used to scan the relative phase $\theta$ between the local oscillator and the signal light (**Fig. 23e and 23f**[279]). The differentiated output is an electrical signal proportional to the fluctuation of the signal light's quadrature component $\hat{X}_\theta$ at the phase angle $\theta$. By analysing the noise power spectrum of this electrical signal, $V(\hat{X}_\theta)$ is obtained. **Fig. 23g** depicts the photocurrent noise for coherent state and squeezed coherent state for illustration.

The measurement and calculation begin with calibrating the SQL. This is done by blocking the signal light path, allowing only the local oscillator to incident. In this state, the detector measures the vacuum fluctuations (i.e. the SQL). The amplitude of the noise power spectrum under this condition is recorded and defined as the noise reference (0 dB). Next, with the signal path open, the phase $\theta$ is scanned, and the noise power at different phase angles is recoded. The extremum points are then identified: the point where the noise power is minimum (the squeezed quadrature, $V(\hat{X}_\theta) < 1$) and the point where it is maximum (the anti-squeezed quadrature, $V(\hat{P}_\theta) > 1$). The **measured squeezing** is calculated using the formula (in decibels, dB):

$$S_{meas}(\text{dB}) = -10 \cdot log_{10} \frac{V_{min}}{V_{SQL}} \tag{47}$$

where $V_{min}$ is the measured minimum noise power and $V_{SQL}$ is the shot noise limit power. The measured squeezing, $S_{meas}$, is closely related to the total efficiency $\eta_{total}$ of the entire system. Knowing the total efficiency, the **inferred squeezing** ($S_{inf}$), which represents the ideal squeezing degree of the source itself, can be deduced using the model:

$$S_{meas} = -10 \cdot log_{10}\left[1 - \eta_{total}\left(1 - 10^{-S_{inf}/10}\right)\right] \tag{48}$$

If the system efficiency $\eta_{total}$ is low, even an ideal high-squeezing source will yield a poor measured squeezing degree. **Fig. 23h** depicts the quantum noise level of a squeezed light, from which the measured squeezing level is -4 dB and the anti-squeezing level is 11.85 dB.

### 5.1.2.b Generation of squeezed light sources

The generation of squeezed states mostly follows the three technical pathways: degenerate four-wave mixing in $\chi^{(3)}$ materials[280], degenerate parametric down conversion in $\chi^{(2)}$ materials[2], and self-phase modulation[281]. In $\chi^{(3)}$ materials, silicon nitride and silica platforms have been used for many quantum photonic experiments, thanks to their CMOS compatibility and integrability. Dutt et al. directly observed the photocurrent noise squeezing in a $Si_3N_4$ microring driven above the parametric oscillation threshold[89]. Subsequently, tunable squeezing has been realized by varying the coupling between two microrings[282]. Zhang et al subsequently utilized a structure comprising two adjacent optical microrings to enhance the squeezing factor to -8 dB[283]. Squeezed vacuum in a single-mode degenerate configuration has also been proposed based on dual-pump four-wave mixing in microrings[284], where noise contributions from nonparametric effects and unwanted parametric effects can be effectively avoided without significantly sacrificing efficiency. Microcombs based on SiN Kerr cavities have also been explored for squeezing generation. In Si/SiN, their inherently weak $\chi^{(3)}$ nonlinearity necessitates the use of high-quality factor resonators, which imposes limitations on accessible squeezing bandwidths to megahertz and gigahertz ranges. Moreover, most of the measured squeezing levels have so far remained around -2 dB below the shot noise level. Besides, the small wavelength separation between pump and squeezed light makes it challenging to remove the pump without introducing extra loss on squeezed light, and the pump efficiency is low compared with parametric down-conversion.

In $\boldsymbol{\chi^{(2)}}$ **materials**, originally, high-quality continuous-variable states are generated via parametric processes producing single-/two-mode squeezed vacuum in bulk crystals or large-mode-area waveguides (~10–100 μm²)[160,159,285,286]. The record for the highest level of squeezing, at -15 dB, has been set by optical parametric oscillators utilizing a bulk monolithic cavity design[287]. Nonetheless, there is a strong desire to achieve squeezed states of comparable quality in nanophotonic systems to facilitate the development of large-scale integrated quantum circuits. LN nanophotonics has opened a promising avenue in generating squeezing states due to its strong $\chi^{(2)}$ nonlinearity. Squeezed states are generated via both single-pass and cavity configurations. Periodically poled titanium-indiffused LN waveguides achieve up to −3.17 dB squeezing[288], yet their large bending radii constrain circuit depth[289], limiting practical CV protocol implementation. Alternatively, ZnO-doped ridge PPLN waveguides on $LiTaO_3$ substrates attain −6 dB squeezing across 2.5 THz[290] (**Fig. 24a**), with recent upgrades reaching -8 dB[291] and 6 THz bandwidths[292]. However, weak waveguide confinement and mechanical structuring fabrication methods limit highly dense photonic integration. In contrast, Z-cut TFLN ridge waveguides exploit strong $\chi^{(2)}$ nonlinearity and sub-wavelength field confinement, enabling single-pass parametric down-conversion with measured broadband quadrature squeezing of -0.56 dB (inferred -2.5 dB on-chip) across 7 THz bandwidth[293] (**Fig. 24b**). Quasi-phase matching in PPLN enables near-unity-efficiency squeezed light extraction in single-pass configurations, avoiding the challenging strong over-coupling required in cavities. In X-cut TFLN, femtosecond pulses generated −4.9 dB (inferred -11 dB) squeezing across 25 THz[294]. As **Fig. 24c** shows, the circuit integrates two dispersion-engineered phase-sensitive optical parametric amplifiers (OPAs): the first creates microscopic squeezed vacuum, amplified by the second OPA to macroscopic levels on-chip. This macroscopic field encodes microscopic state information with high loss tolerance.

Using cavities, OPO has also been developed in TFLN, where tuning over the coupling condition could lead to efficient on-chip squeezers[295] (**Fig. 24d**). As shown in **Fig. 24e**, a monolithic quantum photonic circuit containing PPLN was successfully demonstrated, utilizing waveguide second harmonic generation (SHG) to prepare the pump for stimulating an optical parametric oscillator, which in turn generates squeezed states[296]. A squeezing level of 0.55 dB was measured. Additionally, due to the use of a cavity, the bandwidth is limited to 140 MHz. Besides, considering the fabrication expense of poling structures, a compact squeezed light source using modal phase matching has been reported recently[297], and leads to the observation of up to -2.2 dB squeezing (**Fig. 24f**). The summary of typical experimental results of squeezed states in $\chi^{(2)}$ and $\chi^{(3)}$ waveguides and microcavities is provided in **Table 6**.

### 5.2 Quantum state manipulation

#### 5.2.1 Passive components

LNOI facilitates the fabrication of monolithic ridge waveguides with large index contrast and strong mode confinement. The waveguide on LNOI has demonstrated low propagation loss of 0.027 dB/cm[29,30], and beam splitters, MZIs, and their arrays have also been demonstrated[236,298,299] (**Fig. 25a**). Recently, the arrayed waveguide gratings have been demonstrated for the first time on X-cut LNOI. The best insertion loss of 2.4 dB and crosstalk of −24.1 dB are obtained for the fabricated device[300] (**Fig. 25b**). Regarding fiber-to-chip coupling, similar to silicon, there are two types, the end-fire coupling and grating coupling. End-fire coupling is polarization-independent and shows a typical fiber-to-chip coupling efficiency of -4 to -6 dB[47] using lensed fiber. Using a specifically designed mode convertor in the form of a two-step inverse taper, the coupling loss can achieve -1.7 dB[301] (**Fig. 25c**). Moreover, by adding a silicon-on-nitride cladding waveguide to the LNOI adiabatic taper, the coupling loss can be as low as 0.54 dB per facet[302] (**Fig. 25d**). Recently, Guo et al. developed a polarization-insensitive edge coupler comprising three width-tapered full-etched waveguides with silica cladding[303]. The device achieves minimum coupling losses of 0.9 dB ($TE_0$) and 1.1 dB ($TM_0$) per facet, with polarization-dependent loss below 0.5 dB across the 1260–1340 nm wavelength range. Grating couplers allow larger alignment tolerance but narrower bandwidth and are sensitive to polarization. There have been a number of reports about grating couplers on LNOI[304–308,272], with a record coupling efficiency as low as -3.5 dB per coupler[309] (**Fig. 25e**). Using buried metal layers can significantly improve coupling efficiency by acting as optical reflectors to mitigate downward light leakage[304,305]. Beyond monolithic LN gratings, fabricating hybrid gratings with amorphous silicon atop LNOI waveguides[24,310] (**Fig. 25f**) achieves up to -3 dB coupling efficiency per coupler with 55 nm 1-dB bandwidth[310].

Optical microcavities serve as essential integrated photonic components, enhancing wavelength filtering while significantly boosting nonlinear, electro- and acousto-optic interactions. Microdisk[311], microring, and racetrack resonators[32] (**Fig. 25g - i**) have been fabricated by dry etching[23,29,32,33,46,312,313], rib loading[248,314], CMP[30], as well as using BIC waveguides[250]. Resonators have enabled low-voltage EO modulators[32,33,315], broadband EO frequency comb generation[44], ultra-efficient parametric wavelength conversions[42,43,313,316–319], and Kerr comb generation[45,46,320,321]. Coupled rings with EO modulation form photonic molecules[322], enabling applications in frequency shifting[323] and microwave-to-optical photon conversion[324,325], which will be addressed in later sections.

### 5.2.2 Electro-optic components

Tunable materials enable the electromagnetic properties to be altered by applying an external signal. There are three types of LN tunability, through the EO, TO, and AO effects. EO tunability and accordingly EO modulator is the most distinguished part when comparing the LN platform to the Si platform. The fundamental physical picture of the EO effect describes a static electric field inducing a constant refractive index change ($\Delta n$) in the material. This effect typically exists in crystals lacking an inversion center, and LN is precisely such a crystal. the optical properties of anisotropic crystals are often described using the index ellipsoid. When no electric field is applied, the equation for the index ellipsoid of lithium niobate (a uniaxial crystal) is $\frac{x^2}{n_0^2}+\frac{y^2}{n_0^2}+\frac{z^2}{n_e^2}=1$, where $n_0$ and $n_e$ are the refractive indices for ordinary and extraordinary light, respectively. When an external electric field $E_k$ is applied, the index ellipsoid distorts. This change is described by the Pockels tensor ($r_{ijk}$, or using the contracted notation $r_{Ik}$), and is mathematically represented as an alteration in the coefficients of the index ellipsoid (i.e., the inverse permittivity tensor):

$$\Delta\left(\frac{1}{n^2}\right)_I=\sum_{k=1}^{3} r_{Ik}E_k \tag{49}$$

Here, the contracted subscript $I$ (1 to 6) corresponds to the original tensor subscripts $ij$ (11, 22, 33, 23/32, 13/31, 12/21). For LN, their lattice symmetry reduces the number of independent electro-optic coefficients. The form of tis Pockels tensor matrix is as follows:

$$r_{Ik}=\begin{bmatrix} r_{11} & r_{12} & r_{13} \\ r_{21} & r_{22} & r_{23} \\ r_{31} & r_{32} & r_{33} \\ r_{41} & r_{42} & r_{43} \\ r_{51} & r_{52} & r_{53} \\ r_{61} & r_{62} & r_{63} \end{bmatrix}=\begin{bmatrix} 0 & -r_{22} & r_{13} \\ 0 & r_{22} & r_{13} \\ 0 & 0 & r_{33} \\ 0 & r_{42} & 0 \\ r_{42} & 0 & 0 \\ r_{22} & 0 & 0 \end{bmatrix} \tag{50}$$

Among the several electro-optical coefficients of lithium niobate, two are particularly prominent in the design of integrated photonics, the $r_{33}$ and $r_{42}$. The dominant coefficient $r_{33}$ is the strongest electro-optic coefficient in LN, with a value of approximately 31 pm/V. When the electric field is applied along the z-axis (or the extraordinary axis) of the crystal, it induces a significant change in the refractive index along the z-axis. The approximate refractive index change is given by $\Delta n\approx-\frac{1}{2}n_e^3 r_{33}E_z$. Modulators based on $r_{33}$ that typically employ a transverse modulation structure where the electric field direction is perpendicular to the light propagation direction, are core components of intensity and phase modulators. The off-diagonal coefficient $r_{42}$ also has a large value, about 30 pm/V. When the electric field is applied akig the x-axis (or y-axis), it induces a shear deformation of the index ellipsoid in the xz (or yz) crystal plane. This effect can be utilized to achieve polarization rotation, construct optical isolators, or couple modes of different polarizations within an optical cavity.

The EO effect is one of the most attractive properties of LN. This effect serves as the critical link between electronics and photonics, unlocking widespread applications ranging from communications and computing to sensing and quantum information[326]. In recent years, leveraging the exceptional intrinsic EO coefficient $r_{33}$ of LN (which is~31 pm/V), continuous developments of LN EO modulators, based on both non-resonant and resonant optical structures, have been witnessed, pushing towards better modulation performances and diverse advanced applications, as depicted in **Fig. 26**. Although TFLN-based EO modulators were first demonstrated in as early as 2005 for non-resonant type[233] and 2007 for resonant type[32], they have since undergone relatively sluggish development. This is mainly restricted by the low-quality LN etching and the resulted high-loss LN waveguides. The initial significant advancement occurred circa 2013, establishing rib loading as a viable method to circumvent the need for LN etching[248]. Subsequently, various loading materials have been exploited, rendering rib-loaded modulators a predominant approach[242,38,327–330]. For resonant modulators, the improvement has also been achieved by replacing the microring resonator with a microdisk resonator, which largely reduces the scattering loss from the rough sidewall, leading to a higher quality ($Q$) factor and tuning rate. The subsequent major advancement occurred around 2016, marking the achievement of sub-1 dB/cm waveguide propagation loss in dry-etched LNOI modulators[28]. This monolithic approach has propelled swift advancements in the field, exemplified by the realization of modulators with over 100 GHz bandwidth functioning at CMOS-compatible voltages[35], as well as in-phase/quadrature (IQ) modulators capable of facilitating data rates of up to 320 Gbit/s[331]. At the same time, heterogeneous integration by bonding thin-film LN (either etched[332,333] or unpatterned[334]) on SOI and SiN PICs has also demonstrated considerable promise. Following the breakthrough in the fabrication technology, various designs have been proposed to further improve the modulator performance, such as the utilization of photonic crystal[335] or segmented electrodes[336] to realize high tuning efficiency and large modulation bandwidth simultaneously. Through the optimization of capacitance-loaded traveling-wave electrodes for low microwave attenuation and perfect velocity matching, IQ modulators with 1 V driving voltage and 110 GHz bandwidth were realized[337]. Besides the prosperous development of telecom-wavelength modulators, very recently in 2023, a visible-band modulator with sub-1 V driving voltage was reported[338]. Subsequently, a visible-telecom tunable dual-band optical isolator was demonstrated[339]. For a more thorough and detailed review of TFLN-based EO modulators, readers may refer to Refs. [326,340,341].

On top of the remarkable progress made in the fabrication process and modulator performance, various applications have been explored. Beyond standard telecom EO modulators, an integrated EO Fourier transform spectrometer has been demonstrated on a SiN-loaded TFLN platform[342]. Efficient EO interactions further enable EO frequency combs for spectroscopy and topological photonics applications. Both non-resonant[343] and resonant[44] EO combs have been demonstrated. Non-resonant EO combs offer favorable spectral flatness, while resonant EO combs provide broader spectral bandwidth (i.e., larger number of comb lines). Flat-top waveguide EO combs were demonstrated via cascaded amplitude and phase modulators[344]. The coexistence of EO and Kerr effects in TFLN enables on-site nonlinear interactions through Kerr effects within synthetic EO lattices, facilitating novel comb generation that synergizes both effects. This hybrid approach is exemplified by EO-Kerr combs in coupled-microring resonators[345] and EO-tunable Kerr frequency combs[346], capitalizing on integrated EO reconfigurability and Kerr-based nonlinear processes. Furthermore, the integration of new gain technologies with TFLN is promising and could lead to innovative developments. For example, a high-peak-power, electrically pumped actively mode-locked laser was developed in nanophotonic LN based on its hybrid integration with a III-V semiconductor optical amplifier[347].

EO modulation in coupled resonators introduces a plethora of intriguing physical phenomena, paving the way for innovative and practical applications. A photonic molecule with two distinct energy levels was achieved using a pair of coupled LNOI microring resonators[348]. Programmed microwave signals enable precise control of the frequency and phase of light. Additionally, Photonic molecules can be programmed into bright-dark mode pairs for photon storage and retrieval. Coupled resonator modulators enable GHz-scale frequency shifting and beam splitting. Complete frequency conversion is achieved via generalized critical coupling[349]. Unlike conventional single-sideband modulators, coupled-microring shifters mediate bidirectional photon frequency exchange, forming tunable frequency-domain beam splitters. More recently, a frequency circulator was demonstrated based on three coupled microring resonators, enabling 40 dB isolation with only 75 mW microwave power[350].

Time modulation enables synthetic dimensions in the frequency domain by arranging distinct optical frequencies into lattice structures. This photonic synthetic dimension framework is of particular interest to explore novel physics in complex systems like non-Hermitian, high-dimensional, and topological regimes. The simplest synthetic EO crystal configuration applies microwave-frequency (equal to cavity FSR) EO modulation to a single cavity. Such systems demonstrated four-dimensional frequency crystals on TFLN in 2020[351], achieved via multiple microwave drives with frequencies matched to cavity FSRs. Each drive established separate dimensions for photon hopping, enabling over 100 lattice sites per dimension. Furthermore, frequency-domain mirrors reflecting optical energy were implemented in frequency crystals using coupled-cavity and polarization-crossing methods[352]. By 2023, TFLN-based synthetic dimension research expanded to quantum regimes, leveraging quantum frequency conversion to explore entangled photon correlations and multilevel Rabi oscillations within extended dimensionalities[353]. Quantum applications benefit from EO-based frequency conversion linking superconducting microwave photons and optical network photons. Cavity EO devices integrating electro-optically active optical resonators with low-loss microwave cavities show promise for quantum transducers converting single photons between microwave/optical frequencies. Recent TFLN cavity EO transducers achieved high microwave-to-optical photon conversion efficiency at cryogenic temperatures[354,355].

#### 5.2.3 Control mechanism

Similarly, in TFLN platforms, controlling the degrees of freedom of photons, such as spatial modes, polarization, and frequency, is the key to realizing photonic quantum integration. **Passive components** do not require external control and can perform photon transmission and mode conversion without consuming additional power. LNOI-based passive devices, such as directional couplers, Y-junction splitters[356] (**Fig. 27a**), and multimode interference splitters[357] (**Fig. 27b**), are widely used in quantum optical circuits to manipulate the spatial distribution of photons[23,33,358]. Due to the orientation of its crystal axis, LN waveguides exhibit strong modal birefringence, allowing quantum circuits to exploit polarization degrees of freedom. Specially designed directional couplers have successfully demonstrated wideband polarization routing of photons[359]. Additionally, with an optimized waveguide design, modal birefringence in LN waveguides can be eliminated, ensuring both polarization modes have the same propagation constants[360]. For spectral control, LN-based Bragg reflectors – fabricated through periodic indentations on the waveguide surface – have been utilized in quantum photonics[306,361–363] (**Fig. 27c**). Due to the demands for higher extinction ratios and narrower spectral linewidths, resonators and interferometric filters are commonly integrated into quantum photonics to effectively filter out strong classical pump light[364] (**Fig. 27d**). On the LN platform, high-quality resonators and interferometric filters have been realized[23], particularly in the near-visible spectrum, where pump photons are situated for photon-pair generation in the communication wavelengths. Another important function of LN-based passive components is enabling nonlinear phenomena at the single-photon level. Leveraging the high nonlinearity of LN and incorporating microring resonator designs, researchers demonstrated near 1% single-photon anharmonicity[313]. These results indicate that with material and process improvements, stronger nonlinear effects and even photon blockade effects could be realized in the future, paving the way for deterministic all-optical quantum operations.

The high-speed and high-fidelity state manipulation is crucial to realizing photonic quantum integration, this requires **active control** of the quantum states. The above-discussed EO modulation is the most widely adopted technique in LNOI platforms, which induces changes of the material refractive index upon application of electric voltages. TO modulation - the widely used method in silicon photonics – has also been implemented in LN devices. However, due to the low TO coefficient, the modulation speed is limited by the heat transfer in LN[24,365]. Compared to conventional indiffused waveguides, EO modulators in LNOI waveguides can be operated at low voltages and high speeds due to the small waveguide dimensions and high field confinement. EO modulators are integrated on interferometers[35,366,367] (**Fig. 27e**), resonators[33,313] (**Fig. 27f**), or Bragg reflectors[363]. Based on the principle of interferometers, general modulators capable of controlling both phase and amplitude have been demonstrated[24], which form the heart for integrated quantum photonic circuits. Advanced features have been achieved in LN modulators, such as small modulation voltages compatible with CMOS electronics, modulation speed up to 100 GHz and even 400 GHz[37,367,368], extinction ratio as high as 53 dB in a cascaded design[369], and compact structures[370,371]. All these functionalities enable the realization of tunable photonic quantum gates for the spatial modes[298,372], which is the basis for integrated quantum circuits using the path degree of freedom.

#### 5.2.4 Frequency conversion

In photonic quantum systems, usually different functionalities require different photon frequencies: the communication band near 1.5 μm is suitable for long-distance transmission needs; photons in visible to near-infrared range can meet the needs of high-performance, miniaturized silicon single-photon detectors; quantum memory requires photon wavelengths to be close to the corresponding atomic energy levels. Therefore, efficient single-photon frequency conversion is crucial to the needs of providing different functionalities in constructing photonic quantum networks. The field of using LN to realize efficient frequency conversion originated from bulk LN optical superlattices[373,374]. Integrated LN devices based on optical superlattice waveguides have become mature, and significant progress has been made in single-photon frequency manipulation. Roussev et al. successfully demonstrated single-photon frequency upconversion from the communication band to the visible band on a proton-exchanged LN waveguide platform, achieving an on-chip conversion efficiency of 82% with a 4.8 cm device and an 88 mW pump power[375]. Recent studies show that PPLN waveguide can achieve quantum frequency conversion from the communication band to near-visible light, with an internal conversion efficiency approaching 50% and a noise level of just $10^{-4}$ photons per time-frequency mode[376]. In 2023, Wang et al. developed a single photon frequency upconversion device based on TFLN. This 5.3 mm long device achieved an on-chip conversion efficiency of 73% with a 310 mW pump power[377] (**Fig. 27g**), further demonstrating the feasibility of TFLN devices in single-photon frequency conversion applications. The frequency converter has also been demonstrated on PPLN microring resonator on chip, generating photon-pair sources[265] (**Fig. 27h**).

Besides nonlinear frequency conversion which can achieve large frequency difference, EO modulation can precisely tune frequencies within a smaller range, such as shifting frequencies across grid channels of dense wavelength division multiplexing[378]. Coupled microring resonators combined with the EO effect have achieved frequency shifts in the GHz range with near 100% efficiency and tunability, supporting the realization of frequency beam splitters[323]. Additionally, electro-optically induced gratings can be used to change the frequency of light by adjusting their structure, serving as tunable spectral filters for dynamically selecting and filtering optical signals[22]. The possibilities for controlling the spectrum of quantum signals in LNOI may be unrivalled by any other integrated photonic platform.

### 5.3 Single-photon detectors

Thin-film LN single-photon detectors open up the possibility to construct fully integrated quantum photonic chips, which incorporate on-chip photon sources, spectral filters, reconfigurable quantum gates, and detection modules. Similar to Si, LN-based single-photon detectors also have the two categories of avalanche photodiodes[379] and SNSPDs. Diamanti et al.[380,381] compared the performance of quantum-key-distribution systems using up-conversion single-photon detectors on LN waveguides and silicon avalanche photodiodes, with the LN demonstrating higher communication rates and longer transmission distances. Ma et al.[382] implemented a single photon detector and spectrometer based on frequency up-conversion technology for high-efficiency detection and spectral measurement in communication bands. Alternatively, an amorphous silicon photodetector has been also integrated on LNOI[383](**Fig. 28a**). Although the sensitivity of this device is not yet sufficiently high to detect single photons, the results underline the potential to realize avalanche photodiodes or fast detectors suitable for the characterization of squeezing. More recently, Wang et. al. [377] advanced the field by demonstrating quantum frequency conversion and single-photon detection with LN nanophotonic chips, showcasing the feasibility of upconversion single-photon detectors on-chip with a high detection efficiency of 8.7% and low noise count rates of 300 cps.

Superconducting nanowire single-photon detectors (SNSPDs) have emerged as the leading technology for efficient and broadband single-photon detection, particularly in the telecom wavelength range[384]. These detectors have been successfully integrated into on-chip optical waveguides across various material platforms[385], including recent implementations on LNOI waveguides[50,51,386]. For instance, Ref. [50] demonstrated a detection efficiency of 46% at 1560 nm using an NbN superconducting nanowire, which was only 250 μm in length and lithographically defined on a 125 μm-long LNOI ridge waveguide (**Fig. 28b**). This setup achieved a low dark-count rate of 13 Hz and a timing jitter of 32 ps. Furthermore, Ref. [51] integrated two SNSPDs and an EO switch onto the same LNOI chip. The successful operation of this integrated system at cryogenic temperatures, as shown in **Fig. 28c**, represents a significant step toward fully on-chip, fast reconfigurable photonic quantum circuits on the LNOI platform.

Despite the fact that the measured efficiencies in Ref. [50] were significantly underestimated due to the lack of correction for photon losses between the input port and the SNSPD on-chip, it is anticipated that SNSPDs on the LNOI platform will soon achieve the common efficiencies of over 90% that are already realized in other integrated platforms. For instance, periodic nanobeam structures have enabled near-unity detection efficiencies with recovery times below 10 ns on other platforms[387], and these structures could potentially be applied to the LNOI platform as well, thereby supporting fast repetition rates for the source. Overall, these threshold detectors typically exhibit a low sub-100 ps latency in their response[388]. In terms of number-resolving capability, transition-edge sensors (TES) have been implemented on conventional titanium-indiffused LN waveguides[389] (**Fig. 28d**). However, TES detectors have a relatively long response time, on the order of a few microseconds, which limits their suitability for fast multiplexing applications. To achieve a fast photon-number-resolving response, multiple threshold SNSPDs can be multiplexed together[390]. Although these schemes have a probabilistic response and require more detectors for more accurate number-resolving measurements, high detection efficiencies of up to 86% have also been reported for such detector systems[391].

# 6 LITHIUM NIOBATE INTEGRATED PHOTONIC QUANTUM COMPUTING

LN has been a burgeoning material in the field of photonics, showing significant promise for various applications due to its strong $\chi^{(2)}$ nonlinearity, EO effect, and low-loss linear optical characteristics. However, the majority of development efforts have been concentrated at the device level, with integrated works on LNOI still having significant room for expansion. In this section, we not only review the current state of integration efforts but also explore the broader implications of LNOI integration. This exploration is initially hinted by silicon photonics development and is further guided by a comprehensive review that reveals intriguing combinations and potential advancements. We will discuss the innovative ways in which LNOI can be fused into more complex systems, the challenges that need to be addressed, and the future prospects for integrated LN-based photonic platforms.

## 6.1 Integrated lithium niobate circuits

Rapid development in LNOI-based quantum computing circuits is seen, thanks to its excellent performance in EO modulation, integrated photonic circuits, and rapid control of photonic states. Recent research progress indicates that the LNOI platform can be used to implement programmable quantum interferometers, which are very important in quantum information processing and quantum computing. For example, Bonneau et al.[392] demonstrated electro-optically modulated quantum inference circuit based on titanium-diffused LN waveguides, achieving dynamic manipulation of photon state polarization and path dimensions. Jin et al.[59] integrated a quantum light source with a path interference device on the same proton-exchanged LN chip, creating the first LN-based photonic quantum chip. As shown in **Fig. 29a**, the chip, through the monolithic integration of entangled light sources, photon interferometers, and EO modulators, demonstrated the efficient generation and rapid switching of two-path entangled photon states. Subsequently, Luo et al.[393] achieved the generation of entangled photon pairs with electrical control of polarization states on titanium-diffused LN waveguides (**Fig. 29b)**. By leveraging on-chip photonic state polarization switching and transmission, the team achieved dynamically tunable optical delays with an adjustable range of up to 12 ps, and demonstrated a fully integrated on-chip two-photon interference experiment.

A key advantage of LNOI quantum computing circuits lies in their EO modulators. Compared to the TO modulators in traditional SOI platforms, EO modulators offer faster modulation speeds, which is crucial for special applications such as rapid simulation of quantum Hamiltonians. Recently, Sund et al.[394] demonstrated a programmable quantum interferometer using the LNOI platform, capable of rapid control functions of photonic states, including on-chip two-photon interference, deterministic single-photon path conversion, and more complex two-photon quantum bit interference (**Fig. 29c**). Zheng et al.[299] demonstrated a 6-mode arbitrary unitary matrix on LNOI chips, and benchmarked the chip with several deep learning missions including in situ training (**Fig. 29d**). It proves the potential of LN for demonstrating large-scale computing circuits, although currently operating in the classical light regime rather than quantum operations. Assumpcao et al.[395] demonstrated an integrated TFLN photonic platform operating across the visible to near-infrared (VNIR) wavelength range (**Fig. 29e**). They employed high-bandwidth electro-optic modulators (>50 GHz) to achieve high-efficiency (>50% at 15 GHz), continuous-wave (CW)-compatible frequency shifting, along with simultaneous laser amplitude and frequency control. With this architecture, they showcase a multiplexing framework for quantum memories and explain how this platform can achieve a hundredfold increase in entanglement rates compared to individual memory nodes. These findings illustrate that TFLN is capable of meeting the required performance and scalability standards for large-scale quantum nodes.

The development of this part relies on the concept of linear optical quantum circuits, which can be inherited from SOI quantum computing circuits, as we described in Section 3. Since they rely on MZIs with modulators to manipulate photonic states, the availability of EO modulators instead of TO modulation in SOI further provides the possibility of enhancing modulation speeds. LN and Si share a similar development path in quantum circuits based on linear optics, and many circuits developed for silicon can be realized on LN, with the added benefits of higher-performance light sources, faster programming speeds, and the capability for integrating modulators and integrated detectors at cryogenic temperatures. Therefore, migrating quantum algorithms that have already been developed on silicon to the LN platform will offer a lot of room for performance improvement. Furthermore, in the next sections, we would like to provide some implications of where LN holds advantages over silicon that are unparalleled.

### 6.2 Cluster states and measurement-based quantum computing

Many demonstrations of integrated silicon photonics utilize discrete variables. However, scalability encounters a bottleneck because generating entanglement between qubits encoded in discrete variables of single photons is challenging. This difficulty stems from facilitating interactions between individual photons on photonic chips. Devices that operate with continuous variable (CV) offer a promising alternative, as they enable the deterministic generation and entanglement of qumodes. In CV scheme, information is encoded in continuous amplitude and phase values of the quantized electromagnetic field, necessitating the generation of squeezed sources. Moreover, highly entangled quantum states (i.e., cluster states) created from squeezed sources serve as a universal resource for one-way computing. Ideally, we would like the squeezed sources to be broadband to facilitate time-multiplexed CV-MBQC schemes, as the bandwidth determines the minimum width of the temporal mode. Additionally, a high level of squeezing is required to achieve the fault tolerance threshold for error correction[396–398]. For a quantum processor, not only broadband light but also high-level squeezing is required to obtain the quantum effect, such as at least 3 dB for entanglement swapping[399,400] and 4.5 dB for the generation of 2D cluster states[160]. Theoretical studies have suggested that quantum error correction and fault-tolerant quantum computing are possible in photonic CV-based approaches[396] when squeezing reaches 10 dB[397].

Leveraging the squeezed states generated on TFLN, this platform shows great promise for the creation of cluster states[401–404], which are essential for CV computing protocols. Although there have been few explicit demonstrations of such works in TFLN to date, the field is on the brink of such advancements. The quantum correlation form of the CV cluster state is given by

$$\hat{P}_a - \sum_{b \in N_a} \hat{X}_b \equiv \hat{\delta}_a \to 0, a \in G \tag{51}$$

where $a \in G$ represents the node in the graph $G$ that describes the cluster state, $b \in G$ represents the neighbouring modes connected to mode $a$, and $\hat{\delta}_a$ represents additional noise. A multipartite entangled state possessing the form of **Eq. (26)** is referred to as a CV cluster state. Under the condition of ideal squeezing, an N-mode CV cluster state is the zero-eigenvalue state of the linear combination operator of the orthogonal components of each mode, i.e.,

$$\left(\hat{P}_a - \sum_{b \in N_a} \hat{X}_b\right)|\psi\rangle = 0 \tag{52}$$

for all $a \in G$. Cluster state exhibits entanglement persisting properties, since measuring one mode only disrupts the quantum correlation between it and its adjacent nodes, leaving the entanglement among the remaining modes intact. Till now, experimental preparation of cluster states at different scales have been achieved utilizing multiplexing technique for spatial, temporal and frequency degrees of freedom of the optical filed.

**Spatial multiplexing** couples squeezed light fields through a network of beam splitters. Using this method, Su et al. generated CV four-partite GHZ states and cluster states[405]. This method allows local operations to be applied to each individual model. However, as the number of entanglement modes increases, the beam splitter network structure becomes significantly more complex. **Temporal multiplexing** was first proposed in 2011 by Menicucci et al. to generate CV cluster states at arbitrarily large scale[406]. By 2013, Yokoyama et al. experimentally realized a cluster state comprising 10,000 modes, which was subsequently scaled up to a million entanglement modes[407,408]. In 2019, Asavanant et al. implemented the generation of time-domain-multiplexed 2D cluster states, generated and verified a universal cluster state with a 5-by-1240-site 2D square lattice structure[160] (**Fig. 30a**). Larsen et. al. proposed a scalable scheme for generating photonic cluster states using temporal multiplexing of squeezed light modes, delay loops, and beam-splitter transformations, demonstrating the deterministic generation of a 2D cluster state suitable for universal quantum information processing[159] (**Fig. 30b**). Recently, Fukui et al. proposed a scheme for creating 3D cluster states[409]. They demonstrated that these entangled states exhibit robustness against analog errors arising from finite squeezing in measurement-based topological quantum computation, highlighting their potential as a critical quantum resource for achieving fault-tolerant quantum computation. **Frequency-multiplexed** generation of continuous-variable cluster states typically relies on optical frequency combs that reuse the resonant modes of a single optical parametric amplifier[410]. Using this approach, Chen et al. produced a 60-mode cluster state[411], whereas Roslund et al. and Cai et al. demonstrated 10- and 13-mode versions, respectively[412,413]. Recently, integrated silicon-nitride photonic chips have demonstrated eight-mode continuous-variable entanglement[414]. The chip generates a microcomb producing multimode squeezed-vacuum states via optical frequency combs operating below threshold (**Fig. 30c**). The inseparability of the resulting eight-mode state was experimentally verified, demonstrating multipartite entanglement across supermodes spanning hundreds of megahertz. The maximum number of frequency combs is limited by the influence of phase-matching bandwidth constraints, and the individual entangled modes are not easily separable.

Based on the generated CV cluster states, measurement-based computing can be executed[169,415–420]. Based on the cluster generation in **Fig. 30d**, Larsen et al. proposed a universal continuous-variable quantum computation that incorporates linear-optics-based continuous-variable cluster state generation and cubic-phase gate teleportation[421] (**Fig. 30e**). This work demonstrated the deterministic implementation of a multi-mode set of measurement-induced quantum gates in a large 2D optical cluster state, marking a significant step towards universal scalable quantum computing. Continuous-variable systems currently face limitations due to their reliance on free-space optical networks. The growing need for greater complexity, reduced loss, high-precision alignment and stability, as well as hybrid integration, calls for a different approach. Masada et al. previously demonstrated an integrated photonic platform that realized the core capabilities for continuous-variable quantum technologies, specifically the generation and characterization of Einstein-Podolsky-Rosen (EPR) beams[399] (**Fig. 30f**). They use bulk OPO to generate squeezed light, and coupled it into a silicon photonic chip with four variable-reflectivity beam splitters. If combined with integrated squeezing in LN and non-Gaussian operations, these results will open the way to universal quantum information processing with light. Leizini et al. have achieved the generation, manipulation, and interferometric stage of homodyne detection of nonclassical light on a single device, marking a key step toward an integrated approach to quantum information processing with continuous variables[289]. As depicted in **Fig. 30f**, the chip incorporates PPLN waveguides that generate squeezed vacuum states. These states are subsequently manipulated in reconfigurable directional couplers for entangled state generation, signal isolation, and homodyne detection. This work is based on a z-cut LN substrate by reverse proton exchange. If transferred to low-loss, high-confinement ridge waveguides in LNOI, the squeezing factor can potentially reach more than 10 dB, and the footprint can be reduced to be comparable to the SOI platform, enabling integration of more functionalities on the same chip.

Based on these protocols, the integration of CV measurement-based quantum computing on TFLN platforms holds immense potential. TFLN's ability to generate high levels of squeezing and its compatibility with high-speed modulation techniques make it an ideal candidate for the construction of large-scale cluster states through time-multiplexing schemes. These capabilities empower the demonstration of a variety of quantum algorithms, heralding a new era in quantum computing. Future developments in this field will likely focus on optimizing squeezing levels, expanding bandwidth, and enhancing the integration of these systems with existing quantum computing algorithms. The ongoing efforts in this area promise to unlock the full potential of TFLN for quantum technologies. We can anticipate a significant leap forward in the practical implementation of CV quantum computing, leading to more efficient and scalable quantum information processing solutions.

### 6.3 Low-loss delay line and on-chip photon source multiplexing

Both silicon's SFWM and LN's SPDC are probabilistic; This leads to a rapid decline in success rates when scaling up to multiple sources, thereby restricting the generation of resource states. However, photon source multiplexing provides a straightforward solution to achieve determinism. Fast, low-loss switches dynamically direct a single photon from numerous multiplexed photon pair sources to a single output. The scheme for photon source multiplexing is as shown in **Fig. 31a**. The photon sources are identical to each other, and each is heralded by another photon from the same pair. One photon from each heralded single-photon source is directed to a single-photon detector, while the other photon is sent to a routing network of switches. These switches are electrically controlled by the combined signals from the detectors. The signals from all detectors are analyzed to determine which heralded single-photon source has successfully produced a pair. Subsequently, a photon from only one of the corresponding sources is routed to the output. This process increases the likelihood of obtaining an output while preserving high fidelity, which reflects the degree of overlap between the output and a pure single-photon state.

From the scheme we see that the photon source multiplexing scheme requires various components, including SPDC sources, single-photon detectors, delay lines, and fast switches, along with fast electronics to synchronize the operations. To date, no fully integrated system with all components on the same chip has been realized, which has impeded the achievement of performance levels necessary for large-scale quantum computing and simulation protocols. We think that the unique strengths of the nanostructured LNOI platform could pave the way for a fully integrated multiplexed source. This is because it enables efficient and tailored photon pair generation, as well as low-loss and rapid switching[422], both of which are crucial for any multiplexing approach. To successfully implement the multiplexing scheme and feed-forward quantum operations on a chip, the key lies in developing high-repetition-rate, low-latency electronic devices that are closely integrated with optical chips. This latency refers to the time difference between the detection of a photon by an SNSPD and the processing of the corresponding electrical signal to activate a switch. To synchronize single photons in different channels, precise delays need to be introduced on the chip, which is typically achieved using long waveguides, but this method increases photon transmission loss. For example, to achieve a 1-ns delay on the LNOI platform, a ridge waveguide as long as 14 cm may be required, resulting in a loss of about 0.4 dB.

To reduce this loss and achieve efficient multiplexing, Zhou et al. utilized the PLACE processing technology to create a meter-scale photonic delay line with ultra-low loss on TFLN, with a transmission loss of about 0.03 dB/cm[371,423] (**Fig. 31b**). Song et al. demonstrate a continuously tunable delay line in X-cut TFLN driven by graphene electrodes, featuring a continuously tunable delay range from 0 to 100 ps[424] (**Fig. 31c**). At the time, the technology for fabricating long electrodes was not yet available, so the dynamic switching of the photon delay line's delay amount was achieved through on-chip optical switches, which precluded continuous adjustment of the delay amount. Recently, the efficient fabrication technology for on-chip long electrodes has been realized, and when combined with low-loss optical waveguides, it enables photon delay lines with continuous tuning capabilities and a broad range of delay tuning[425] (**Fig. 31d**). The delay line features a racetrack-shaped waveguide configuration, with gold electrodes distributed on both sides of the straight waveguide section. It consists of a 30 cm long waveguide delay line integrated with multiple segments of micro-electrodes totalling 24 cm in length. The measured on-chip transmission loss is approximately 0.025 dB/cm. The photon delay line is able to continuously and finely adjust the time delay, with a delay tuning efficiency of about 3.146 fs/V. By raising the voltage to 70 V (limited by the breakdown voltage of the TFLN), a broad range of time delay tuning exceeding 220 fs was achieved. Recently, continuous tuning of on-chip optical delay with a microring resonator is demonstrated[426] (**Fig. 31e**). By incorporating an electro-optically tunable waveguide coupler, the coupling between the bus waveguide and the resonance can be effectively adjusted from the under-coupling regime to the over-coupling regime. The optical delay is experimentally determined by measuring the relative phase shift between lasers. It exhibits a wide dynamic range of delay from −600 to 600 ps, and efficient tuning of the delay from −430 to −180 ps and from 40 to 240 ps can be achieved with just a 5 V voltage [426].

Furthermore, to achieve fast photonic switch modulation to route the selected photon, low-loss, high-bandwidth modulators need to be implemented on the TFLN platform. These modulators require CMOS-compatible driving voltages and have already been realized on the LNOI platform. Unlike silicon-based and indium phosphide modulators that use nonlinear modulation methods, LN modulators utilize linear EO effects, enabling higher modulation speeds, linearity, and lower power consumption. By designing traveling-wave electrodes, TFLN devices can easily achieve a 3 dB bandwidth of over 100 GHz[35,37,340,427] and even terahertz[39], which is significant for high-speed optical communication, optical interconnects, and on-chip optical computing.

Finally, it must be mentioned that all current multiplexing schemes suffer from electronic delays on the order of hundreds of nanoseconds, which necessitates the use of off-chip optical fibers as a practical solution to introduce such long delays, requiring 20 meters of fiber to achieve approximately 100 nanoseconds of delay. Therefore, reducing electronic latency is essential. A significant portion of this electronic delay is due to the time required for radiofrequency (RF) signals to propagate from the detectors to off-chip circuits for analysis and then back to the switches. On the LNOI platform, it is possible to integrate all optical components onto the same chip, allowing for the placement of electronic and optical components in close proximity, such as by flip-chip bonding the electronic chip to the optical chip, which significantly reduces the communication delay between the optical and electronic chips[428], as shown in **Fig. 31f**. It is crucial to use fast and low-latency RF and electronic components to minimize this delay as much as possible. Importantly, these components should also be compatible with the cryogenic temperatures required for the operation of SNSPDs, and further development of electronic and RF components for quantum technology applications is needed.

### 6.4 Microwave-to-optics interface for quantum computing

To outperform classical computers in practical computational tasks, state-of-the-art quantum computers must significantly increase the number of high-quality qubits, potentially by several orders of magnitude[429]. As superconducting qubits and photonic photons emerge as the two primary platforms for providing qubits, with photons being the most suitable long-haul quantum information carrier, efficient interconnection between these platforms is paramount. There is a critical demand for efficient bidirectional conversion between microwave and optical photons at the quantum level[430–432]. In recent years, the development of microwave-to-optical frequency conversion technology[433] has garnered significant attention in the field of quantum computing and networking and is considered essential for the long-distance transmission of quantum information, as it facilitates the efficient conversion between microwave signals in a cryogenic environment and optical signals at room temperature.

At present, there are three primary approaches for microwave-to-optical conversion: electro-optomechanical (EOM), acousto-optic (AO), and electro-optical (EO). The comparison of the three approaches is shown in **Table 7**. The EOM approach utilizes a mechanical vibration resonator as an intermediary. Microwave signals drive the mechanical resonator to vibrate (generating phonons) via either the piezoelectric effect or capacitive gradient forces. These mechanical vibrations then couple to an optical cavity or waveguide, modulating the optical signal through the optomechanical effect. While this approach leverages mature MEMS/NEMS technology and enables bidirectional conversion, the mechanical resonator is susceptible to environmental thermal noise and difficult to achieve quantum ground-state operation, and requires cryogenic temperatures to mitigate thermal photons. The use of low-frequency mechanical resonators fundamentally limits the conversion bandwidth due to their mechanical Q-factor constraints[434].

The AO approach employs propagating acoustic waves as an intermediary. As LN is a strong piezoelectric material, when a GHz microwave signal is applied to interdigital transducers (IDTs) integrated on the TFLN platform, the IDTs convert the microwave electric field into mechanical strain, exciting either surface acoustic waves (SAWs) or bulk acoustic waves (BAWs)[435]. The acoustic wave frequency is inversely proportional to the IDT electrode spacing, necessitating sub-micron electrodes for GHz operation. The propagating acoustic waves generate a moving refractive index grating via the photoelastic effect. As light propagates through the LN waveguide and interacts with this moving acousto-optic grating, Brag diffraction occurs, leading to a frequency shift in the diffracted light through stimulated Brillouin scattering (SBS). This approach inherently enables frequency conversion (shift magnitude equals the acoustic wave frequency). However, its conversion efficiency is typically low, requiring intense pump laser light to enhance the SBS effect.

The EO approach exploits the intrinsic electro-optic effect (mainly Pockels nonlinearity) to enable direct conversion between GHz microwaves and optical photons without intermediate excitations. Microwave signals directly interact with the nonlinear optical material, generating a time-varying electric field. This field directly modulates the optical refractive index of the material. By confining the optical field within the same electro-optic resonant cavity, the refractive index changes enable direct, intermediary-free energy conversion between microwave photons and optical photons (via frequency modulation or sideband generation). The EO approach principally enables high-speed conversion with low noise. The absence of low-frequency mechanical resonators or propagating acoustic photons eliminates additional thermal noise channels, facilitating operation at the quantum limit. It also offers a compact structure, particularly suitable for integration on TFLN platforms.

In literature, bidirectional EO conversion at the microwave ground state has been demonstrated with bulk LN[436] and integrated AlN resonators[437]. Targeting at higher conversion efficiency, Xu et al. demonstrated a bidirectional EO conversion between microwave and light using a TFLN platform[52], because of its strong Pockels nonlinearity and large vacuum EO coupling rate. As shown in **Fig. 32a, b**, they fabricated a pair of strongly coupled ring resonators patterned from X-cut TFLN and a superconducting microwave resonator of NbN. Due to the cavity-enhanced EO interaction under the triply-resonant condition, the input microwave field modulates the optical pump and creates an optical sideband at the signal mode frequency. Conversely, a microwave field output can be produced through optical frequency mixing within the lithium niobate LN cavity. A substantial increase in conversion efficiency is achieved, reaching an on-chip conversion efficiency of $1.02 \pm 0.01\%$ (internal efficiency of $15.2 \pm 0.4\%$) recorded with a peak optical pump power adjusted to 13.0 dBm in the waveguide (**Fig. 32c**). By employing an air-clad device architecture, this design effectively suppressed the photorefractive effect that typically limits the cryogenic performance of TFLN, and establishes the TFLN-based EO system a competitive platform for future quantum network applications.

Weaver et al.[53] introduced an integrated microwave-to-optical interface device that uses a planar superconducting resonator coupled to a silicon photonic cavity through a LN-on-silicon mechanical oscillator (**Fig. 32d**). The device achieved a peak transduction efficiency of 0.9% using 1 μW of continuous optical power and had a spectral bandwidth of 14.8 MHz (**Fig. 32e**). When using short optical pulses, the added noise was kept to just a few photons, and the repetition rate could go as high as 100 kHz. This device can connect directly to a 50 Ω transmission line and can be scaled up to include a large number of transducers on a single chip. This sets the stage for distributed quantum computing. These studies indicate that by utilizing LN and specific designs, it is possible to enhance the efficiency of microwave-to-optical conversion, which is vital for scaling up quantum computers and quantum networks. As technology continues to advance, we can anticipate the realization of more efficient and larger-scale quantum information processing and transmission systems in the future.

## 7 PERSPECTIVES AND OUTLOOK

**Key milestones (Fig. 33).** Silicon photonics established foundational quantum capabilities, starting with on-chip CNOT gates and quantum walks (2008). By 2014, entanglement generation and high-efficiency quantum dot sources enabled scaled photonic circuits. The platform matured with multi-dimensional entanglement (2016), near-ideal photon sources (2018), and programmable quantum processors (2022). Recent breakthroughs include fusion-based quantum computation and large-scale graph-state simulations (2024), solidifying SOI's roles in complex quantum algorithms. SOI still remains indispensable for the long term due to its mature CMOS ecosystem, which uniquely enables ultra-large-scale integrated quantum circuits.

Compared to CMOS-compatible fabrication of SOI platforms, the LNOI platform required significantly more time during its early development to optimize device fabrication, due to the intrinsic hardness and brittleness of LN that created a thin-film processing bottleneck. Another challenge is the chemical inertness along the Z-axis, which necessitated the creation of specialized plasma dry etching methods rather than leveraging mature silicon deep reactive-ion etching processes. LNOI progressed from foundational material innovations, such as low-loss waveguides (<0.03 dB/cm) and monolithic wafer-scale fabrication (2017) to functional quantum devices. Critical milestones include polarization modulators (2010), photon-pair generation in PPLN waveguides (2019), and cascaded electro-optic (EO) combs (2020). Recent advances feature microwave-photonic processing engines (2022), inverse-designed circuits (2023), and adapted poling techniques (2024), positioning LNOI as a potential leader in high-speed, low-noise quantum photonics.

Based on the review of the silicon and LN platform - spanning basic components, manipulation tools, and integrated circuits - LN demonstrates some clear performance advantages in photonic quantum computing, which are manifested in three aspects:

a) EO Effect: LN exhibits a strong EO effect, which facilitates effective modulation of light phase and amplitude through the application of electric fields. This enables precise control without the disturbance of thermal crosstalk or heat accumulation issues that are commonly associated with silicon. High-speed EO modulation is essential for the realization of key components like photonic quantum logic gates and for executing rapid quantum operations and multiplexing of quantum light sources. Its carrier-injection-fee operation fundamentally eliminates free carrier absorption noise and thermo-refractive fluctuations inherent to silicon-based modulators, delivering advantages in quantum noise suppression. Moreover, when integrated with superconducting nanowire single-photon detectors, EO modulators can provide operation modes at cryogenic temperatures, a versatility that silicon TO devices cannot match.

b) Broad transparency window: LN exhibits broad spectral compatibility, maintains low optical loss across visible to mid-infrared wavelengths (350-5000 nm), while silicon suffers from significantly enhanced absorption (>3 dB/cm) beyond 1550 nm. This unique low-loss transparency enables direct compatibility with diverse quantum memory wavelengths (e.g., 880 nm for $Nd^{3+}$, 1530 nm for $Er^{3+}$) and supports the integration of multi-wavelength entanglement sources critical for quantum repeater networks.

c) Nonlinear Optical Effects: LN exhibits strong second-order nonlinear effects, enabling efficient frequency conversion and the generation of non-classical light states at lower pump powers, such as heralded single photons, entangled photon pairs, and squeezed light. These are vital resources for quantum computing and quantum communication. In terms of frequency conversion, LN's nonlinearity is highly useful for transferring quantum information across different wavelength bands, a functionality that silicon finds difficult to provide.

Building on these advantages, LN is expected to see significant development in the following areas: **Migration of silicon photonic integrated circuits to LN platforms**: By making the shift to LN platforms, we can take advantage of LN heralded single-photon sources that offer higher generation rates at lower pump powers, leading to improved purity. This will allow increased numbers of available photons in probabilistic photon quantum circuits, as well as faster quantum state manipulation, thus promising to elevate the performance of integrated photonic quantum computing to a new level. For example, implementing full-chip quantum simulation based on Gaussian boson sampling is feasible, with a significant advancement being the replacement of low-squeezing photon-pair sources, typically generated through SFWM, with high-squeezing light sources produced via SPDC. This enhancement allows for the expansion of linear optical circuits and the positioning of single-photon detectors at the end of the circuit. Here, the detectors need number-resolving capabilities, which, as previously stated, are principally achievable on the LNOI platform. **Development of squeezed state-based CV quantum computing**: LN is particularly suitable for developing CV quantum computing architectures that rely on squeezed states. It is required to achieve a squeezing factor of 10 dB for the implementation of fault-tolerant CV quantum computing, a milestone that is nearly unattainable with silicon but appears quite feasible with LN, thanks to its strong nonlinear optical properties. **Applications requiring ultra-low transmission loss**: LN is ideal for applications that demand extremely low transmission loss, such as forward control-based multiplexing architectures for achieving deterministic single-photon sources. These are essential for advancing the development of quantum computing and communication technologies.

## 8 CONCLUSIONS

The high-fidelity integration of photonic components and quantum computing circuits holds immense promise for advancing quantum technologies. Research in integrated quantum photonics represents a critical transition from discrete optical elements to multifunctional circuits, providing a robust pathway toward miniaturized, scalable, and high-performance quantum information processing. In this review, we have surveyed fundamental components for photonic quantum computing on SOI and LNOI platforms, as well as the advanced integrated circuits in quantum information processing. The LNOI platform is exceptionally well-suited for large-scale integrated quantum photonics. Generally, every element required by quantum functionalities has now been experimentally demonstrated, though initially for classical applications, enabling direct translation of existing SOI designs to LNOI. Beyond this compatibility, LNOI offers distinctive advantages: high-rate, high-purity heralded single-photon sources, scalable multiplexing architectures, and ultrafast quantum state manipulation capabilities collectively enable large and versatile photonic quantum circuits. Furthermore, as evidenced by squeezed-state generation and CV-MBQC implementations, LNOI provides a viable path toward fault-tolerant quantum systems. Its seamless interface with other quantum platforms further facilitates hybrid quantum networks and expands the quantum technology ecosystem. We thus assert that LN—long a cornerstone of integrated and nonlinear optics—holds exceptional promise for integrated quantum photonics.

Despite significant advancements, quantum photonics remains an emerging field with substantial challenges. Translating current blueprints into practical systems demands resolving key hurdles such as increased component density, enhanced functionality, and monolithic integration of active-passive quantum elements. Addressing these goals will require continuous innovation throughout the development lifecycle. By systematically tackling technical bottlenecks, we can accelerate progress toward scalable, high-performance quantum photonic systems.

## ACKNOWLEDGMENTS

**Funding:** Supported by National Natural Science Foundation of China (62505228, 62405173, 62020106009, 62111530053, 61621001), National Key Research and Development Program of China (2023YFF0613600), Shanghai Committee of Science and Technology (25JD1406000), Chenguang Program of Shanghai Education Development Foundation and Shanghai Municipal Education Commission (24CGA19), and Fundamental Research Funds for the Central Universities.

**Competing interests**: The authors declare no competing interests.

## FIGURES

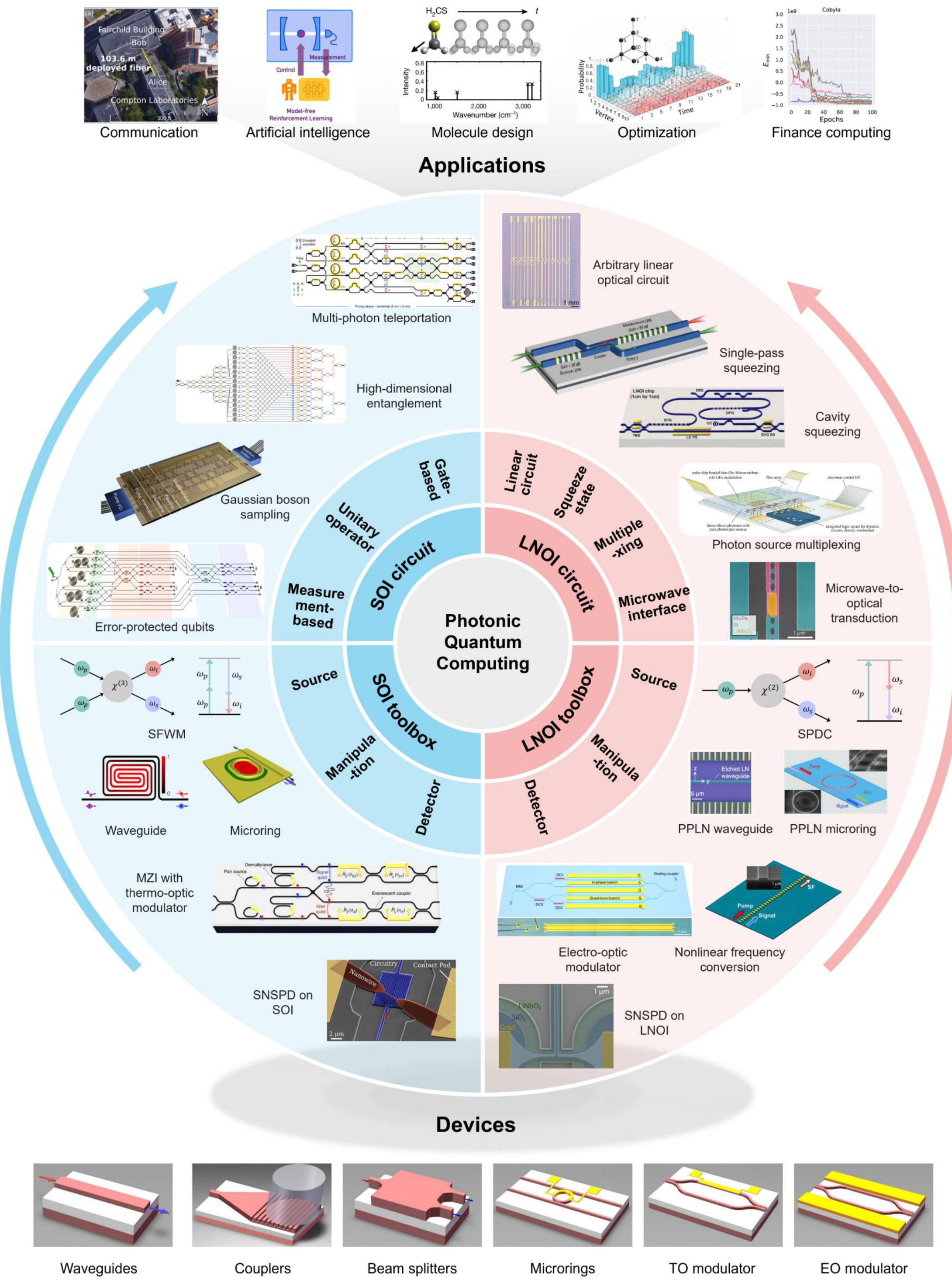

**Fig. 1 | Overview of photonic quantum computing on silicon and thin-film lithium niobate Platforms.** The figure is structured in a hierarchical manner, starting from basic devices at the bottom and culminating in applications at the top. The first tier represents the foundational devices, which are largely shareable between the silicon-on-insulator (SOI) and lithium niobate on insulator (LNOI) platforms. The second and

third tiers, specifically the toolbox and circuitry, are the focal points of this review, with close interconnections between the components and their applications in the quantum computing realm. The second tier is photonic quantum toolbox encompassing light sources, quantum manipulation techniques, and single-photon detectors, which are crucial for both platforms. Moving further up to the third tier, the diversity in the integrated quantum computing circuits becomes apparent, with Si platforms presenting gate-based, unitary operator-based, and measurement-based circuits, and LN platforms presenting linear circuits, squeezing states, multiplexing circuits, and microwave-optics quantum interfaces. The top tier exemplifies the applications of both platforms, underscoring their commonality in quantum computing potential despite architectural differences in component and circuitry. **Abbreviations:** Wg, waveguide; TO, thermo-optic; EO, electro-optic; SFWM, Spontaneous four wave mixing; MZI, Mach-Zehnder interferometer; SNSPD, Superconducting nanowire single photon detector; SPDC: Spontaneous parametric down conversion; PPLN: periodically poled lithium niobate. **References:** Si toolbox – SFWM photon pair source in long waveguides: ref. [66], SFWM photon pair source in microring resonator: ref. [61], MZI with thermo-optic modulator: ref.[127], SNSPD using photonic crystal cavity on Silicon platform: ref. [120]; Si circuit – multi-photon implementation of chip-to-chip gate-based quantum teleportation: ref. [76], high-dimensional quantum entanglement and computing: ref. [146], simulating Gaussian boson sampling using arbitrary unitary operator: ref. [189], error-protected qubits in measurement-based quantum computing: ref. [163]; LN toolbox: SPDC photon pair source in PPLN waveguide: ref. [49], SPDC photon pair source in PPLN microring resonator: ref. [265], electro-optic modulators: ref. [24], single-photon nonlinear frequency conversion: ref. [377]; LN circuit: arbitrary linear optical circuit: ref. [299], high-factor squeezing state generation in single-pass optical parameter amplifier: ref. [294], squeezing state in cavity-based optical parametric oscillator: ref. [296], dense chip layout for photon source multiplexing, microwave-to-optical transduction for quantum computing: ref. [53]; Applications: Quantum key distribution for communication: ref. [54], quantum reinforcement learning: ref. [55], simulating of molecular dynamics for potential drug design: ref. [179], implementation of graph-theory related quantum optimization algorithms: ref. [132], quantum finance computing: ref. [56].

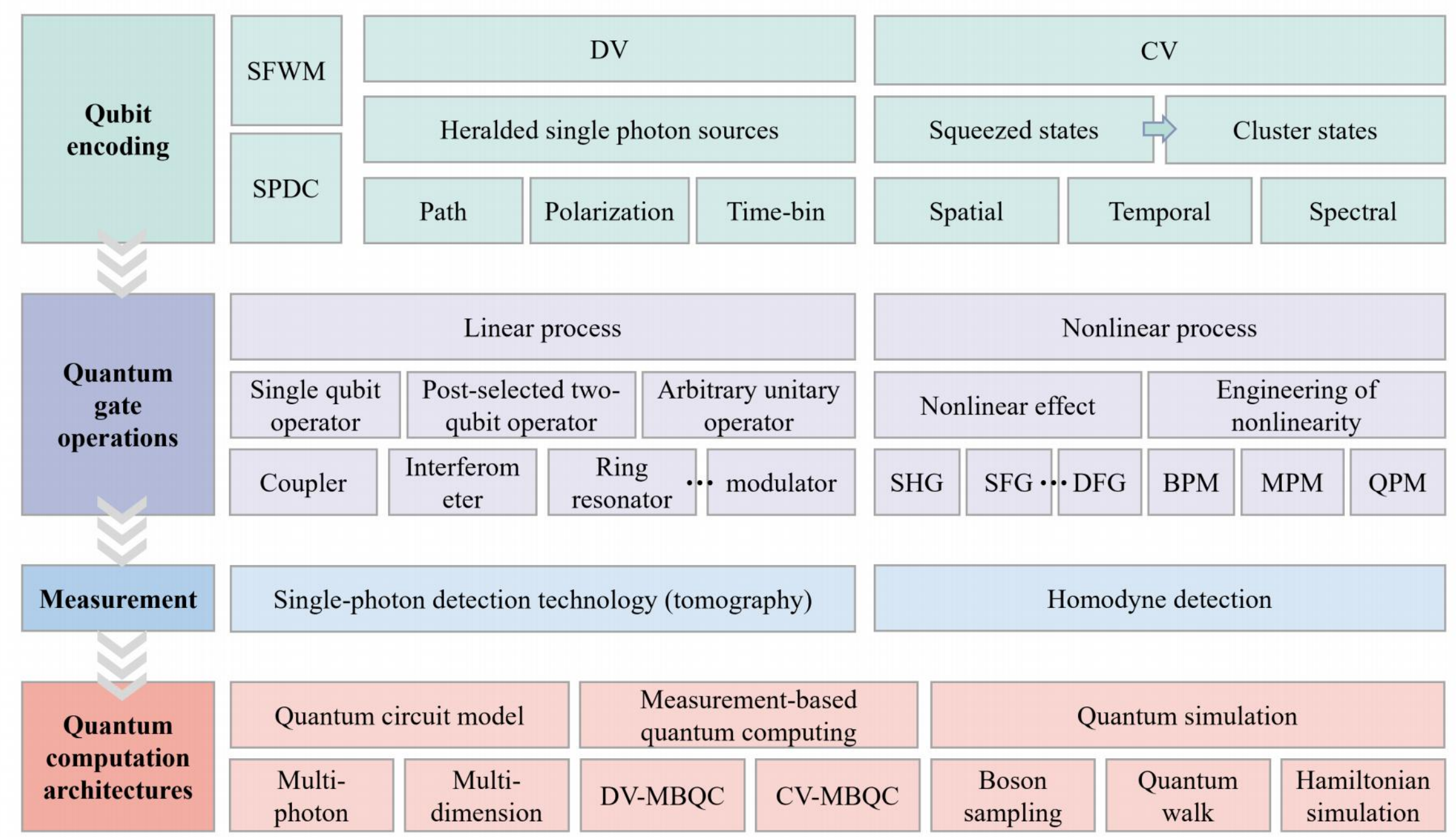

**Fig. 2 | The physical framework of integrated photonic quantum information processing**. The framework is divided into four core layers from top to bottom: Qubit Encoding (photon generation and state encoding), Quantum Logic Operations (gate operations based on linear optics and nonlinear effects), Quantum Measurement (detection techniques for discrete and continuous variables), and the foundational layer of Quantum Computing and Simulation Architectures. The following sections of this review are elaborated under this framework. **Abbreviations:** SFWM, spontaneous fouw-wave mixing; SPDC, spontaneous parametric down conversion; DV, discrete-variables; CV, continuous-variables; SHG, second harmonic generation; SFG, sum frequency generation; DFG; difference frequency generation; BPM, birefringent phase matching; MPM, modal phase matching; QPM, quasi-phase matching; MBQC, measurement-based quantum computing.

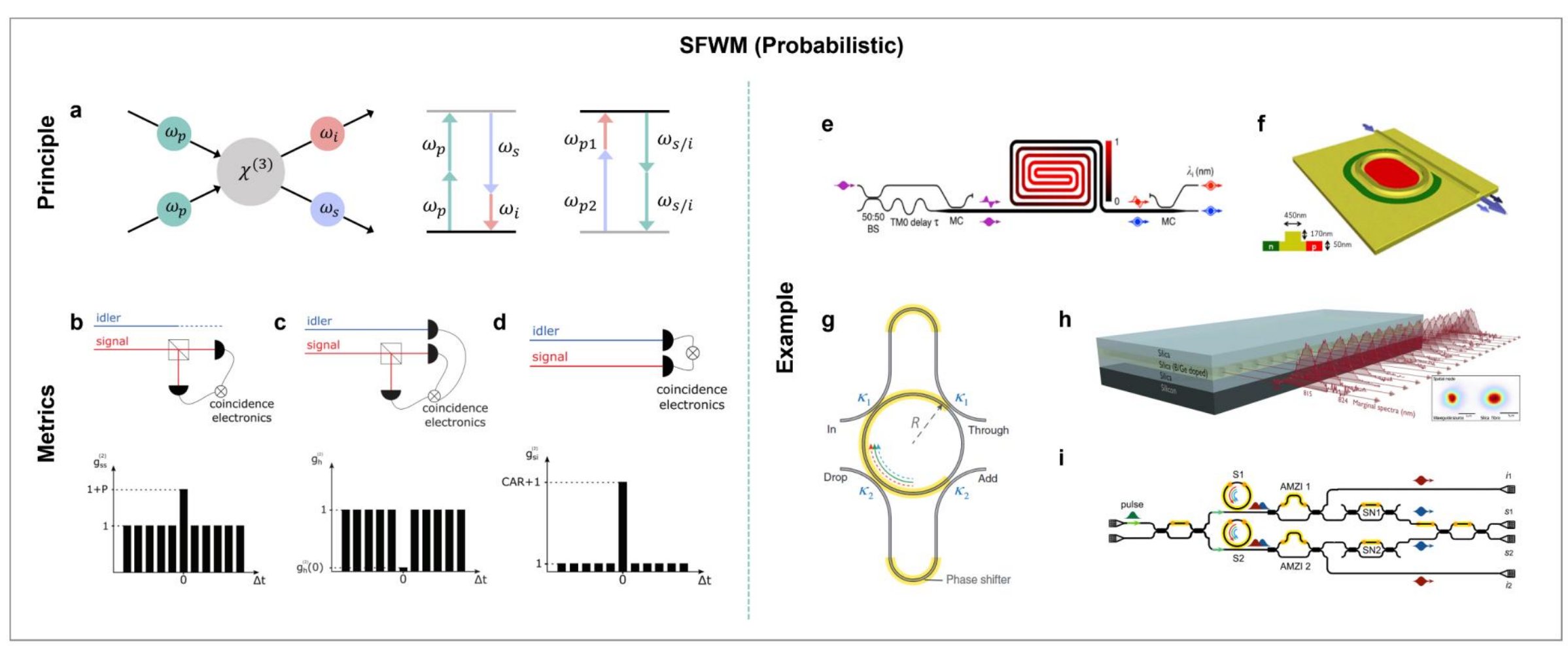

**Fig. 3 | Integrated quantum photon sources in SOI platforms – SFWM sources. a**, The process of spontaneous four-wave mixing (SFWM): two pump photons are annihilated through nonlinear interaction, producing a pair of signal and idler photons with correlated frequencies. Both non-degenerate (two generated photons have different wavelengths) and degenerate (the two photons have identical wavelengths) schemes are shown. **b**, HBT setup for the purity measurement of the heralded single photon via unheralded $g^{(2)}(\Delta t)$, showing the relationship between purity $P$ and $g^{(2)}(0)$. **c**, HBT setup and results of measuring the heralded $g_H^{(2)}(\Delta t)$. **d**, Setup for the CAR measurement, and the experimental relationship of $g_{si}^{(2)}(\Delta t)$ with the CAR is visible at $\Delta t = 0$. **e**, Schematic of the photon pair source using long waveguides. A specialized delayed-pump excitation scheme is employed to suppress residual correlations in the joint spectrum, thereby achieving near-unit spectral purity. **f**, A microring surrounded by a p-i-n junction, which serves to remove free-carriers generated by the pump and overcome the free-carrier-absorption (FCA) associated optical losses. **g**, The dual-MZI-coupled silicon ring resonator, which overcomes the trade-off between spectral purity and brightness in the post-filtering way and ensures the generation of bright pure photons. **h**, Array of heralded single-photon sources on a silica photonic chip, where a series of straight waveguides are fabricated via UV-laser writing in a germanium-doped silica-on-silicon photonic chip. **i**, Array of micro-ring resonator sources in silicon, that serve as sources for quantum entanglement generation and precise quantum manipulation circuits. Panels reproduced from: **b**,**c**,**d**, ref.[60]; **e**, ref. [66]; **f**, ref. [61]; **g**, ref. [70]; **h**, ref. [64]; **i**, ref. [76].

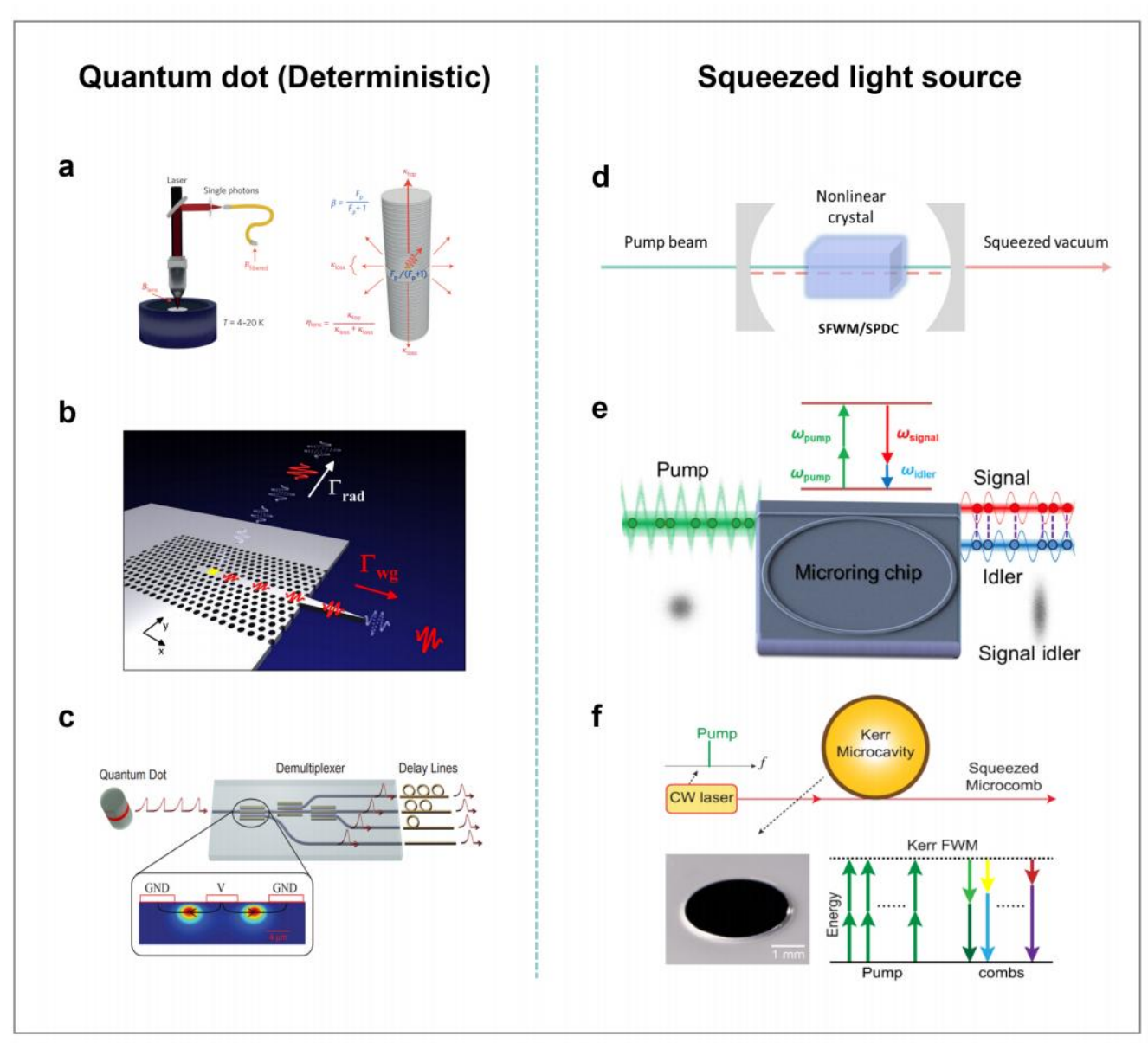

**Fig. 4 | Integrated quantum photon sources in SOI platforms – quantum dot and squeezed light sources.** **a**, Schematic of a QD-based photon source: the QD is inserted in a cryostat typically operating between 4 and 30 K, and the pillar microcavity used to collect single photons. **b**, The quantum dot sources coupled to a photonic crystal waveguide, achieving near-unity coupling efficiency. Red pulses are the train of single-photon pulses emitted from a triggered QD (yellow trapezoid). **c**, Scheme for an ideal active spatial-temporal demultiplexing: A stream of single photons emitted at successive time intervals from a quantum dot are actively routed into different spatial channels by an optical demultiplexer. The optical demultiplexer consists of a network of reconfigurable directional couplers with electro-optically tunable splitting ratio. A set of delay lines at the output can be used to match the arrival times of the single photons. **d**, The principle for generating squeezing vacuum by pumping nonlinear crystal, via either SFWM or SPDC. **e**, the generation of intensity-correlated signal and idler beams via four-wave mixing in the on-chip microring optical parametric oscillator (OPO). **f**, The microring pumped by a continuous-wave (CW) laser, exciting third-order Kerr nonlinearity and generating unconditional entanglements among hundreds of equidistant frequency modes**.** Panels reproduced from: **a**, ref. [83]; **b**, ref. [87]; **c**, ref. [88]; **e**, ref. [89];**f**, ref. [90]

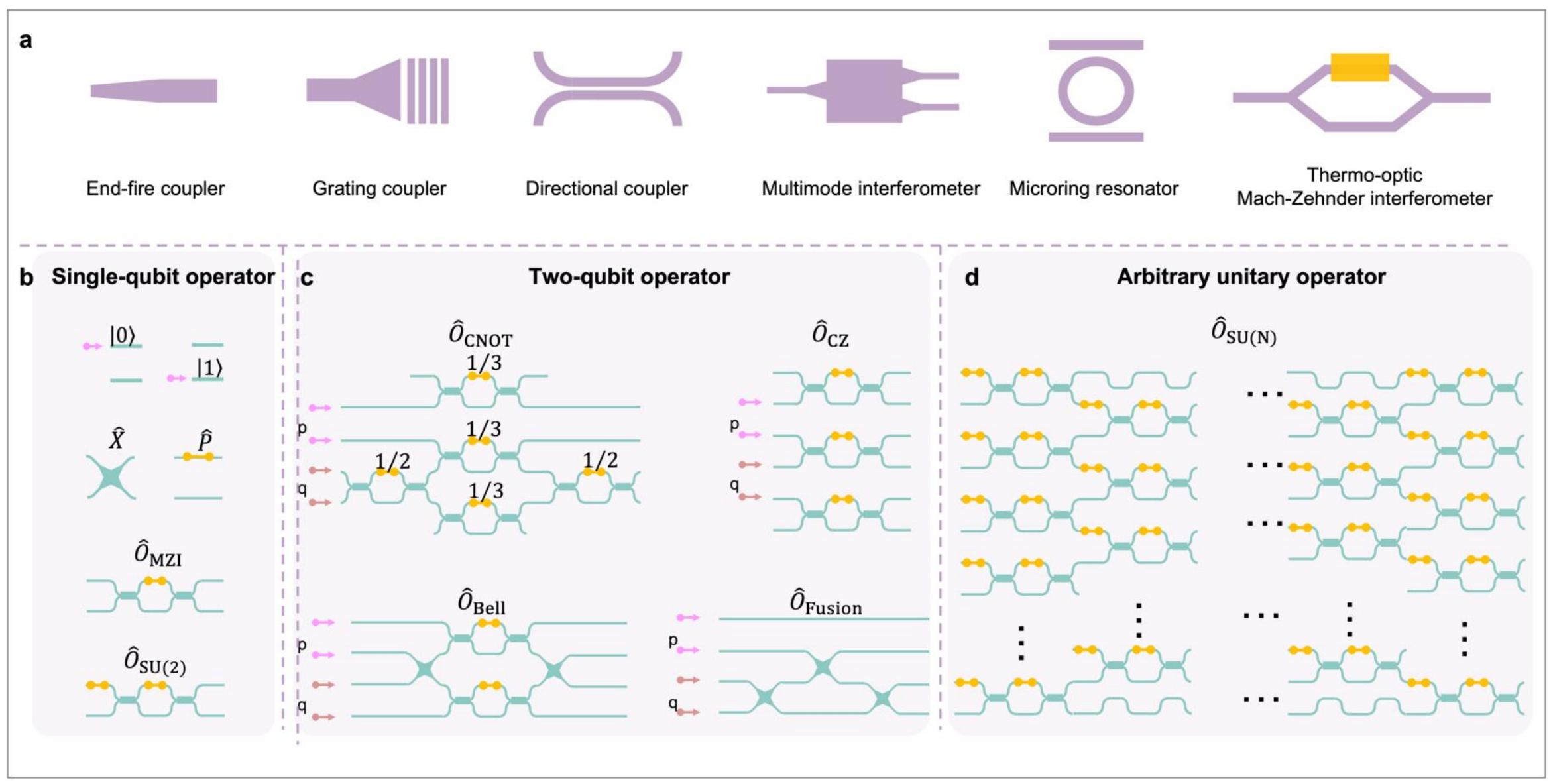

**Fig. 5 | Manipulation techniques in SOI platforms. a**, The foundational elements for integrating photonic circuits include end-fire couplers, grating couplers, directional couplers, multimode interferometers, microring resonators, Mach-Zehnder interferometers, and thermo-optic modulators, which are essential for manipulating light in SOI-based systems. **b**, **c**, **d**, the common quantum operators used in photonic quantum computing. Specifically, **b**, single-qubit operators such as the Pauli X matrix ($\hat{X}$), phase shifter operator ($\hat{P}$), Mach-Zehnder Interferometer operator ($\hat{O}_{MZI}$) and the universal set of single-qubit rotations SU(2) group ($\hat{O}_{SU(2)}$). **c**, Two-Qubit Operators for creating entanglement and performing joint operations on pairs of qubits. The depicted operators include the Bell state operator and the fusion gate, which are used to manipulate the quantum states of two qubits in a correlated manner. **d**, Arbitrary unitary operator ($\hat{O}_{SU(N)}$), describing the ability to implement arbitrary unitary transformations on the quantum states of multiple qubits.

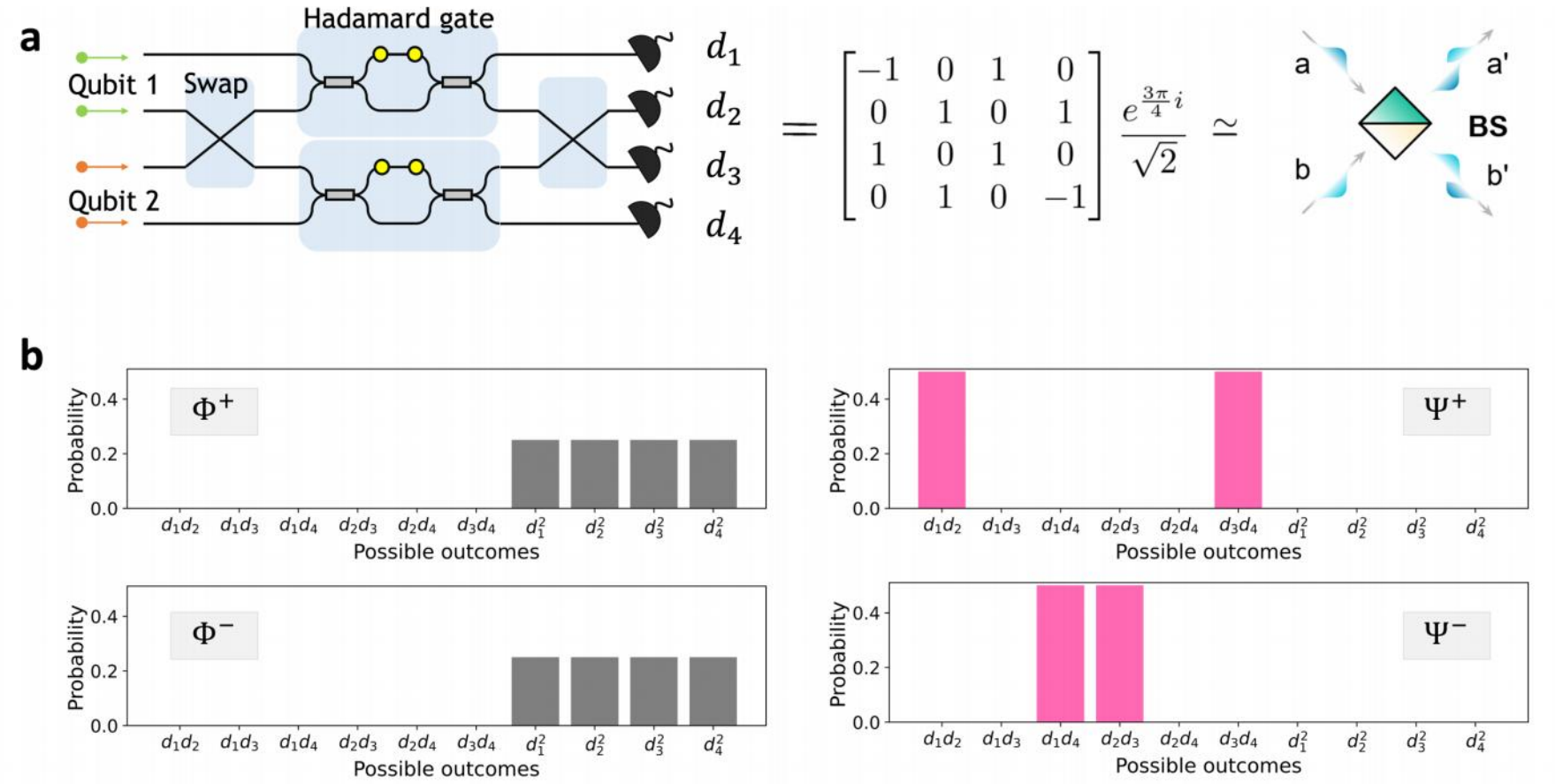

**Fig. 6 | A computation example of Bell state measurement. a**, The linear-optical Bell-state measurement circuit and its matrix representation. Two single-photon qubits arrive at input ports. The middle two modes are swapped, and then each qubit is processed by a Hadamard gate, after which the two output modes are swapped. The output photons are routed to single photon detectors. **b**, the probability distribution obtained when each Bell state of $\{|\Phi^+\rangle, |\Phi^-\rangle, |\Psi^+\rangle, |\Psi^-\rangle\}$ is injected into the circuit. The horizontal axis lists the measurement bases, while the vertical axis gives the probability of every possible two-detector coincidence outcome. According to the result, $|\Phi^+\rangle$ and $|\Phi^-\rangle$ cannot be distinguished, while $|\Psi^+\rangle$ and $|\Psi^-\rangle$ observed coincidence pattern unambiguously distinguishes the input Bell state.

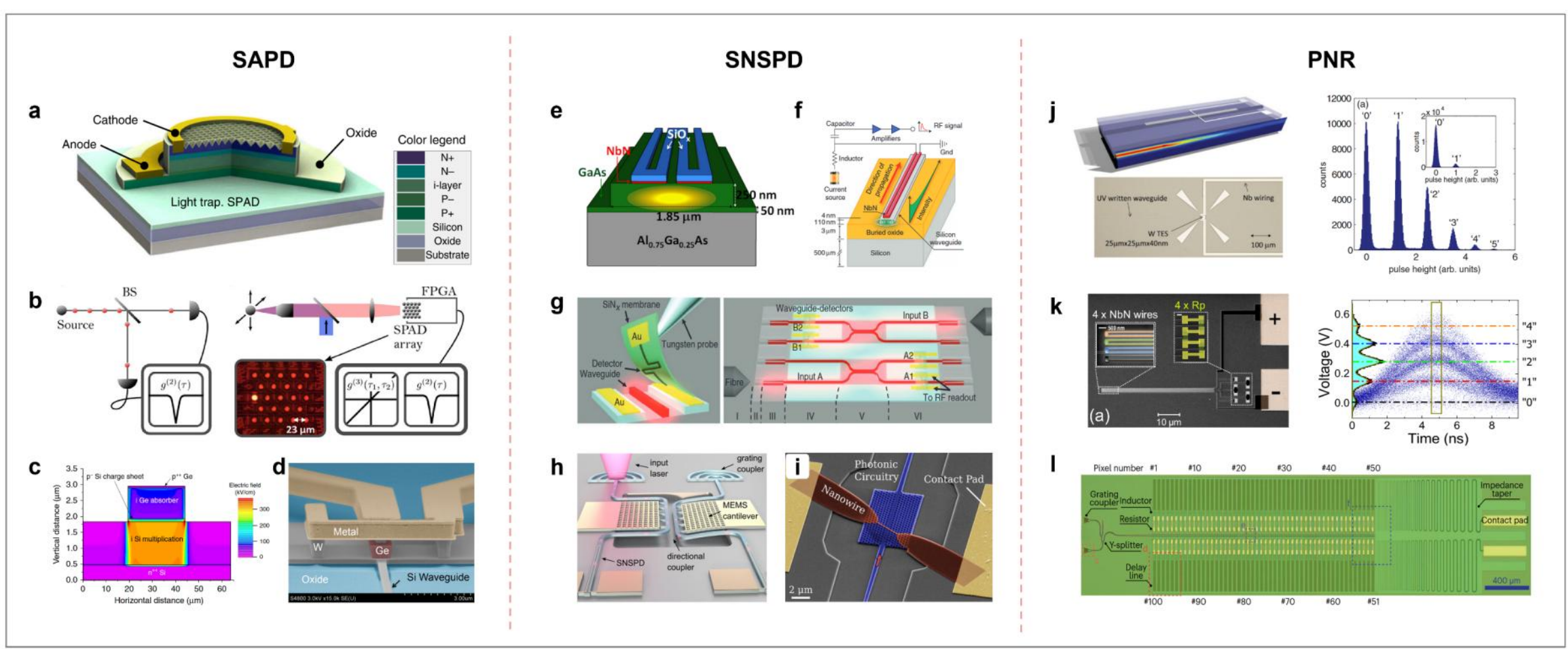

**Fig. 7 | Integrated single-photon detectors in SOI platform. a,** Schematic of the light trapping single-photon avalanche detector (SPAD). The nano-structure is etched as an inverse pyramid, and the color legend shows the names of all the layers in the devices. **b,** Application of SPAD array into the testing setup of quantum correlation measurement. **c**, Single-photon avalanche diode modelling and current-voltage behaviour. The electric field profile at 5% excess bias for the planar SPAD is shown. **d**, Integrated waveguide-coupled Ge-on-Si lateral avalanche photodiode. **e**, schematic of the waveguide superconducting nanowire single-photon detector (SNSPD), where four NbN nanowires are placed on top of a GaAs waveguide, and a deep ridge is etched to provide 2D confinement. **f**, The travelling wave SNSPD: a sub-wavelength absorbing NbN nanowire is patterned atop a silicon waveguide to detect single photons. **g**, The scalable integration of SNSPD onto photonic waveguide using membrane transferring, achieving unity with photonic integrated circuits. **h**, Low-power micro electromechanics with SNSPDs: the same NbTiN layer is used to build the MEMS actuators, electrical connections, contact pads, and single-photon detectors. **i**, the SNSPD implemented in a 2D photonic crystal cavity. **j**, Schematic of an evanescently coupled photon counting detector based on transition-edge sensors (TESs). Up to five photons are resolved in the guided optical mode via absorption from the evanescent field into the TES. **k**, PNR detector consisting of 4 NbN wires on GaAs ridge waveguides, that can resolve up to four photons. **l**, The 100-pixel quasi-PNR detector structure based on the spatial-temporal multiplexed SNSPD array. Panels reproduced from: **a**, ref. [112]; **b**, ref. [113]; **c**, ref. [114]; **d**, ref. [115]; **e**, ref. [116]; **f**, ref. [117]; **g**, ref. [118]; **h**, ref. [119]; **i**, ref. [120]; **j**, ref. [121]; **k**, ref. [122]; **l**, ref. [123].

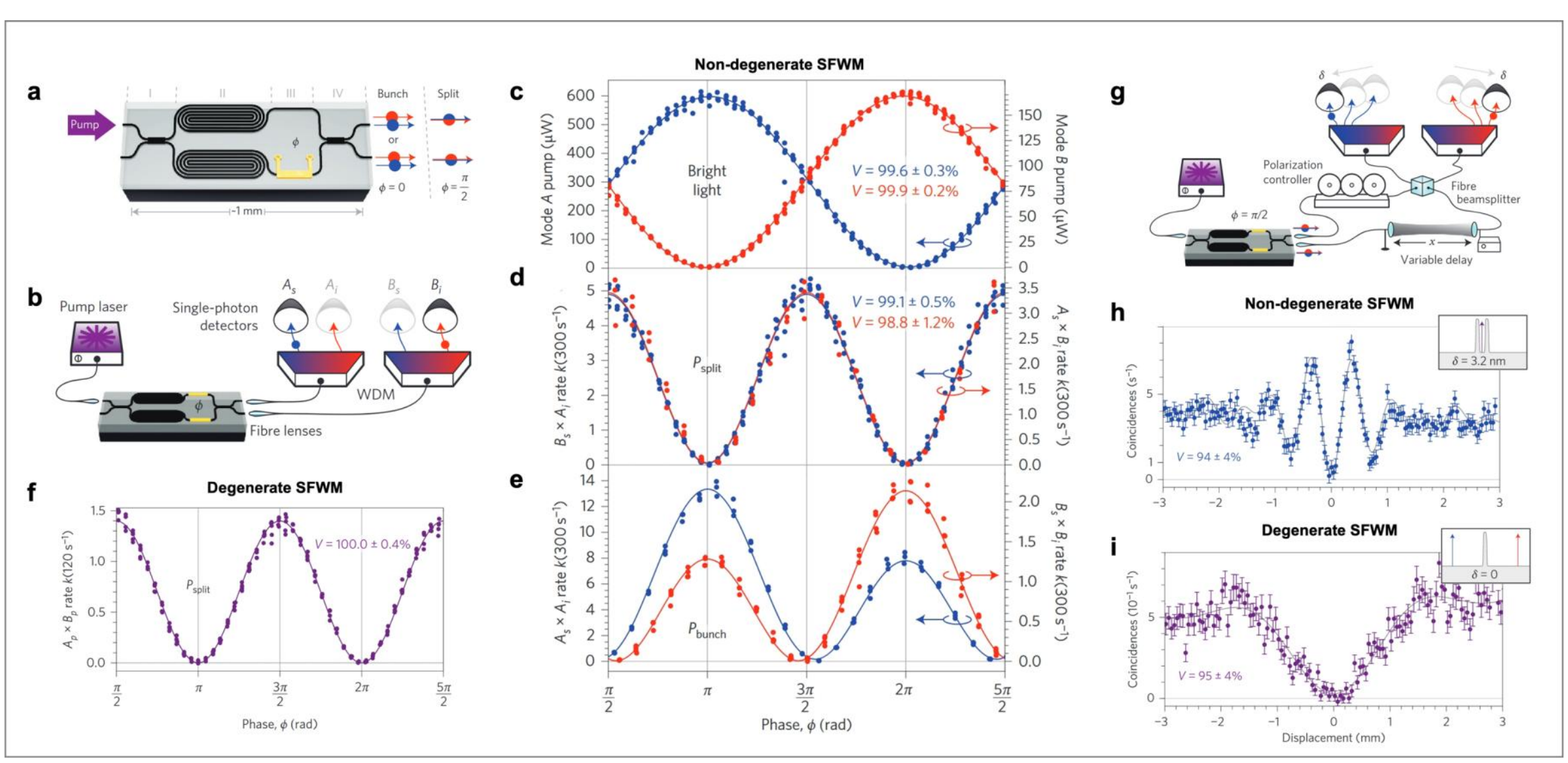

**Fig. 8 | Quantum-classical interference differentiation in integrated degenerate and non-degenerate SFWM sources. a,** Waveguide circuit schematic with dual SFWM sources, reconfigurable phase shifter ($\phi$), and multimode interferometer (MMI) couplers. **b,** Pump laser excitation and single-photon detection with wavelength-division multiplexing (WDM) filters. **c-e,** Non-degenerate SFWM interference results: **c,** Classical 2π-period pump interference, **d,** Quantum π-period coincidence fringes for $A_sB_i/B_sA_i$ pairs, **e,** Quantum π-period coincidence fringes for $A_sA_i/B_sB_i$ pairs showing asymmetry from spurious photons (error bars denote measurement uncertainty). **f,** Degenerate SFWM coincidence rates. **g,** Non-degenerate SFWM experimental implementation with pulsed laser and programmable delay.**h,** Non-degenerate and **i**, Degenerate HOM interference. Panels reproduced from: ref. [106].

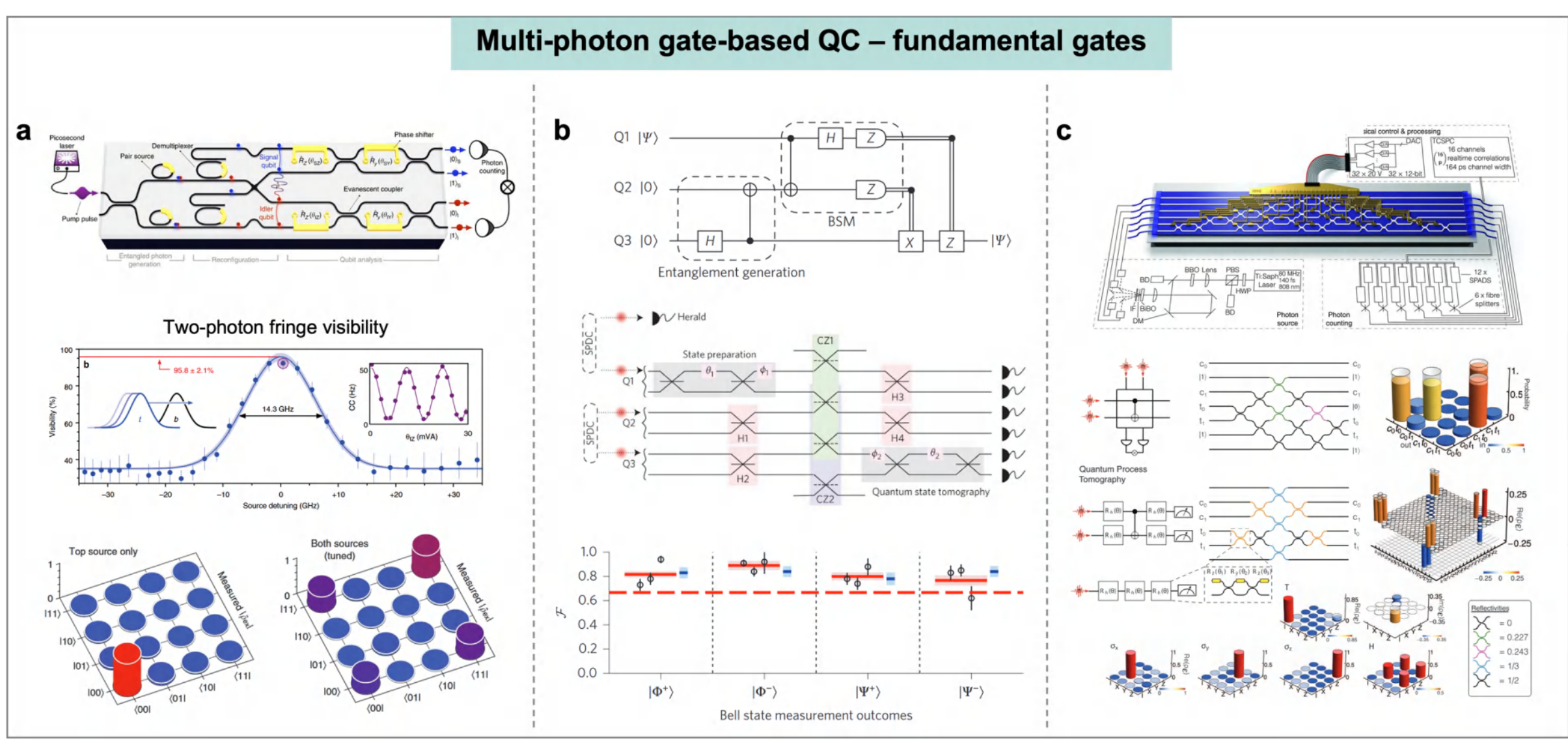

**Fig. 9 | Fundamental gate implementation in multi-photon gate-based quantum computing circuits. a**, The circuit for generating a path-entangled two-qubit state: a picosecond pump pulse is coupled into the silicon chip where it generates a superposition of photon pairs via SFWM in microring resonator. This superposition is separated into signal (blue) and idler (red) path qubits by the demultiplexer, which are then analysed by two MZIs. **b**, The circuit for implementing CNOT gate and quantum teleportation: Three qubits are encoded using dual-rail logic in a silica-on-silicon chip. Specifically, a Bell state is encoded on qubits [Q2,Q3], and a Bell state measurement is performed on qubits [Q1, Q2]. Local Hadamard operations (H1 to H4) are performed using beam splitters of reflectivity 1/2 (solid lines) and the two cascaded C-phase gates (CZ1, CZ2) are implemented using four beam splitters of reflectivity 1/3 (dashed lines). After the process, information on Q1 is teleported to Q3. **c**, A six-mode fully re-programmable universal linear-optic circuit: the device consists of a cascade of 15 MZIs and 30 TOPSs, and up to 6 photons were injected into the chip from bulk optic SPDC photon pair sources. The device was reprogrammed to implement multiple information tasks. Inset: the circuit reconfiguration realizing heralded and unheralded CNOT gate. Panels reproduced from: **a**, ref. [127]; **b**, ref. [140]; **c**, ref. [141].

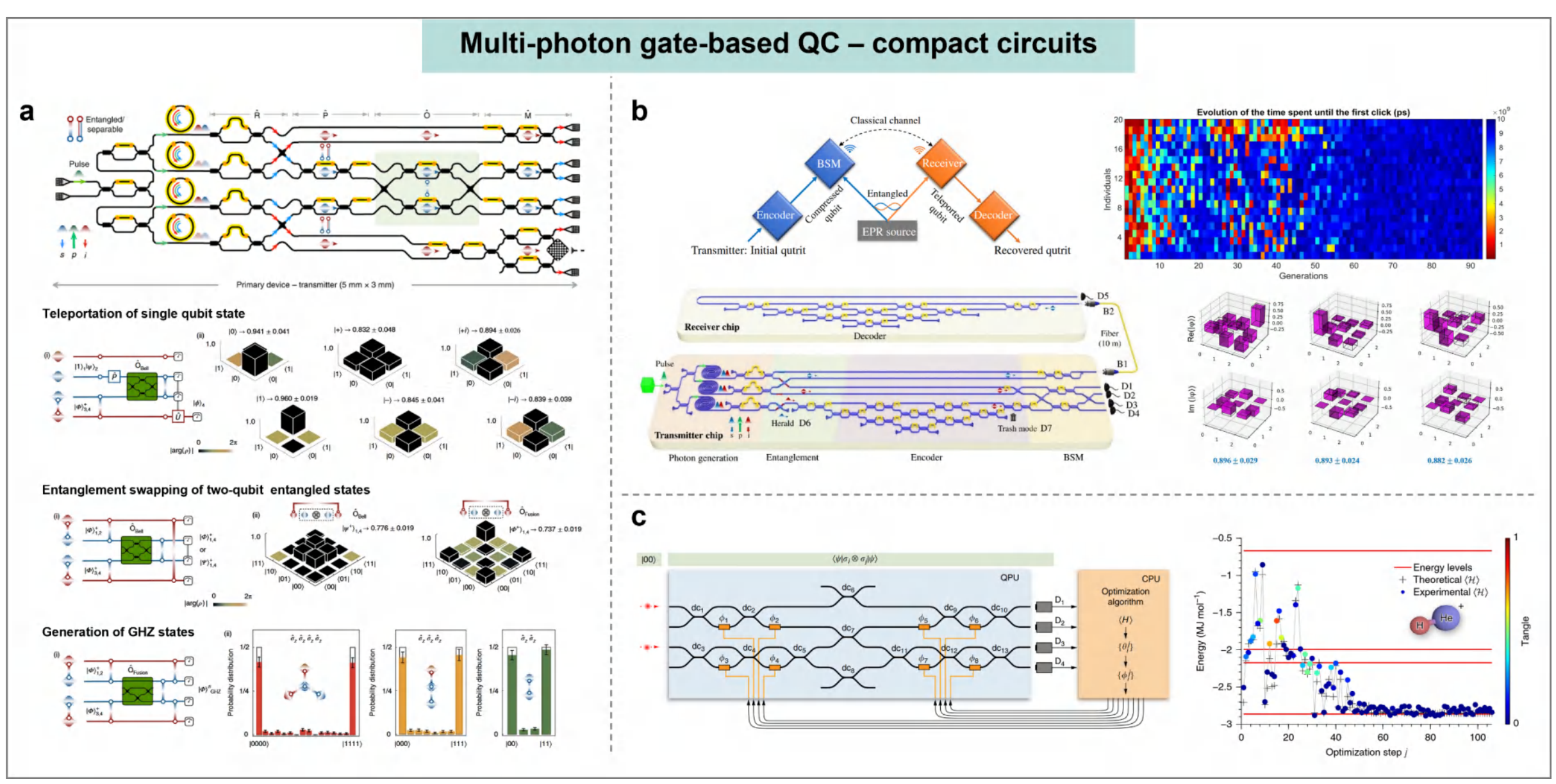

**Fig. 10 | Compact multi-photon gate-based quantum computing circuits. a**, Multi-qubit entanglement generator and teleportation circuit: An array of four identical microring resonators is integrated to generate two pairs of photons (red idler, blue signal), and a linear optical quantum circuit is programmed to work as Bell operator and a fusion entangling operator on the two blue photons. **b**, Resource-efficient teleportation of qudit states: the device is composed of a transmitter chip and a receiver chip that are coherently linked by a single-mode fiber. The input qutrit state is compressed to qubit state by an on-chip trainable quantum autoencoder integrated on the transmitter chip, and then teleported using the Bell operator. **c**, The circuit implementing a quantum variational eigensolver: The state is prepared using thermal phase shifters $\phi_{1-8}$ (orange rectangles) and one CNOT gate, and measured using photodetectors. Coincidence count rates from the detectors D1–4 are passed to the CPU running the optimization algorithm. This computes the set of parameters for the next state and writes them to the quantum device. Panels reproduced from: **a**, ref. [76]; **b**, ref. [143]; **c**, ref. [144].

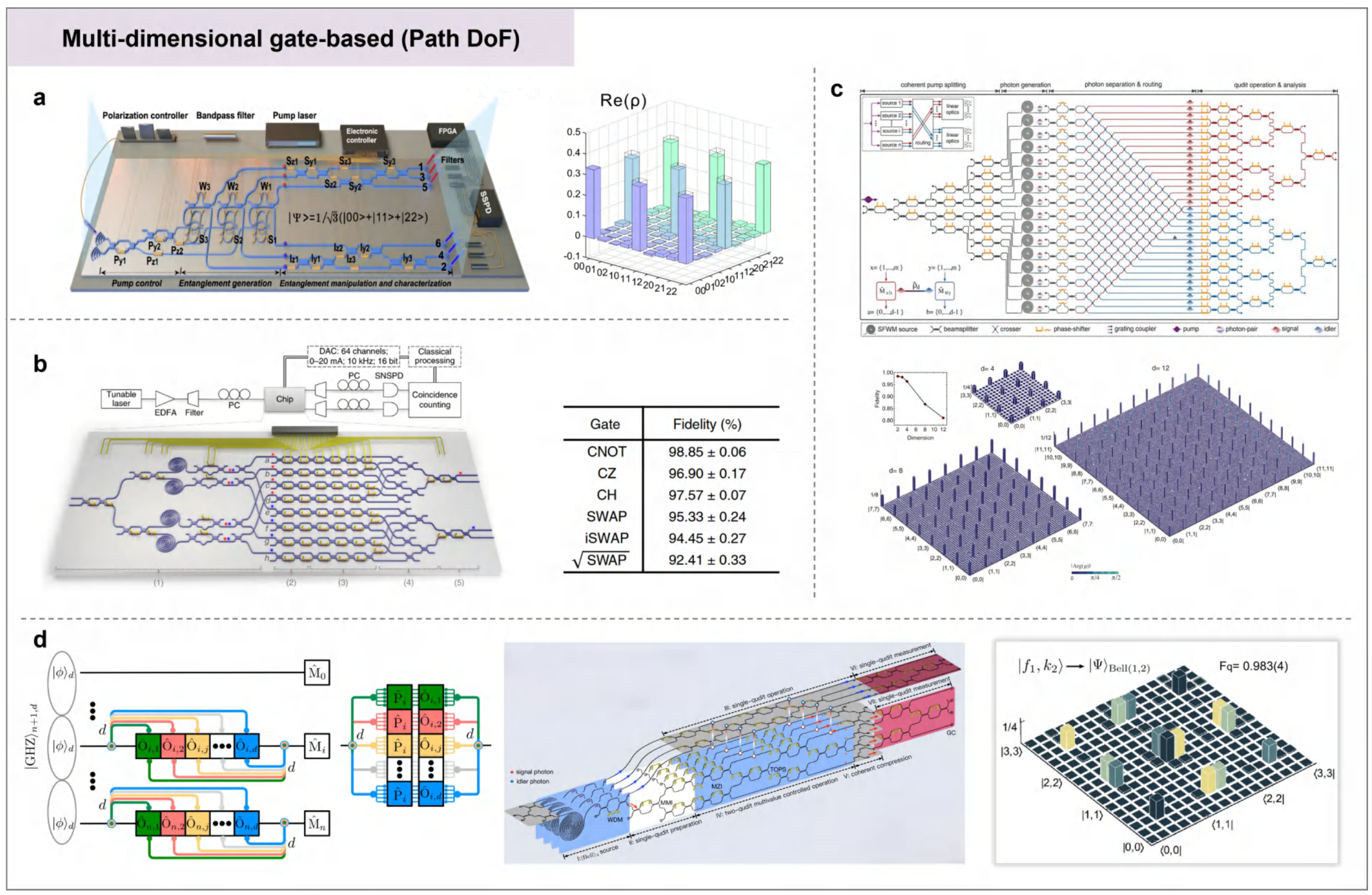

| Gate | Fidelity (%) |
|---|---|
| CNOT | 98.85 ± 0.06 |
| CZ | 96.90 ± 0.17 |
| CH | 97.57 ± 0.07 |
| SWAP | 95.33 ± 0.24 |
| iSWAP | 94.45 ± 0.27 |
| $\sqrt{\text{SWAP}}$ | 92.41 ± 0.33 |

**Fig. 11 | Multi-dimensional gate-based quantum computing circuits using path degree of freedom (DoF). a**, The generation of 3D entanglement on chip: a photon pair is created in a superposition between three coherently pumped dual MZI microring sources (S1-S3). A tunable qutrit entangled state is generated by adjusting the four phase shifters. The signal (red) and idler (blue) photons are separated by W1-W3 and routed to two 3D multiport interferometers that enable the performing of unitary transformations on the entangled qutrits. **b**, The realization of two entangled ququarts (4D) in a re-programmable Si chip: Post-selecting when signal and idler photons exit at the top two output modes (qubit 1) and the bottom two (qubit 2), respectively, yields a path-entangled ququart state. Spatial modes *a–h* are each extended into two modes to form path-encoded single-qubit states with arbitrary amplitude and phase. Then single-qubit and two-qubit unitary operations have been implemented on the circuit with high fidelity. **c**, A programmable bipartite entangled state with dimensions up to 15D. The device integrated 16 SFWM sources (there is a lossy mode that reduces the dimension from 16 to 15) and more than 550 components on an individual quantum photonic chip. Previously unexplored high-dimensional quantum applications are reported, such as certifying multidimensional quantum randomness expansion and entangled state self-testing, have been investigated. **d**, The circuit of a two-ququart qudit processor: It generates a four-level entangled state in an array of four integrated identical SFWM sources, then expands the Hilbert space and prepares arbitrary single-qudit. Arbitrary single-qudit operations and two-qudit multivalue controlled operations are implemented subsequently. **a**,**b**,**c**,**d**, are all high-dimensional qubits encoded on path degree of freedom. Panels reproduced from: **a**, ref. [145]; **b**, ref. [131]; **c**, ref. [146]; **d**, ref. [147].

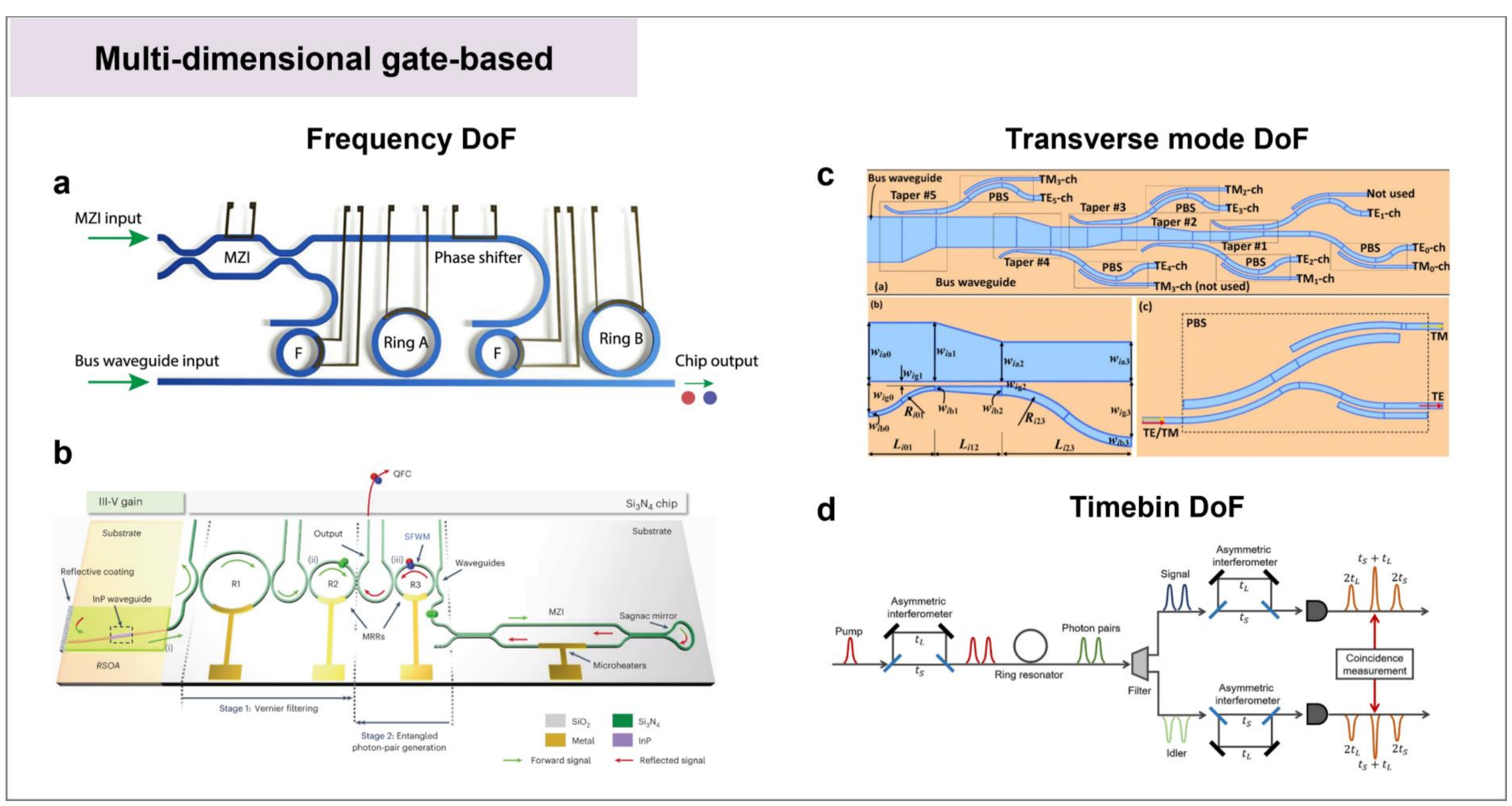

**Fig. 12 | Multi-dimensional gate-based quantum computing circuits, using frequency, transverse mode, and timebin DoFs. a**, programmable frequency bin quantum states: an MZI is used to route optical pumping power to the two generating rings (ring A and ring B) via two add-drip filters (F). The resonances of the ring A and ring B are programmable by a TO phase shifter, to encode different states of the frequency bins. **b**, Frequency-bin entangled photons pairs: An InP source is coupled to a $Si_3N_4$ chip containing three microring vernier filters and an MZI with a Sagnac mirror. The backward-propagating laser field acts as the excitation signal for an SFWM process within the third ring, R3, producing a quantum frequency comb (QFC). The QFC excited by a continuous-wave source dwells in high-dimensional Hilbert space, because the idler and signal photons are in a superposition of multiple frequency modes. **c**, Transverse mode encoded high-dimensional quantum states based on a 10-channel mode demultiplexer. **d**, Generation and measurement of time-bin entangled states. Panels reproduced from: **a**, ref. [150]; **b**, ref. [151]; **c**, ref. [152]; **d**, ref. [153].

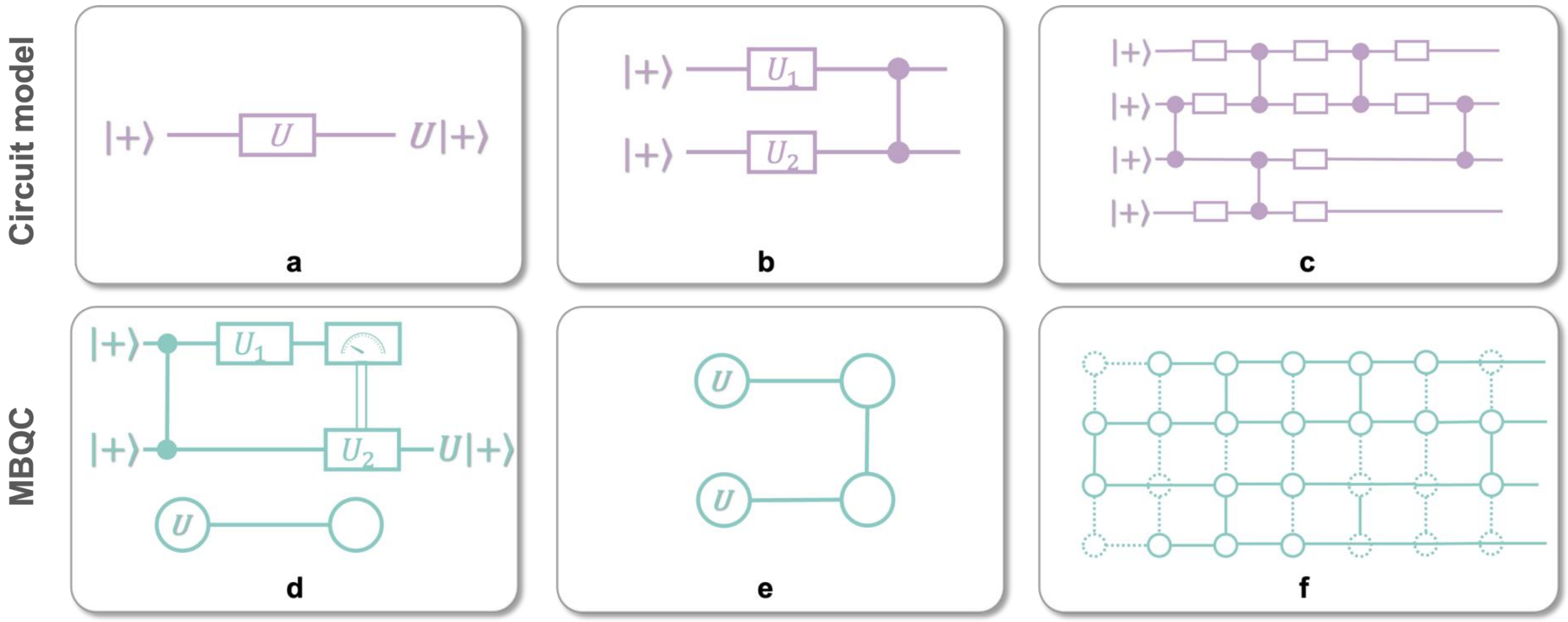

**Fig. 13 | The comparison of the principles of quantum circuit model and measurement-based quantum computing (MBQC). a-c,** single qubit gate, two qubit gate, and general multi qubit circuit under quantum circuit model. **d-f,** single qubit gate, two qubit gate, and general multi qubit circuit under MBQC scheme.

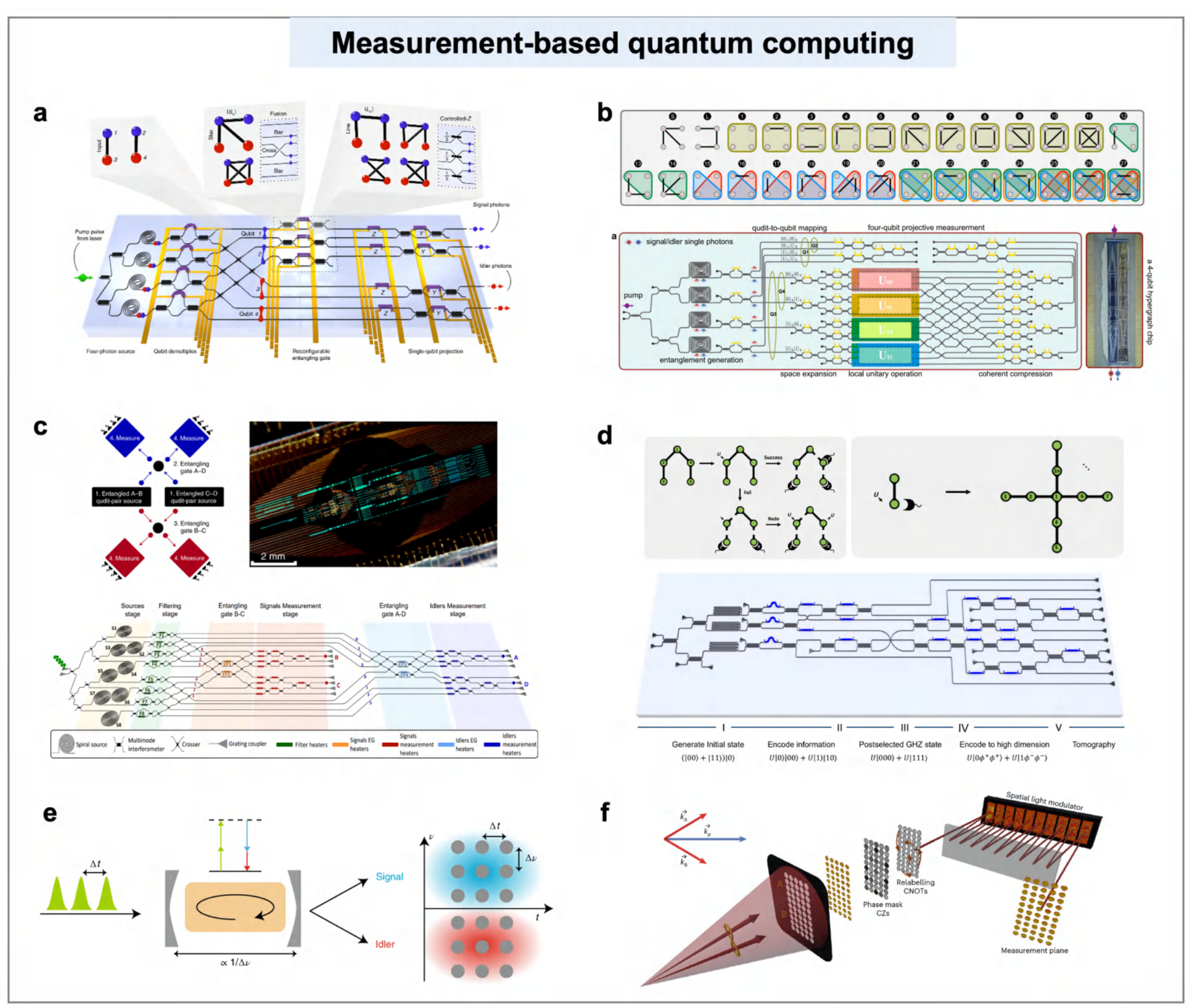

**Fig. 14 | Measurement-based quantum computing circuits. a**, Generation of all types of four-photon graph states and a basic measurement-based protocol on chip: four photons in two pairs are generated in superposition over four sources, and demultiplexed by wavelength and rearranged to group signal and idler photons. The signal-photon qubits are operated upon by a reconfigurable post-selected entangling gate, to perform either a fusion or controlled-Z operation, to generate star- or line-type entanglement. **b**, Generation and manipulation of all types of four-qubit hypergraph states. In previous graph-type entanglement, only the nearest qubits interact; in hypergraph entanglement, any subset of qubits can be arbitrarily entangled via hyperedges. **c**, Error-protected qubits based on entanglement generation and projective measurement: two entangled qudit-pair modules, each consisting of four photon-pair sources, generate two pairs of maximally entangled qudits (A–B and C–D). Each qudit is mapped to a two-qubit system. Photons A–D and B–C are entangled with fusion-type gates to produce entangled states of eight qubits among four parties. **d**, The generation of a five-qubit linear cluster state and the demonstration of fault-tolerant MBQC on chip. The scheme uses multiple qubits to protect one qubit, thus realizes fault tolerance in the manner that: if the teleportation succeeds, the remaining qubit can be disentangled from the states, otherwise the faulty qubit will be removed, and the remaining qubit for reperforming the teleportation. **e**, An optical pulse train excites a nonlinear cavity generating a simultaneously time- and frequency-bin entangled photon pair (that is, a d-level hyper-entangled state). **f**, High-dimensional cluster states created off-chip with transition to chip potential. Photon pairs entangled via SPDC pass through a mask with 50 apertures, conserving momentum. Operations like mode-dependent phases and relabelling correspond to CZ and CNOT gates for qudits encoded in each photon's Hilbert space. Panels reproduced from: **a**, ref. [162]; **b**, ref. [164]; **c**, ref. [163]; **d**, ref. [165]; **e**, ref. [81]; **f**, ref.[166].

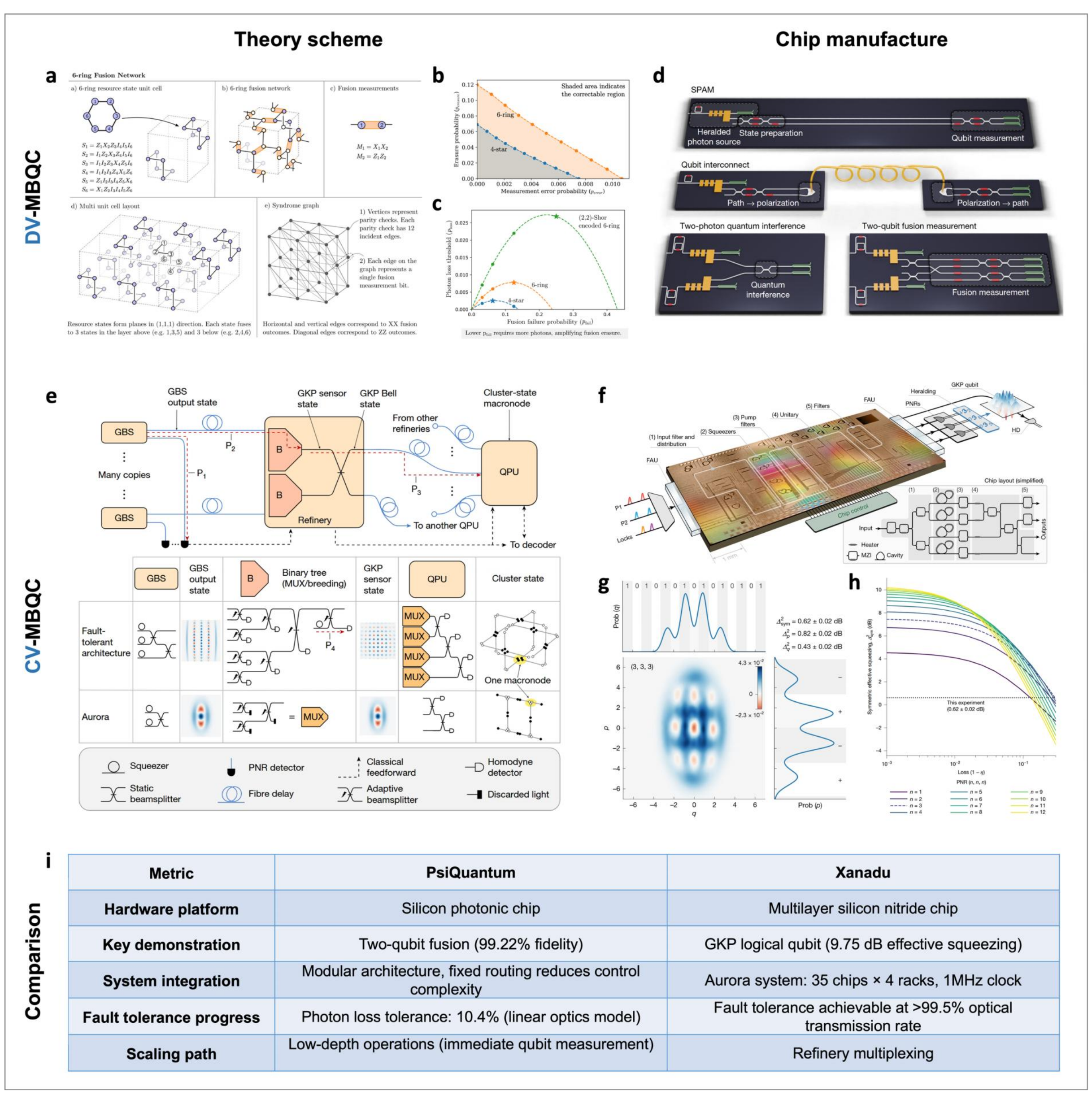

| Metric | PsiQuantum | Xanadu |
|---|---|---|
| Hardware platform | Silicon photonic chip | Multilayer silicon nitride chip |
| Key demonstration | Two-qubit fusion (99.22% fidelity) | GKP logical qubit (9.75 dB effective squeezing) |
| System integration | Modular architecture, fixed routing reduces control complexity | Aurora system: 35 chips × 4 racks, 1MHz clock |
| Fault tolerance progress | Photon loss tolerance: 10.4% (linear optics model) | Fault tolerance achievable at >99.5% optical transmission rate |
| Scaling path | Low-depth operations (immediate qubit measurement) | Refinery multiplexing |

**Fig. 15 | MBQC towards million-qubit universal quantum computing. a**, DV-MBQC scheme from Psiquantum, representing a "6-ring" fusion network. **b,** The correctable region for the two fusion networks, i.e., six-ring (orange line) and 4-star (blue line). **c,** Photon loss threshold for the three fusion-networks: 4-star, 6-ring, and (2,2)-Shor encoded 6-ring. With the (2,2)-Shor encoding, the 6-ring fusion network provides a significantly larger marginal failure threshold of 43.2%. **d,** The manufacturalble platform for the fusion-based quantum computing architecture, such as quantum state preparation and measurement, point-to-point qubit network, two-photon quantum HOM interference, and two-qubit fusion measurement. **e,** The CV-MBQC architecture (including the Aurora system) of Xanadu, consisting of three stages: Gaussian boson sampling for preparing the initial states; Adaptive interferometers entangling initial states into two-mode GKP Bell pairs; Quantum processing unit that creates spatiotemporal cluster state and implement gates by homodyne measurements. **f,** The chip layout for implementing the scheme. **g,** The Wigner function of the resulted rectangular GKP state. **h,**

Symmetric effective squeezing for simulated output states of the four-mode GBS device as a function of the transmission of the heralding and heralded paths. For transmissions in the range of 70–82% that the (3, 3, 3) outcome (dashed line) is best. As transmission crosses 99.5%, the device is sufficient for making approximate GKP states compatible with fault tolerance. **i,** the comparison of the two state-of-the-art MBQC demonstrations of PsiQuantum and Xanadu. Panels reproduced from: **a-d**, ref.[167,170]; **e-f**, ref.[171,172].

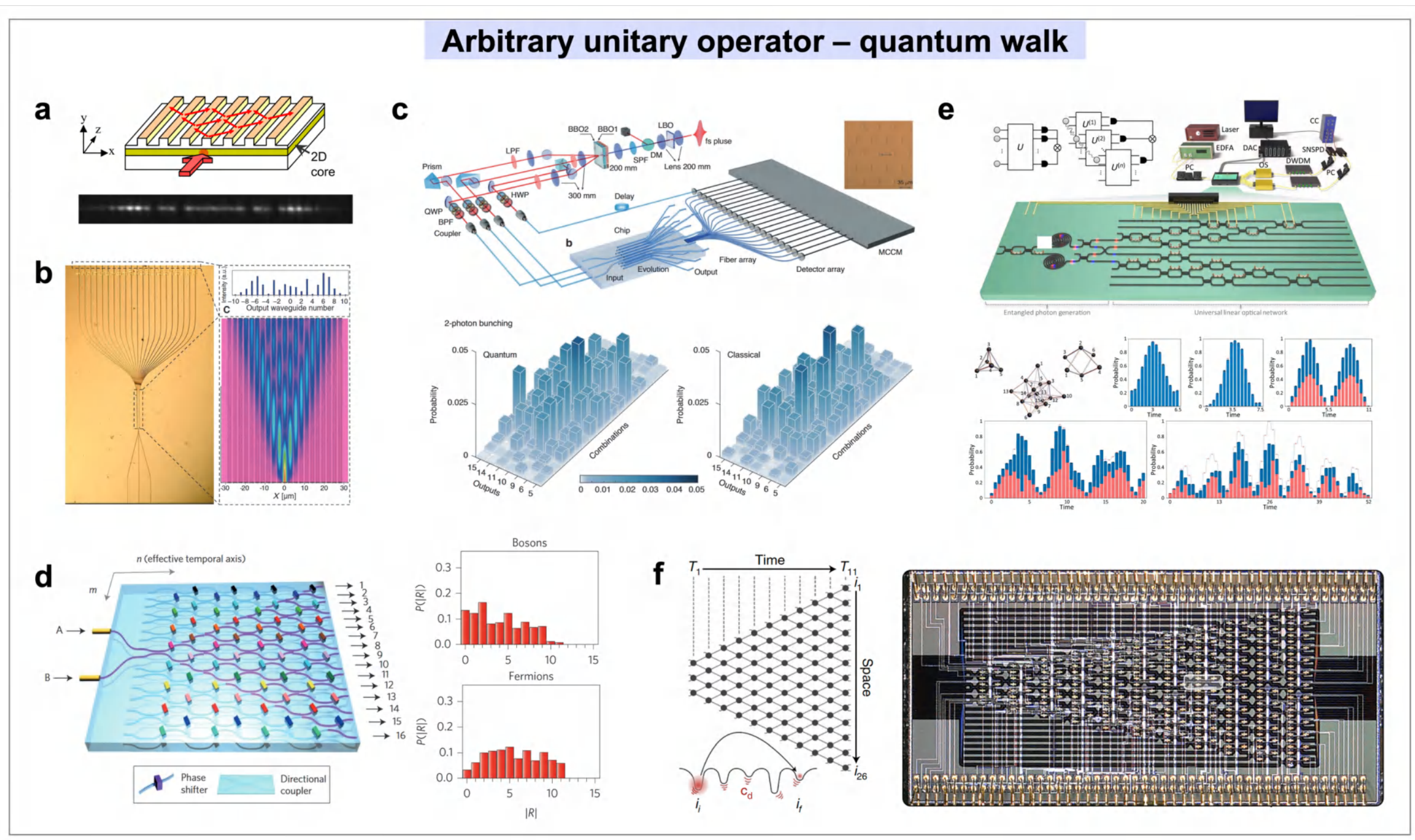

**Fig. 16 | Arbitrary unitary operator-based quantum computing circuits for quantum walk. a,** Two-dimensional single-photon quantum walks via fiber networks but was limited by the inability of classical light sources to exhibit quantum characteristics. **b**, Dual-photon quantum walks on $SiO_xN_y$ waveguide chips, observing nonclassical correlation phenomena in a 21-node 1D array-experimental data violated classical limits by 76 standard deviations, confirming that multiparticle interference exponentially expands state space. **c,** multidimensional quantum walk implementations using femtosecond laser direct-writing technology, which enable three-photon walks in triangular lattices, mapping to a $19^3$ state space. **d**, The interplay between the Anderson localization mechanism and the bosonic/fermionic symmetry of the wavefunction. **e, T**he silicon photonic quantum walk processor: This chip integrates dual reconfigurable five-mode optical networks and on-chip entanglement sources, achieving full parametric control of dual-particle walk Hamiltonians, particle statistical properties and indistinguishability. **f**, Programmable nanophotonic processor for simulating the interaction between a particle undergoing quantum transport. The processor is composed of 88 MZIs, 26 input modes, 26 output modes, and 176 phase shifters, and is electronically programmed using a 240-channel biasing system. Panels reproduced from: **a,** ref. [174]; **b**, ref. [173]; **c**, ref. [175]; **d**, ref. [176]; **e**, ref. [132]; **f**, ref. [177].

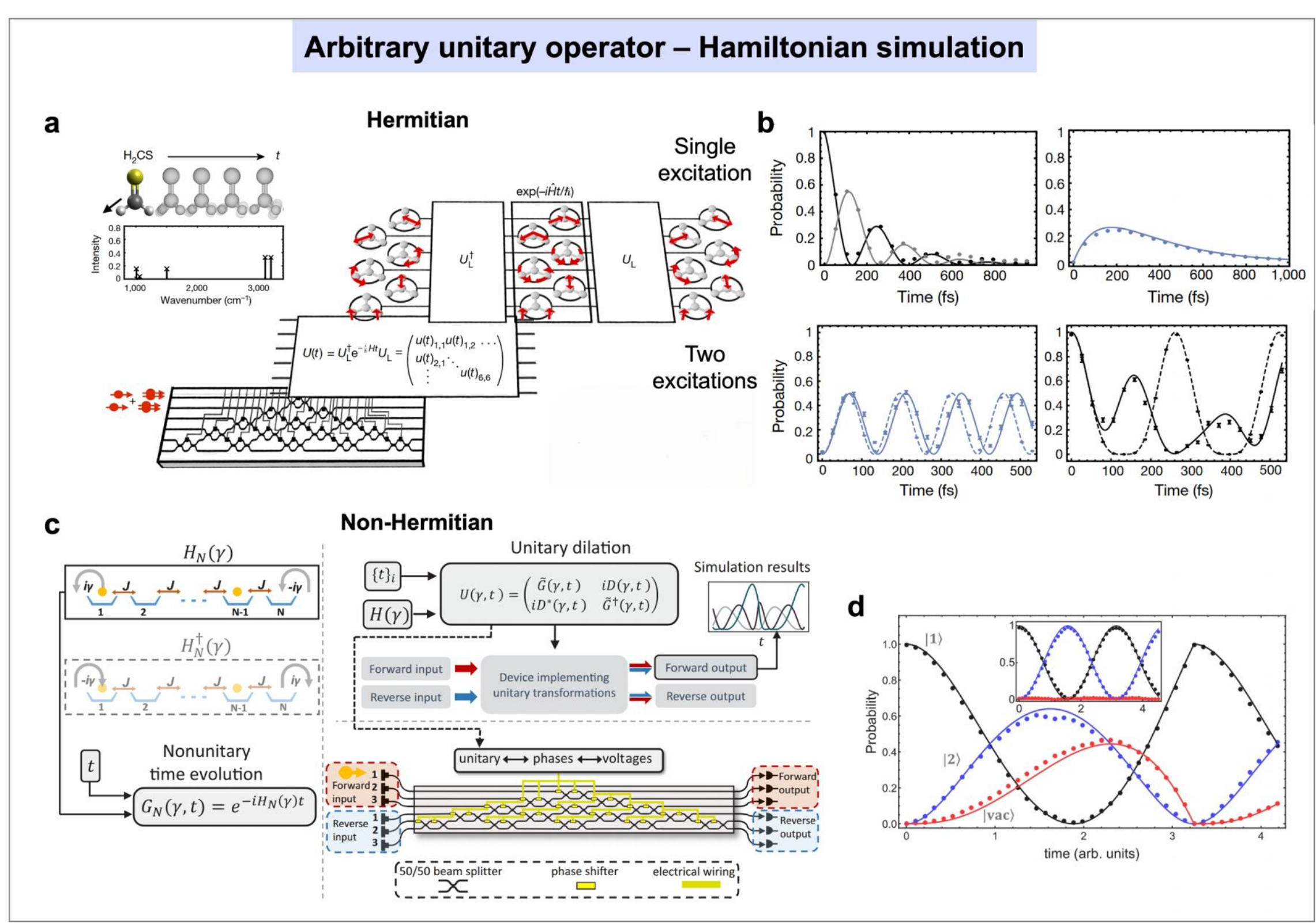

**Fig. 17 | Arbitrary unitary operator-based quantum computing circuits for Hamiltonian simulation. a**, Simulating the vibrational quantum dynamics of molecules using photonics. The evolution of normal modes of molecules is unitarily mapped to a set of local vibrational modes (top layer), and then mapped to a time-dependent unitary transfer matrix (middle layer). Simulations of photonic states under this evolution are then implemented by a reconfigurable photonic chip (bottom layer). **b,** Results of vibrational relaxation and anharmonic evolution in $H_2O$, which demonstrate the simulated probability evolution for both single-excitation and two-excitation states. **c,** The simulation of non-Hermitian PT-symmetry on integrated photonic chip, by employing unitary dilation to embed the $G_N(\gamma, t)$ into an expanded $2N$-mode unitary matrix. **d,** Experimental data that validates the non-Hermitian features: vacuum-state probability emerges due to excitation tunnelling into the time-reversed subsystem, while phase-dependent behaviors—oscillatory (unbroken), polynomial decay (EP), and exponential saturation (broken)—underscore the system's sensitivity to γ. Panels reproduced from: **a,b,** ref. [179]; **c,d,** ref. [180].

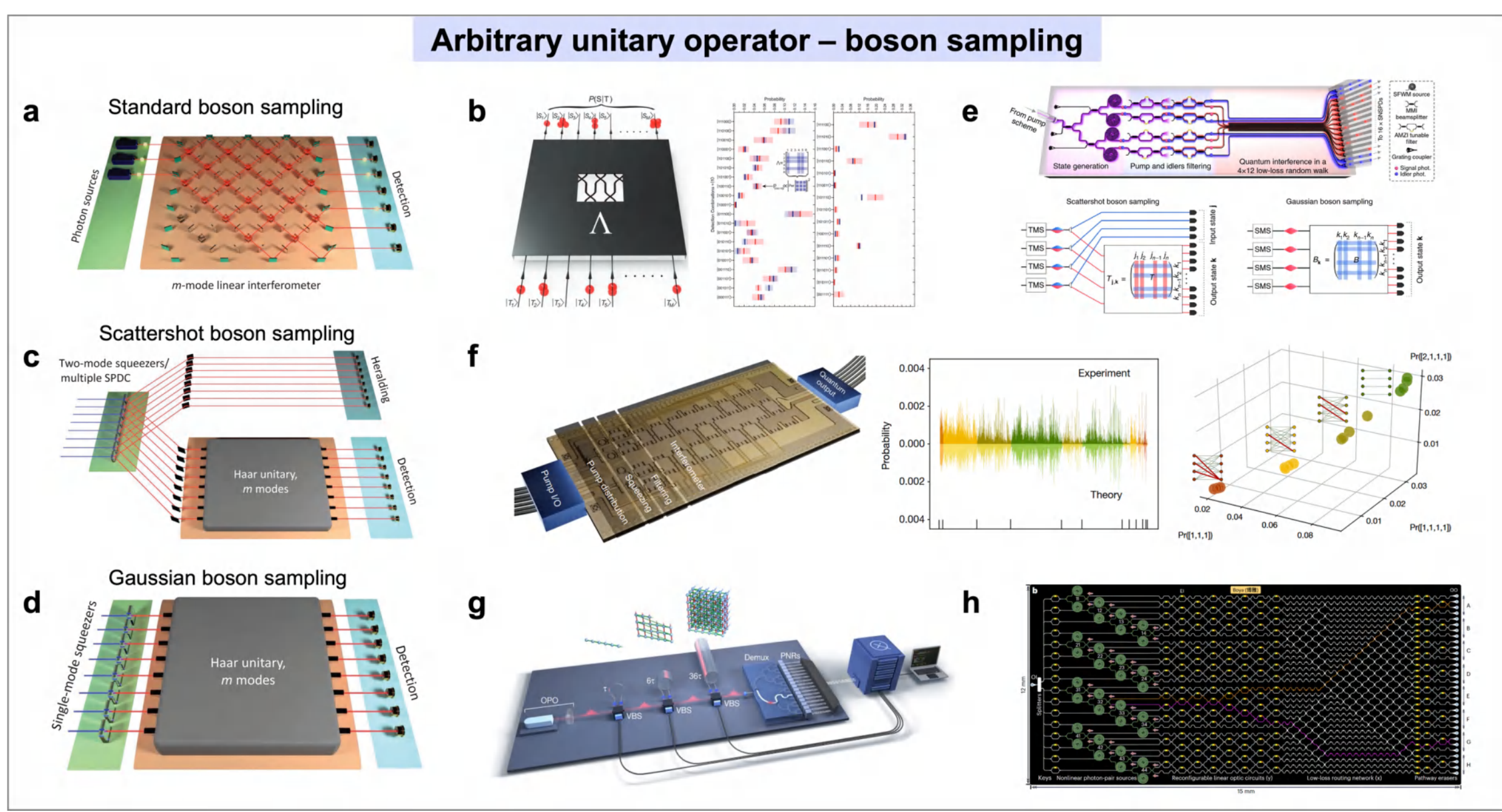

**Fig. 18 | Arbitrary unitary operator-based quantum computing circuits for boson sampling. a**, Standard boson sampling: Single-photon Fock states enter interferometer inputs, undergoing unitary transformation before detection. **b**, Simulating boson sampling on a six-mode photonic chip. **c,** Scattershot boson sampling: Two-mode squeezers or SPDC sources generate photon pairs across multiple modes. Detection of heralding photons triggers post-selection of computational modes, enhancing sampling rates by exploiting parallelization. **d**, Gaussian boson sampling. Compared to standard boson sampling, GBS replaces Fock states with squeezed vacuum states. Continuous-variable inputs enable measurement of photon number distributions without requiring single-photon sources, significantly enhancing experimental feasibility. GBS naturally encodes dense graphs for combinatorial optimization tasks. **e**, Sampling of eight quantum states of light on chip: the silicon chip integrates four SFWM spiral photon sources and twelve continuously coupled waveguides with a network of MMIs. Two pump schemes (i.e., single-pump and dual-pump) and three boson sampling schemes (i.e., standard boson sampling, scattershot boson sampling, SBS, and Gaussian boson sampling, GBS) are demonstrated. **f**, Gaussian boson sampling on chip: eight modes initialized as vacuum are squeezed and entangled to form two-mode squeezed vacuum states. Programmable four-mode rotation gates (SU(4)) are applied to each four-mode subspace. **g**, High-dimensional GBS on 216 squeezed modes entangled with 3D connectivity, using a time-multiplexed and photon-number resolving architecture. It is impactful, although integration has not yet been achieved. **h**, A very-large-scale quantum processor, structured with inspiration from graph theory, monolithically integrates 2,446 components. Two genuine examples are demonstrated: the multipartite multidimensional quantum entanglement with different entanglement structures, and the measurement of probability distributions proportional to the modulus-squared hafnian (permanent) of the graph's adjacency matrices. Panels reproduced from: **a,** ref. [209]; **b**, ref.[192]; **c**, ref.[209];**d**, ref. [209];**e**, ref. [137]; **f**, ref. [189]; **g**, ref. [199]; **h**, ref. [200].

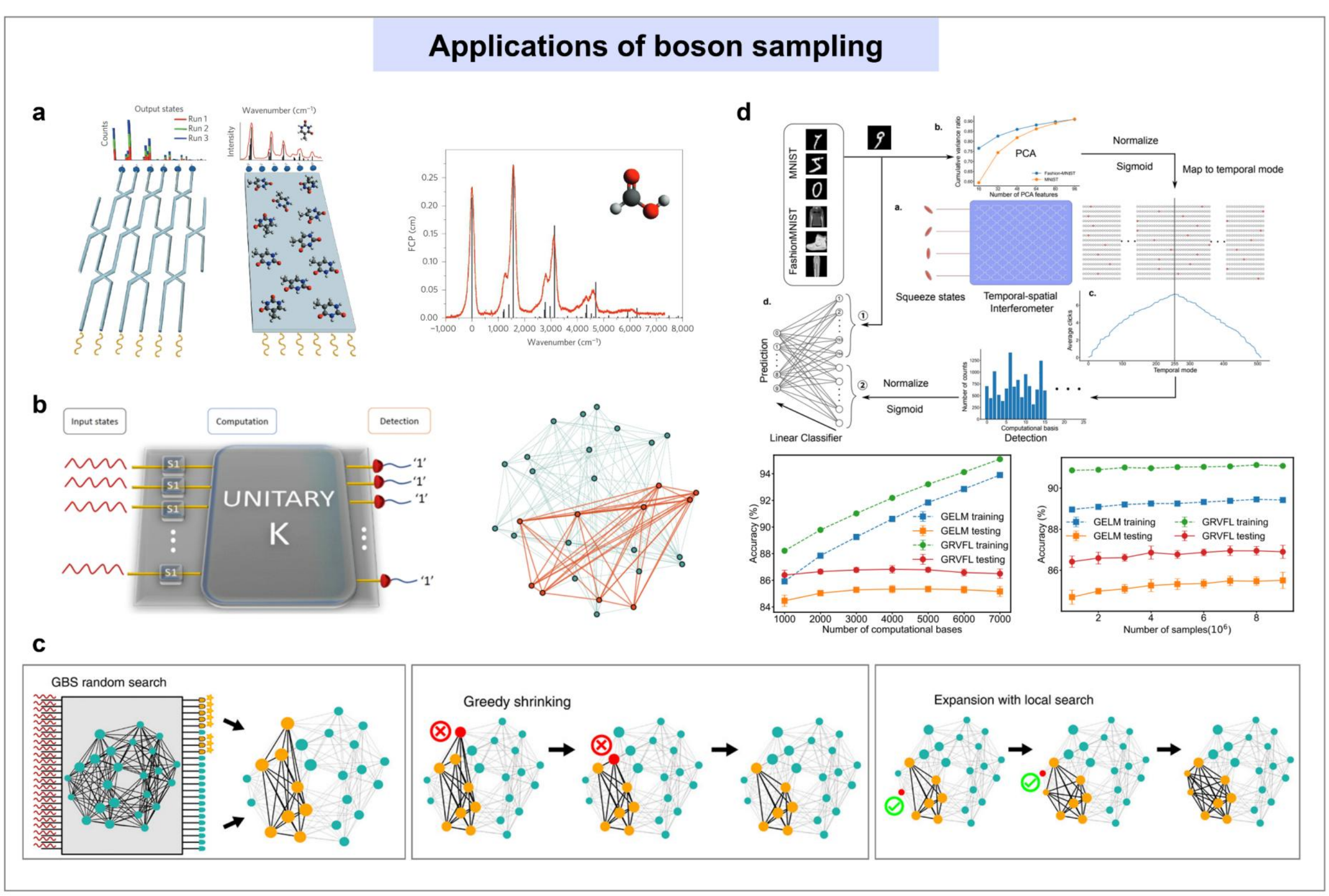

**Fig. 19 | Practical applications of boson sampling. a,** The computational equivalence between boson sampling and molecular vibronic spectrum simulation. Boson sampling involves probabilistically sampling photon output distributions generated by quantum interference in a linear photonic interferometer. This process shares mathematical homology with vibronic spectroscopy, where ensembles of identical molecules are electronically excited by coherent light sources to probe their vibrational energy spectra through re-emitted radiation signatures. **b,** Dense graph problem simulated by Gaussian boson sampling process to find the maximum perfect matching numbers of a subgraph. **c,** Clique-finding protocol using gaussian boson sampling, which is applied to molecule docking. Squeezed light enters a programmable GBS device configured for a vertex-weighted graph. Photon detections identify vertices/edges, forming a candidate subgraph. The three state process include GBS random search, greedy shrinking and local search expansion. **d,** The application of gaussian boson sampling in enhancing image recognition. Radom photon detection defines an initial subgraph, followed by iterative greedy vertex removal and neighbourhood expansion to maximize solutions for clique detection. The hybrid quantum-classical interface processes squeezed states through mode-selective transformations to extract features, which are then fed into multimodal neural networks. Panels reproduced from: **a**, ref.[201]; **b**, ref.[205,206]; **c**, ref.[207]; **d**, ref.[208].

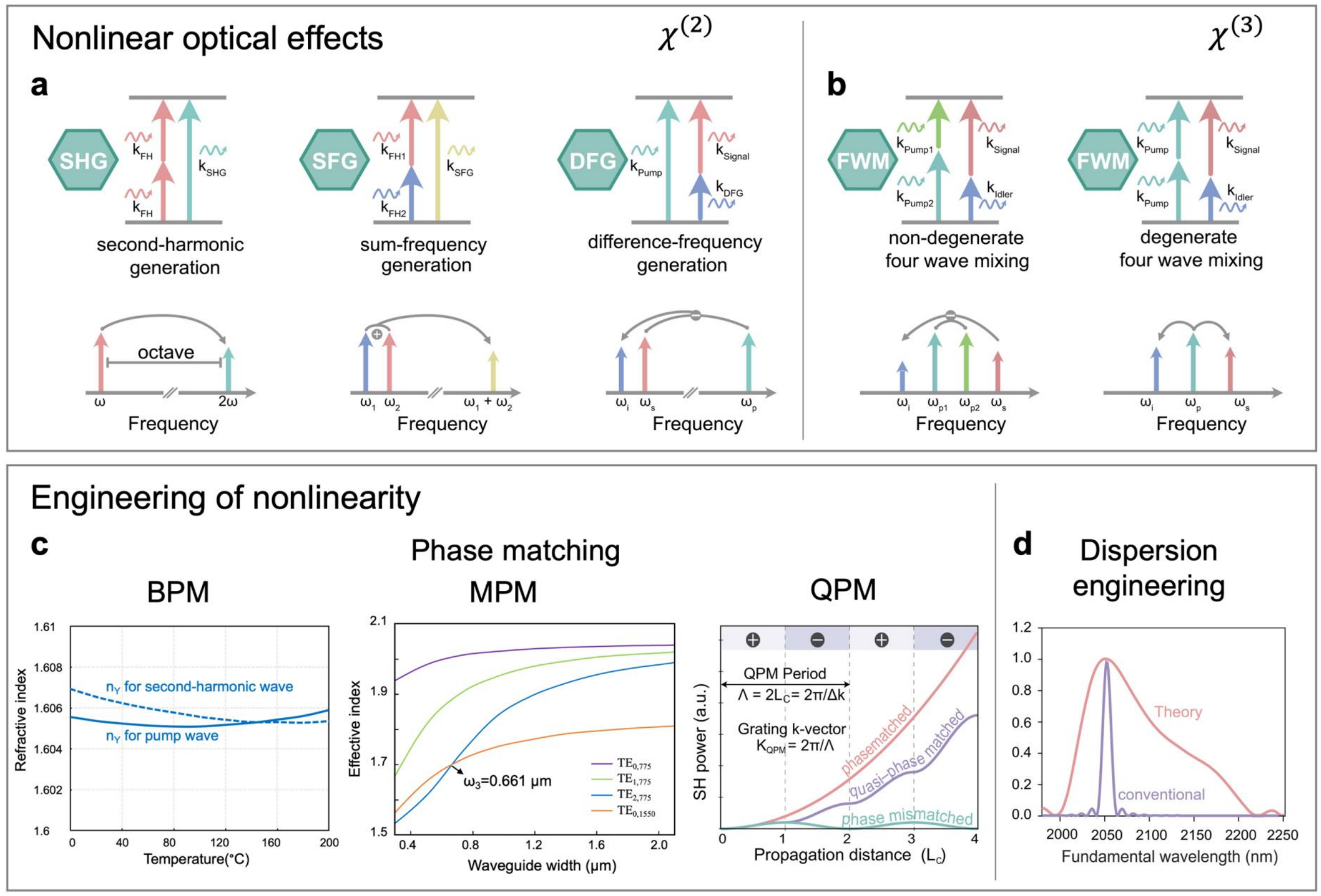

**Fig. 20 | Nonlinear optical effects and engineering of nonlinearity in LN. a**, Illustration of the second-order and third-order nonlinear optical effects. Using the largest nonlinear coefficient $d_{33}$, Common second-order nonlinear processes include SHG, sum-frequency generation (SFG), and difference-frequency generation (DFG). Using nonlinear refractive index of $1.8 \times 10^{-19}$ m²/W, common third-order nonlinear processes include non-degenerate/degenerate four-wave mixing processes, for applications like optical frequency combs. **b,** The common nonlinear engineering techniques in nonlinear processes, which include phase matching and dispersion engineering. The phase matching methods typically include birefringent phase matching (BPM), modal phase matching (MPM), and quasi-phase matching (QPM). Panels reproduced from: ref. [31,438].

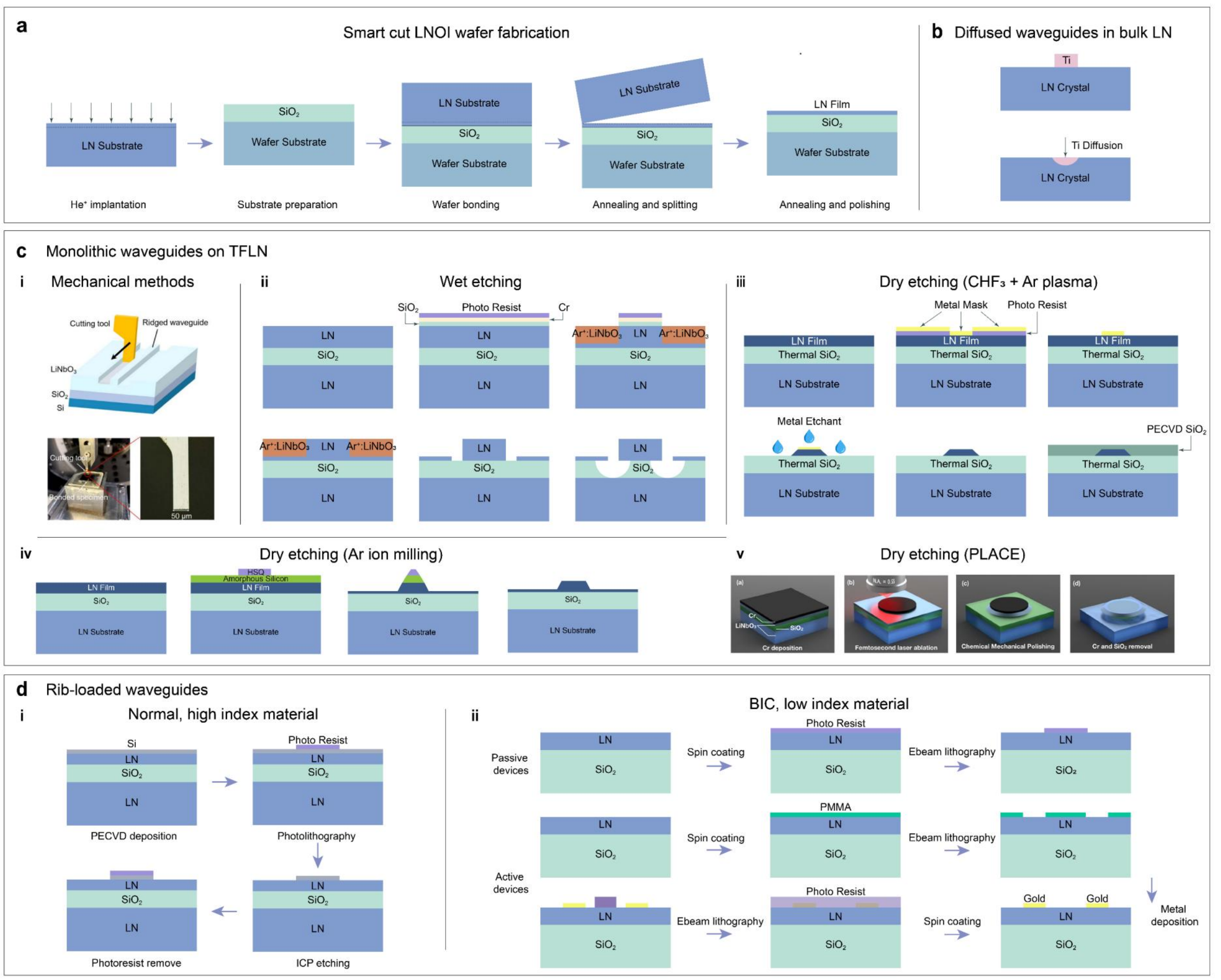

**Fig. 21 | LNOI fabrication processes. a**, Smart cut LNOI wafer fabrication. **b**, Diffused waveguides in bulk LN. **c**, Monolithic waveguides on TFLN, including mechanical methods of ultra-precision cutting, wet etching methods of argon (Ar) ion implantation with KOH wet etching, and dry etching methods. Dry etching can be achieved through RIE using a mix of fluorine ($CHF_3$) and Ar plasma, Ar ion milling, and Proton-exchange assisted chemical etching (PLACE). **d,** Rib-loaded waveguides, showcasing amorphous silicon-lithium niobate thin film strip loaded waveguides, and photonic integrated circuits with bound states in the continuum (BIC). Panels reproduced from: **a**, ref. [326]; **b**, ref. [439]; **c**, ref. [33,223,228,232,440]; **d**, ref. [241,250].

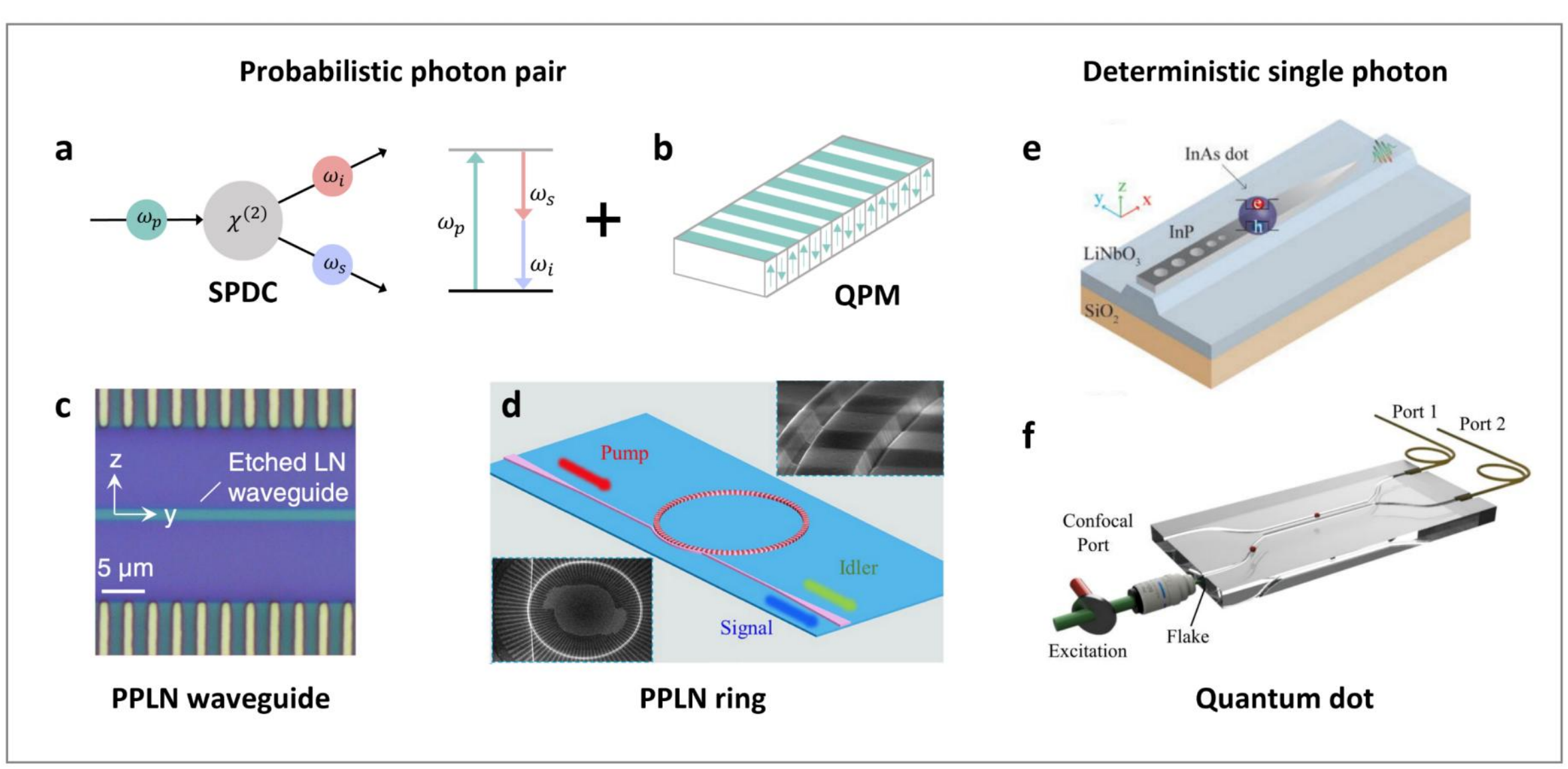

**Fig. 22 | Integrated quantum photon sources in LNOI platforms. a**, SPDC process for generation of photon pairs: a high-energy photon converts into two lower-energy photons within a second-order nonlinear material. **b**, Quasi-phase matching (QPM), essential for achieving efficient frequency conversion leveraging the second-order nonlinearity of LN. This technique involves creating a periodic structure within the nonlinear crystal to compensate for phase mismatch. It is commonly implemented using PPLN, where the crystal's polarization is periodically reversed to facilitate this matching. **c**, Entangled photon pair generation in thin-film PPLN waveguides. **d**, quantum photon sources in PPLN microring resonator at high rates. **e**, A hybrid platform consisting of the LN waveguide and the InAs quantum dot embedded in an InP nanobeam. **f**, Ti in-diffused LN directional coupler with a $WSe_2$ flake at the input facet. Photoluminescence was measured by exciting the emitters in a confocal microscope, while emission was detected in confocal geometry (Confocal Port) and through the two fiber output ports (Port 1 and Port 2). Panels reproduced from: **c**, ref. [49]; **d**, ref. [265]; **e**, ref. [277]; **f**, ref. [278].

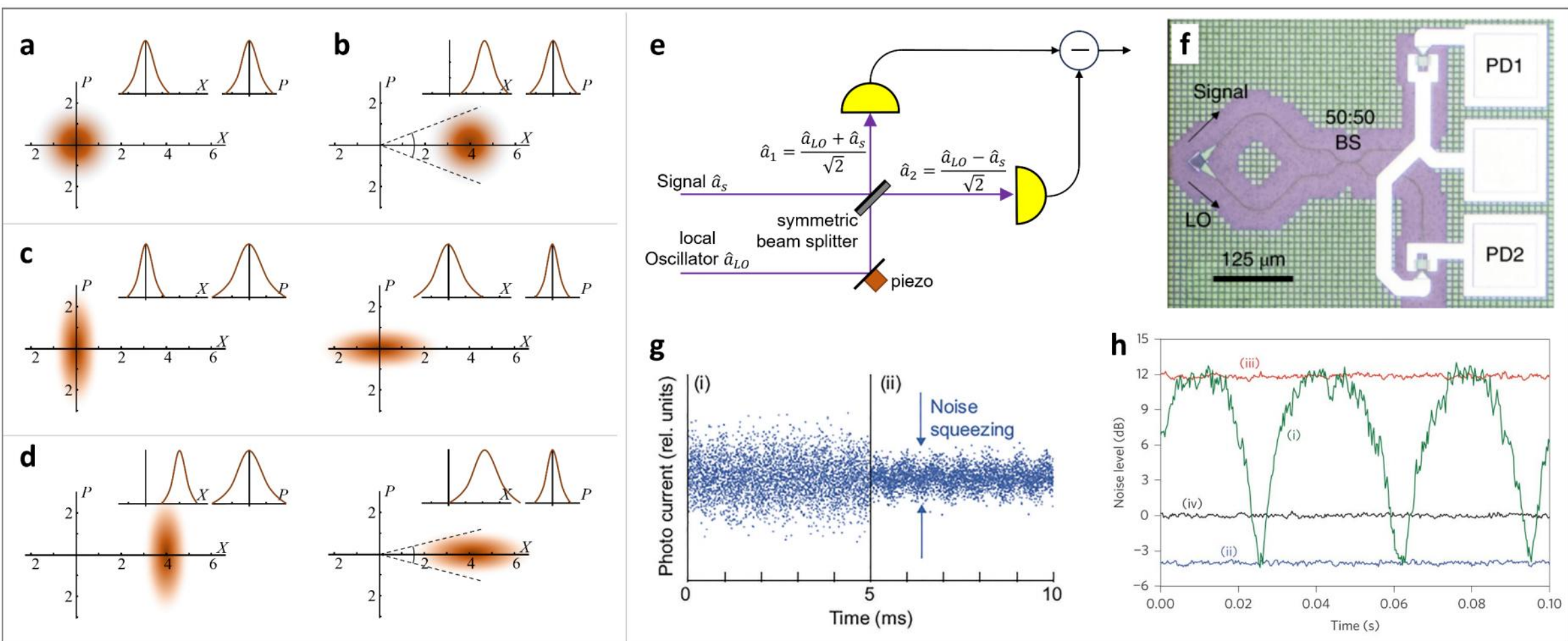

**Fig. 23 | Wigner functions and measurement of quantum squeezed states.** The Wigner functions of **a,** vacuum state; **b**, coherent state; **c,** Squeezed vacuum state with quadrature-amplitude squeezing and quadrature-phase squeezing, respectively; **d,** Squeezed coherent state with quadrature-amplitude squeezing and quadrature-phase squeezing. **e, f,** the off-chip and on-chip integrated homodyne detection setup. **g**, The photocurrent noise for coherent state and squeezed coherent state for illustration. **h,** The quantum noise level of a squeezed light, from which the measured squeezing level is -4 dB and the anti-squeezing level is 11.85 dB. Panels reproduced from: **f,** ref. [279]; **g**, ref. [441]; **h**, ref. [399].

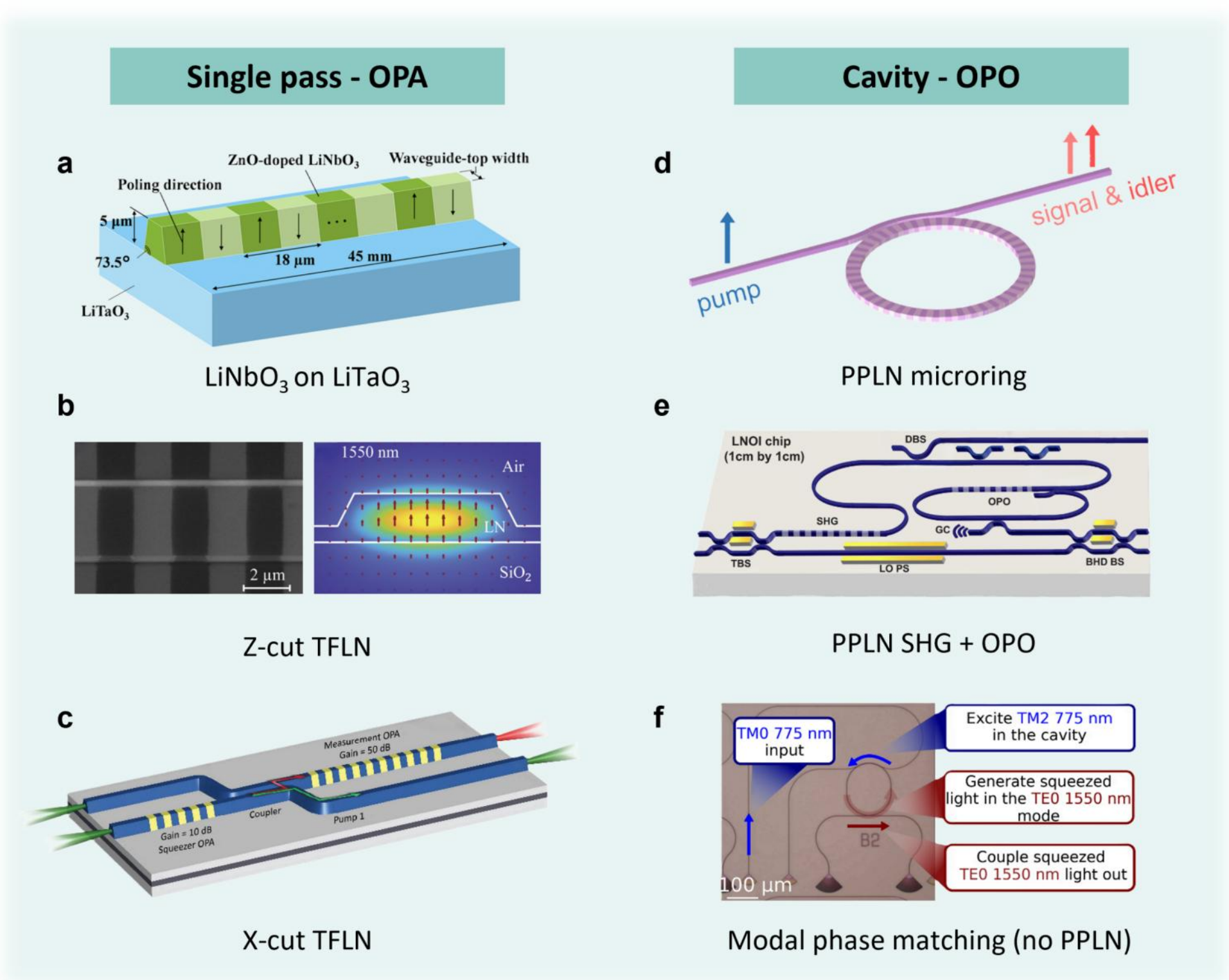

**Fig. 24 | Generation of quantum squeezed states in LNOI platform.** Generation of squeezed state in PPLN waveguide and PPLN ring resonators. **a,** CW-pumped optical parametric amplifiers (OPAs) using a single-mode PPLN waveguide directly bonded onto a $LiTaO_3$ substrate. **b**, CW-pumped quadrature squeezing in a PPLN waveguide fabricated from a z-cut TFLN wafer. **c**, The generation and all-optical measurement of squeezed states on the same LN-based nanophotonic chip. The structure includes a squeezer OPA, a tapered adiabatic coupler, and a measurement OPA. When pumped, the squeezer OPA generates a squeezed vacuum state and is subsequently amplified by the measurement OPA to macroscopic power levels, which is sufficiently above the vacuum noise, thereby making the measurement insensitive to losses due to off-chip coupling and imperfect detection. **d**, A quasi-phase-matched optical parametric oscillator (OPO) on PPLN microring resonator. **e**, A subthreshold OPO for generating and analysing squeezed light. The circuit integrates directional couplers, tunable beamsplitter (TBS), waveguide second harmonic generation (SHG), dichroic beam splitter (DBS), OPO, grating coupler, local oscillator phase shifter (LO PS), and balanced homodyne beamsplitter (BHD BS). SHG section prepares the pump at the second harmonic frequency for the OPO. The pump field only makes a single pass through the cavity due to the presence of the wavelength-dependent output coupler. As a result, below the oscillation threshold, a squeezed vacuum state at the fundamental harmonic emerges. The squeezed vacuum state and the prepared LO are combined by the BHD BS for a balanced homodyne measurement. **f**, Squeezed light generation using modal phase matching. **a**, **i**, ref. [290]; **ii**, ref. [293]; **iii**, ref. [294]; **iv**, ref. [295]; **v**, ref. [296]; **vi**, ref. [297].

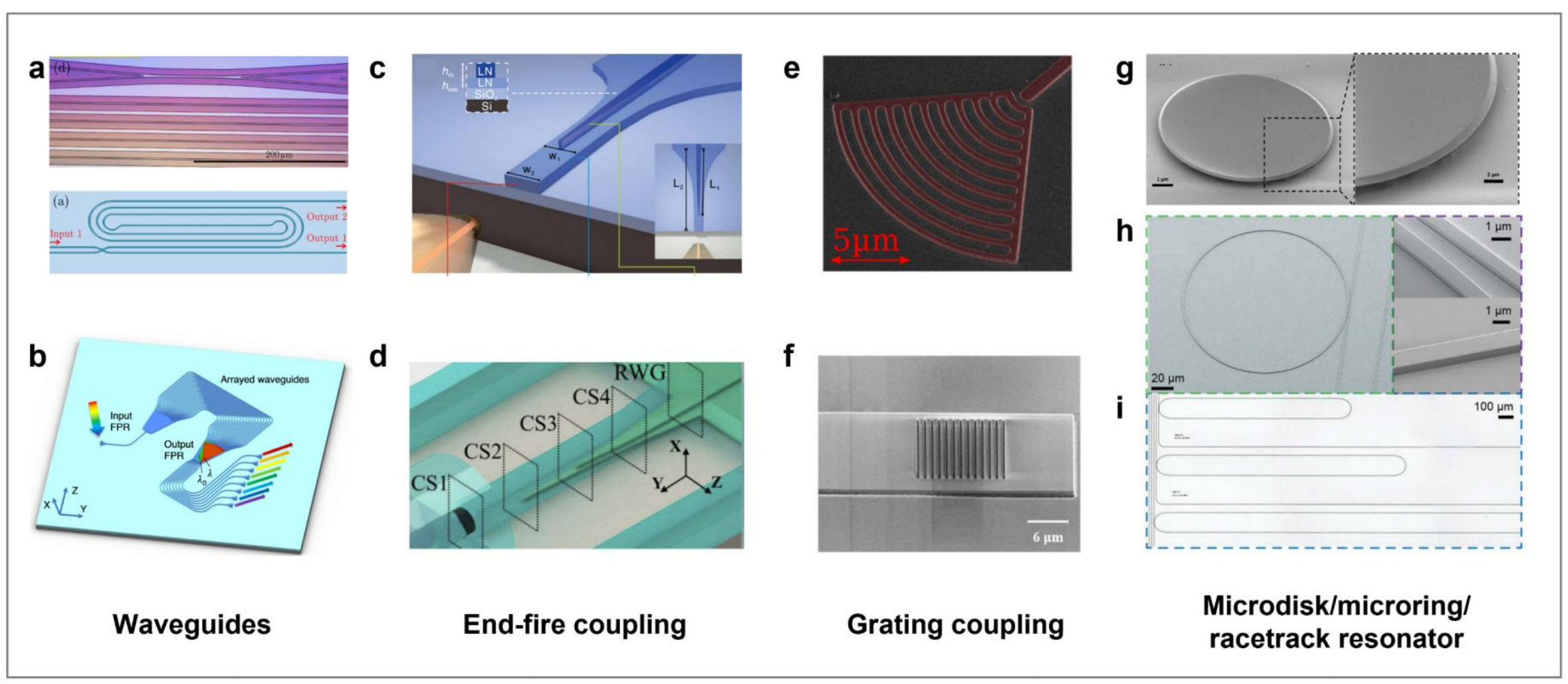

**Fig. 25 | Passive components in LNOI platforms. a**, Optical micrograph of a fabricated beam splitter and a beam splitter connected with two waveguides of different lengths. **b**, schematic of the array waveguide grating (AWG). **c**, Low-loss fiber-to-chip end-fire coupler based on bilayer mode size converter. The measured fiber-to-chip coupling loss is lower than 1.7 dB/facet with high fabrication tolerance and repeatability. **d**, Inverse taper with the widths of the lower LN slab and the ridge reduced simultaneously. **e**, Curved grating coupler in LNOI. **f**, Straight hybrid grating coupler on TFLN using amorphous silicon as grating teeth to achieve high coupling efficiency. **g**, **h**, **i**, Microdisk, microring, and racetrack resonators, respectively. Panels reproduced from: **a**, ref.[371]; **b**, ref. [300]; **c**, ref. [301]; **d**, ref. [302]; **e**, ref. [309]; **f**, ref. [310]; **g**, ref. [311]; **h**,**i**, ref. [32].

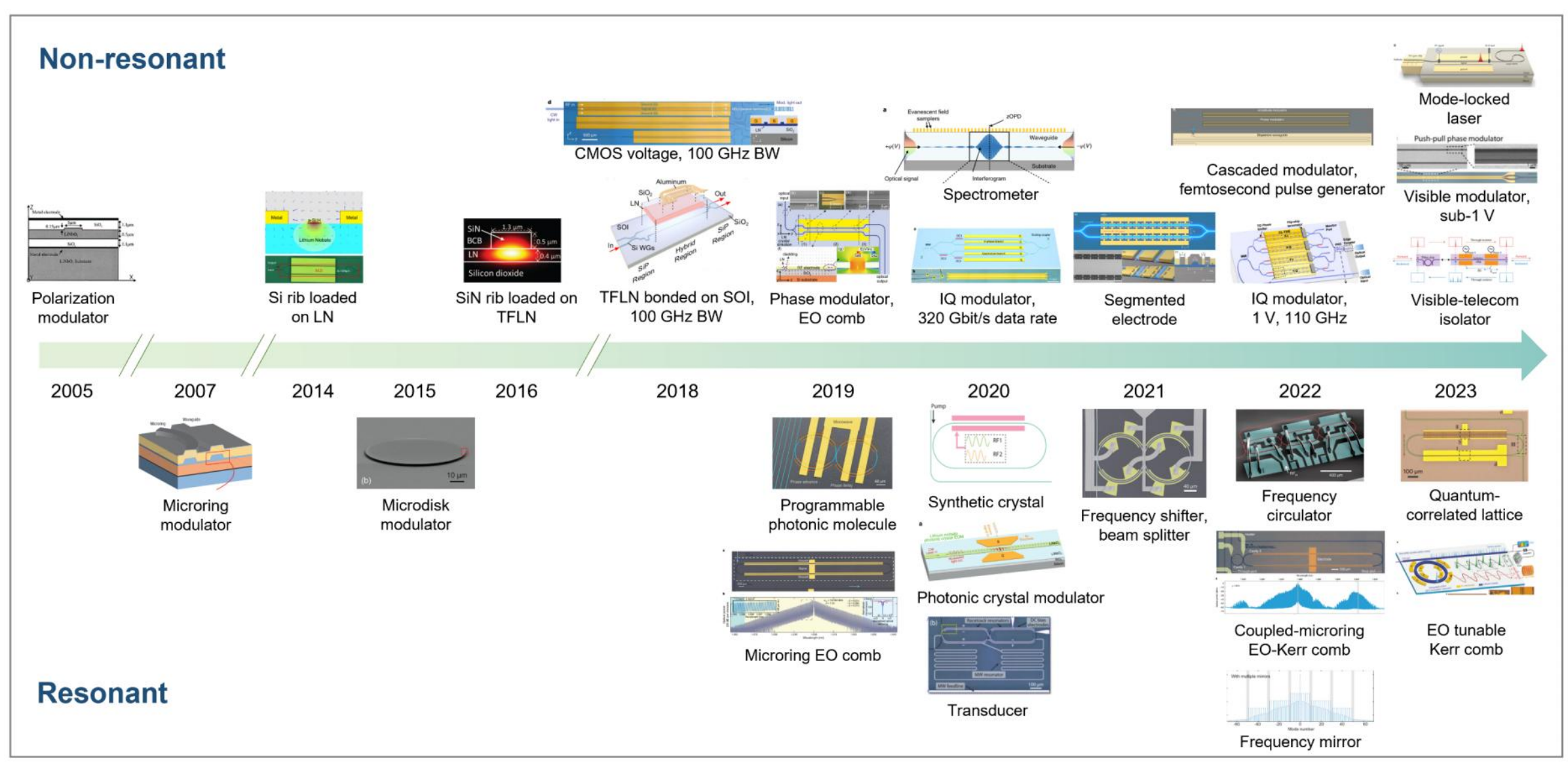

**Fig. 26 | Timeline for the development of lithium niobate electro-optic modulator:** 2005: The first TFLN-based EO modulators for non-resonant type[233]; 2007: the first EO modulator for resonant type[442]; 2014: The Si rib loaded LN waveguide without the need for LN etching[248]. 2015: The microdisk modulator, which largely reduces the scattering loss from the rough sidewall, thus leading to a high quality factor and tuning rate[443]. 2016: sub-1 dB/cm waveguide propagation loss in dry-etched LNOI modulators[28]. 2018: modulators with over 100 GHz bandwidth functioning at CMOS-compatible voltages[35], heterogeneous integration by bonding thin-film LN on SOI[334]. 2019: An integrated low-voltage broadband LN phase modulator[343]. Programmable photonic molecule [322];. Microring (resonant) EO comb[44]. 2020: An integrated EO Fourier transform spectrometer on a SiN-loaded TFLN platform[342]; IQ modulators with 1 V driving voltage and 110 GHz bandwidth[337]; A high-dimensional frequency crystal using a single EO cavity on TFLN[351]; LN photonic-crystal electro-optical modulator[335]; Coupled rings combined with EO modulation for microwave-to-optical photon conversion[35]. 2021: The utilization of segmented electrodes to realize high tuning efficiency and large modulation bandwidth[336]. On-chip electro-optic frequency shifters[349]. 2022: Cascaded modulator as a femtosecond pulse generator[344]. Dual-polarization in-phase quadrature modulators for terabit-per-second transmission[444]. Mirror symmetric on-chip frequency circulation of light[350]. Coupled-ring EO-kerr comb[345]. Coupled-cavity and polarization-crossing for Frequency mirror[352]. 2023: Mode-locked laser which is integrated with a III-V semiconductor optical amplifier[347]. A visible-band modulator with sub-1 V driving voltage[338]. A visible-telecom tunable dual-band optical isolator was demonstrated[339].Quantum related lattice[353]. EO-tunable Kerr comb[346].

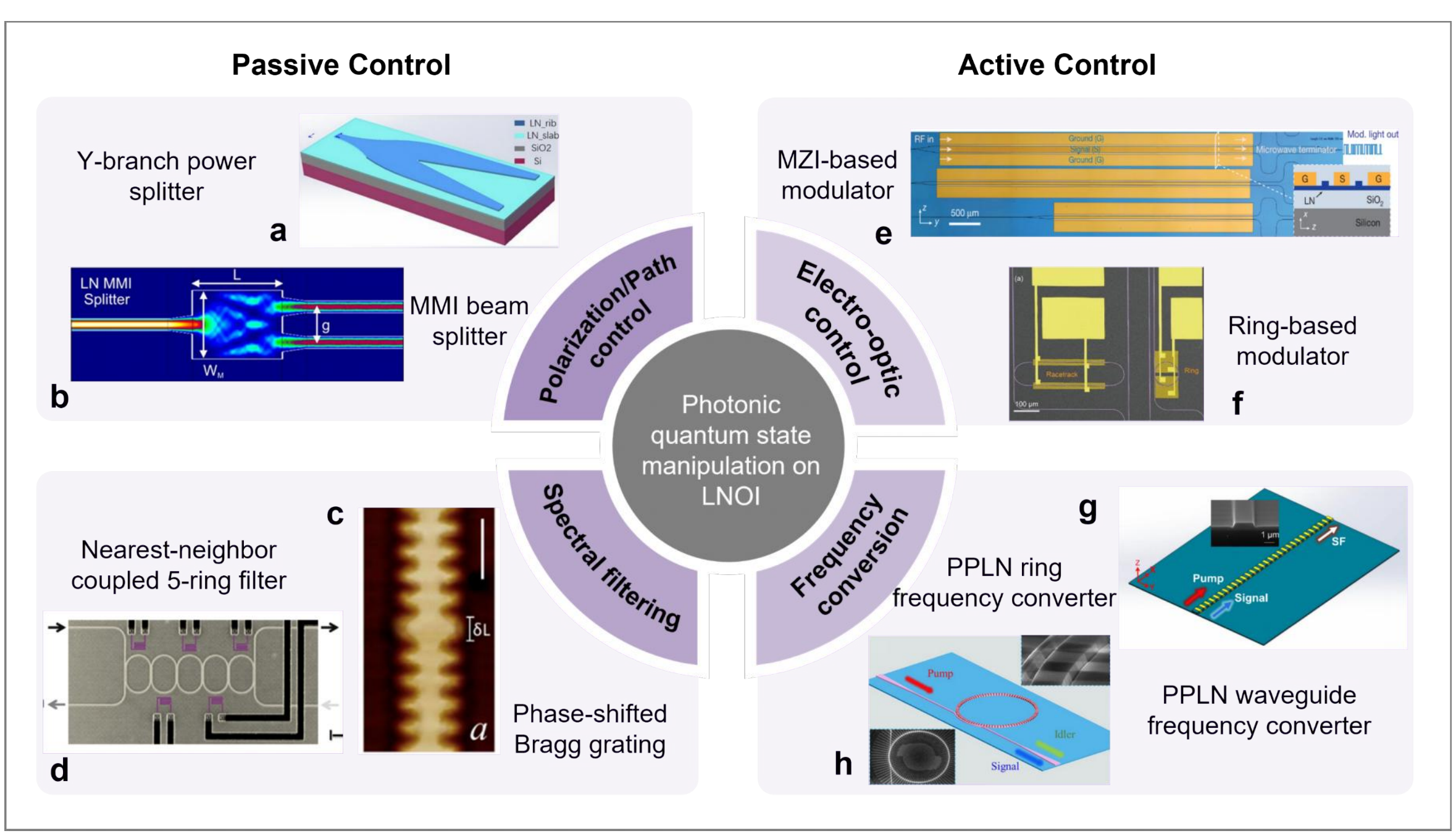

**Fig. 27 | Control mechanisms on LNOI platforms.** **a**,**b**, Y-branch power splitter and MMI beam splitter for polarization and path-based passive control. **c**,**d**, Components for spectral filtering, like phase-shifted Bragg grating, and nearest neighbour coupled 5-ring filter. **e,f,** Electro-optic modulators, featuring non-resonant types for MZI and resonant types for microring resonators. **g,h,** Frequency conversion components, such as PPLN ring and PPLN waveguide, where PPLN is a widely adopted technique for compensating phase mismatching in frequency conversion processes. Panels reproduced from: **a**, ref. [356]; **b**, ref. [357]; **c**, ref. [361]; **d**, ref. [364]; **e**, ref. [35]; **f**, ref. [33]; **g**, ref. [377]; **h**, ref. [265].

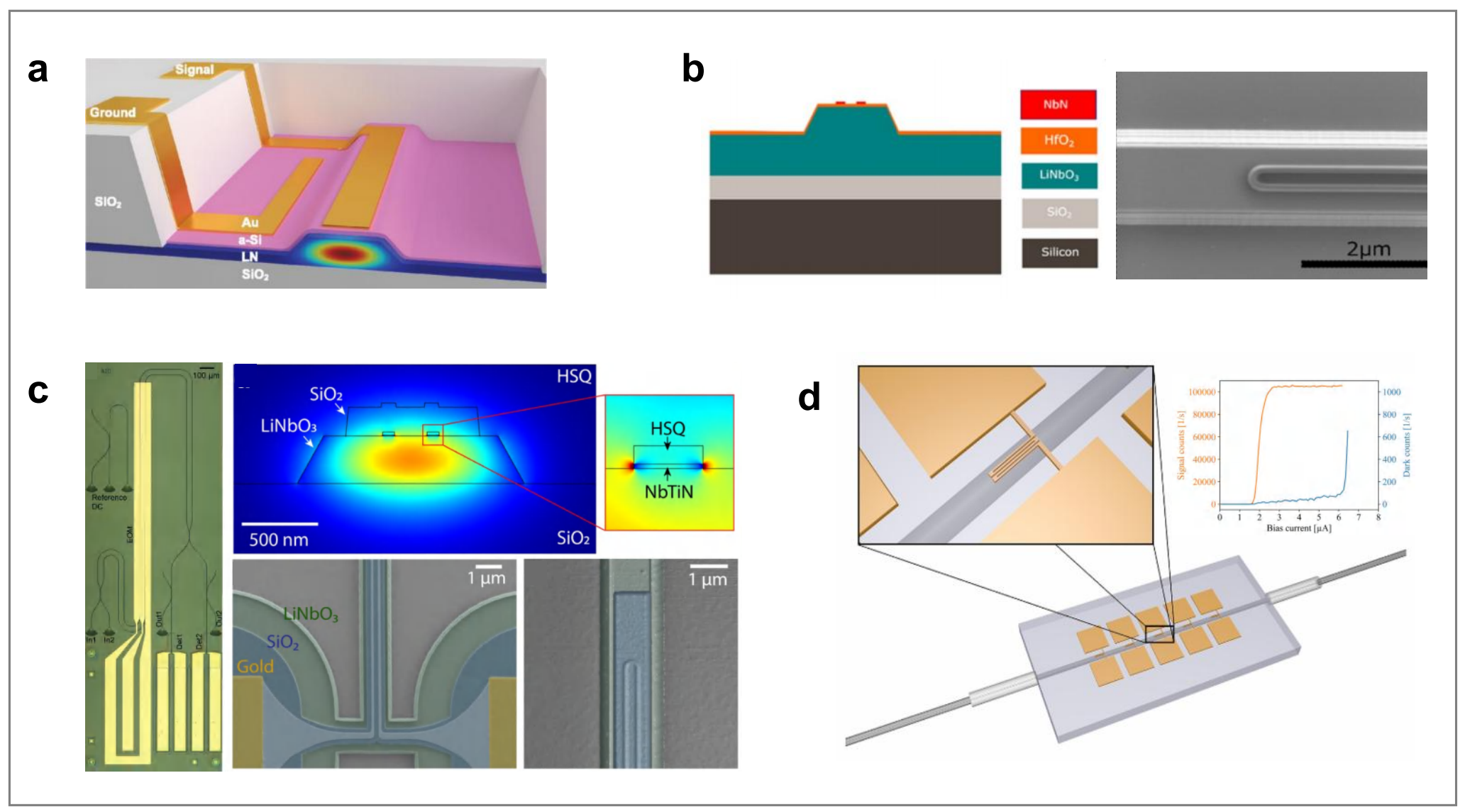

**Fig. 28 | Integrated single photon detectors on LNOI platforms. a**, An integrated photodetector device operating in visible wavelength range, which consists of the LN waveguide, an amorphous silicon absorption layer, and gold contacts. **b**, A waveguide-integrated superconducting nanowire single-photon detectors on TFLN. The U-shape NbN superconducting nanowire is lithographically defined on top of an LN nano-waveguide. **c**, SNSPD integrated with electro-optically reconfigurable LNOI circuits: The reference directional coupler in the left inset enables measurement of the splitting ratio of the DC upstream from the detectors and estimation of the insertion loss of the MZI. The nanowire configuration and the fabricated device are shown in the right inset. **d**, A fiber-coupled waveguide-chip with five in-line SNSPDs on titanium in-diffused LN waveguides. Panels reproduced from: **a**, ref. [383]; **b**, ref. [50]; **c**, ref. [51]; **d**, ref. [389].

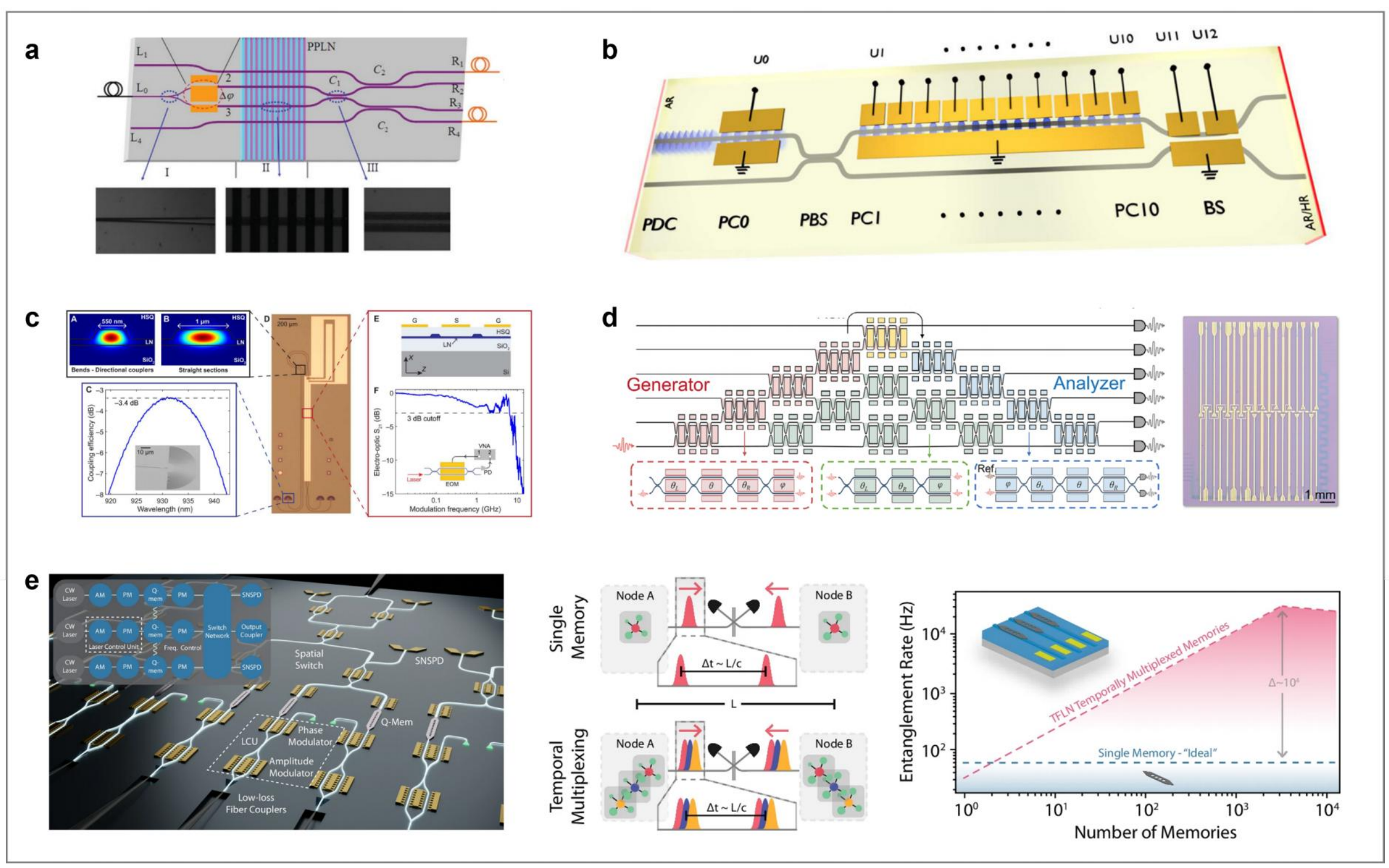

**Fig. 29 | Integrated quantum photonic circuits on LNOI platforms. a**, Entangled photon generation integrated with reconfigurable LN waveguide circuits: By incorporating a periodically poled structure into the waveguide circuits, two individual photon-pair sources with a controllable EO phase shift are generated within a Hong-Ou-Mandel (HOM) interferometer. This setup enables the deterministic production of identical, spatially separated photon pairs. The three insets show the details of the Y-branch, the PPLN waveguide, and the directional coupler, respectively. **b,** A compact quantum circuit for precise manipulation in LN waveguides: The integrated chip includes parametric down-conversion (PDC) sources, EO phase controllers (PCs), polarization beam splitters (PBS), and beam splitters (BS). The Ti-diffused waveguides are indicated by the gray lines. In the periodically poled PDC section, orthogonally polarized photon pairs (horizontal H and vertical V) are generated. The subsequent section demonstrates HOM interference, showcasing the circuit's capability for active and accurate manipulation within LN waveguides**. c,** High-speed quantum processor driven by a solid-state quantum emitter. The LN chip is interfaced with deterministic solid-state single-photon sources based on QDs in nanophotonic waveguides. The generated photons are processed with low-loss circuits programmable at speeds of several gigahertz. **d,** An integrated six-mode universal linear optical circuit on LNOI based on EO tunable MZIs. A specialized MZI mesh has been employed to achieve a high extinction ratio, effectively mitigating the impact of fabrication errors. The chip is designed to perform computations associated with optical neural networks. Although it does not incorporate quantum sources, it offers the potential for precise manipulation of quantum states, laying the groundwork for future advancements of LN in quantum computing. **e**, An integrated TFLN photonic platform operating across the visible to near-infrared (VNIR) wavelength range for quantum memories, demonstrating that TFLN can meet the necessary performance and scalability benchmarks for large-scale quantum nodes. Panels reproduced from: **a**, ref. [59]; **b**, ref. [393]; **c**, ref. [394]; **d**, ref. [299]; **e**, ref. [395].

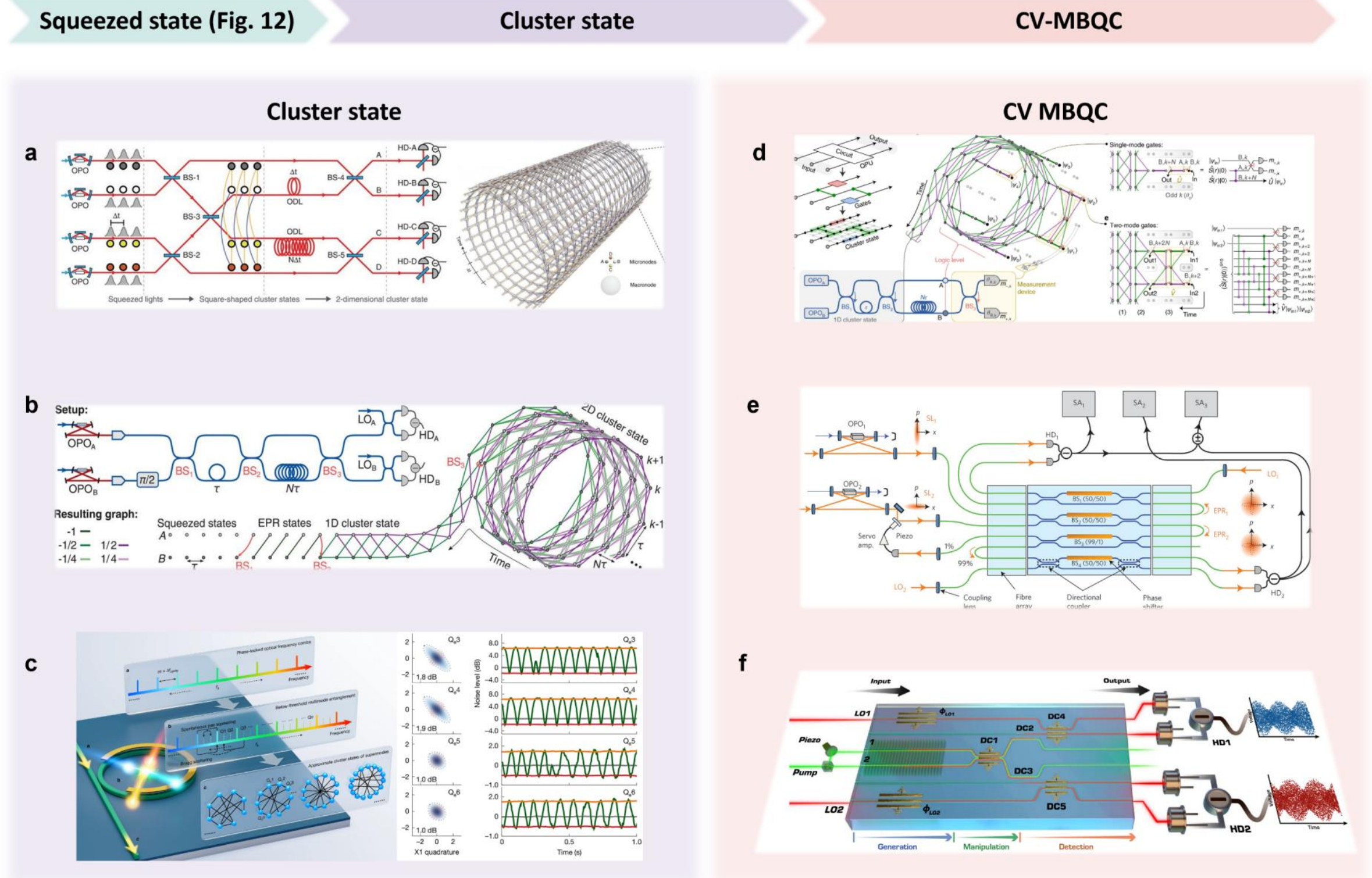

**Fig. 30 | Generation of cluster states and MBQC protocol. a**, Generation of 2D continuous-variable cluster state that consists of 5- by 1240-site square lattice. The cluster state is generated from four squeezed light sources and a linear optical network consisting of five beam splitters and two delay lines. **b**, Deterministic 2D continuous-variable cluster by temporal multiplexing of squeezed light modes, delay loops, and beam-splitter transformations. The generated state consists of more than 30,000 entangled modes arranged in a cylindrical lattice with 24 modes on the circumference, defining the input register, and a length of 1250 modes, defining the computation depth. **c**, Continuous-variable multipartite entanglement in an integrated-optical below-threshold microcomb. **d**, Implementation of MBQC by decomposing the computing circuit into single-mode and two-mode gates. These gates are then implemented through measurements on the generated cluster state. **e**, Generation and characterization of EPR beams within the chip. Two squeezed lights are first combined at BS2. One percent of the weak coherent beams is picked up by BS3 and used for phase-locking between squeezed lights. Output beam EPR1 (EPR2) is combined with local oscillator LO1 (LO2) at BS1 (BS4) and then detected by balanced homodyne detector HD1 (HD2). **f**, The single-chip CV quantum computing. Periodically poled waveguides are used to activate nonclassical light sources. The squeezed vacuum states are manipulated in a reconfigurable directional coupler (DC1) for the generation of two separable squeezed states or a two-mode CV entangled state. Another two directional couplers (DC2, DC3) are used to separate the pump from the signal. Panels reproduced from: **a**, **i**, ref. [160]; **ii**, ref. [159]; **iii**, ref.[414]; **b**, **i**, ref. [421]; **ii**, ref. [399]; **iii**, ref.[289].

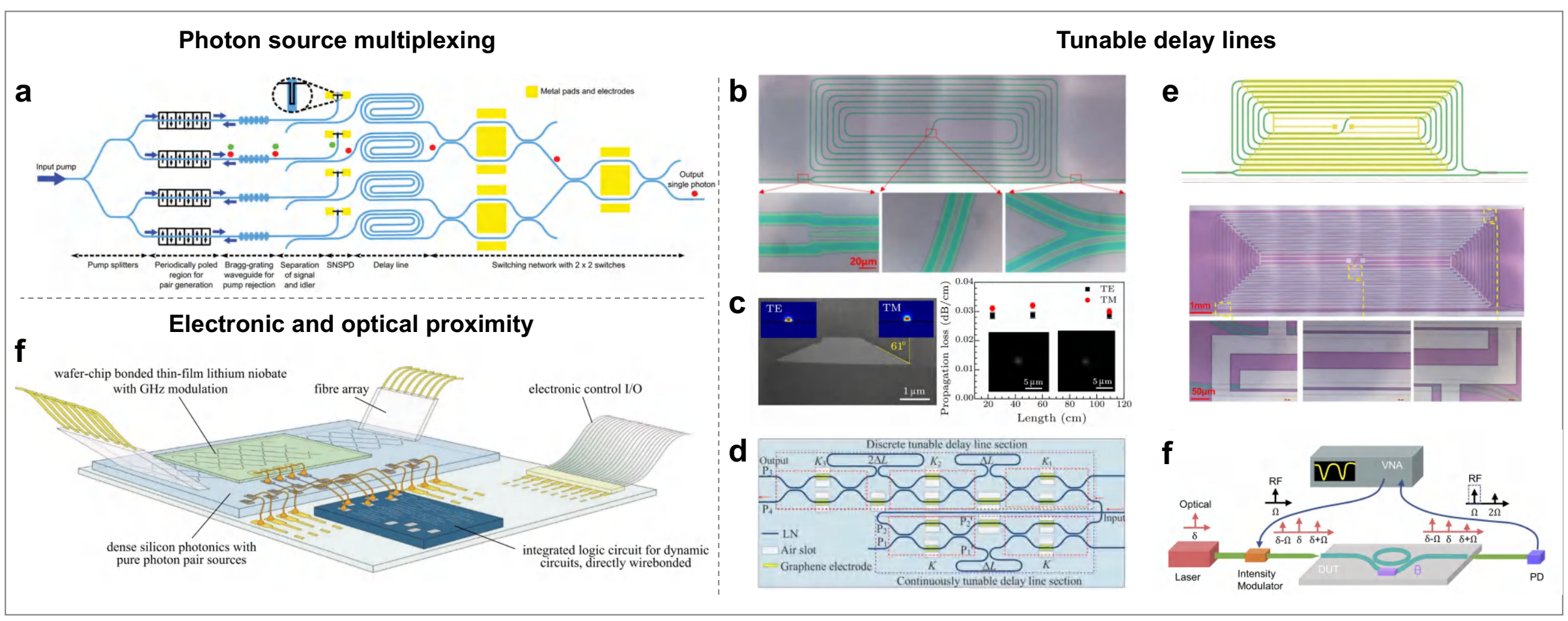

**Fig. 31 | Photon source multiplexing based on low-loss tunable delay lines. a,** Envisioned fully on-chip implementation of a spatially multiplexed single-photon source on LNOI, where four sources are depicted as a simplification. **b,** Micrograph of unbalanced MZI with length difference on LNOI. Insets: zoom-in micrographs of 1 × 2 MMI and Euler bend waveguide. **c,** Cross-sectional scanning electron microscope (SEM) image of the ridge waveguide (Insets: simulated mode field distributions of TE and TM modes); Propagation loss as a function of the length of the optical tunable delay line (Insets: near field distributions of the TE modes at both output ports). **d,** Reconfigurable optical waveguide delay line based on high-performance switches with graphene electrodes. **e,** Schematic and micrographs of a tunable optical delay line on LNOI with microelectrode structures. Insets: zoom-in of the bottom-left, middle, and top-right areas of the device. **f**, Schematic of a tunable delay line based on a self-interference microring resonator with a tunable phase. Panels reproduced from: **a**, ref. [445]; **b**, ref. [423]; **c**, ref. [371]; **d**, ref. [424]; **e**, ref. [425]; **f**, ref. [426].

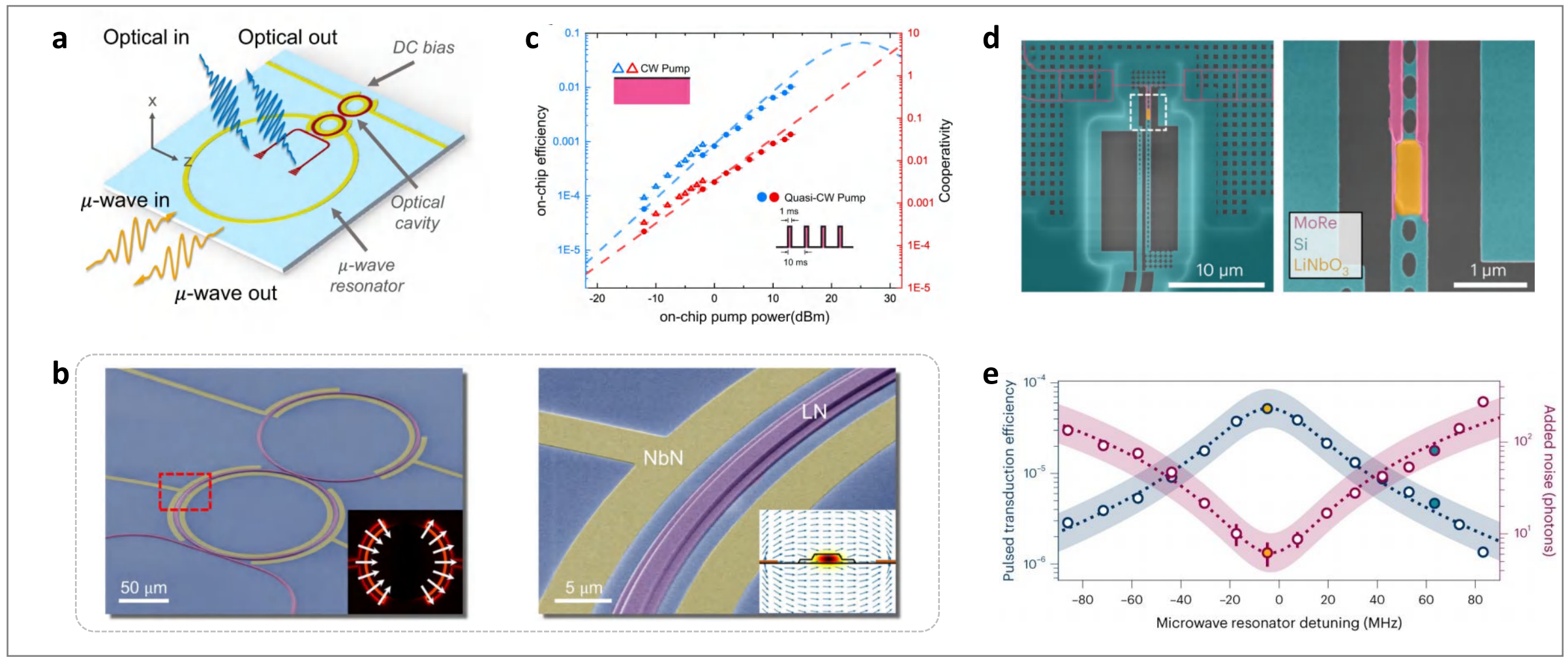

**Fig. 32 | Quantum microwave-to-optic transducers on LNOI. a**, A TFLN cavity EO converter, which consists of a pair of strongly coupled ring resonators patterned from x-cut TFLN and a superconducting microwave resonator made of NbN. The microwave resonator capacitively couples to one of the double rings for EO conversion, while a pair of DC electrodes is coupled to the other microring for electrical tuning of the optical resonance modes. **b**, False-color SEM images of the EO converter device. Insets: the electric field profiles of interacting optical and microwave modes. **c**, On-chip conversion efficiency (blue) and cooperativity (red) as a function of the peak optical pump power in the input waveguide. The maximum conversion efficiency is recorded with a peak pump power of 13.0 dBm. The dashed lines are the theoretical predictions obtained from the data measured with a quasi-CW pump. **d**, SEM images of the microwave-to-optical transduction device: an integrated transducer based on a planar superconducting resonator coupled to a silicon photonic cavity through a mechanical oscillator made of LN on silicon. **e**, Waveguide-to-waveguide transduction efficiency (blue) and added noise (magenta) as a function of microwave resonator detuning. The shaded regions indicate uncertainties arising from the microwave input attenuation. A peak transduction efficiency of $(5.21 \pm 0.03) \times 10^{-5}$ and minimum added noise of $6.2 \pm 1.8$ photons. Panels reproduced from: **a**,**b**,**c**, ref. [52]; **d**,**e**, ref. [53].

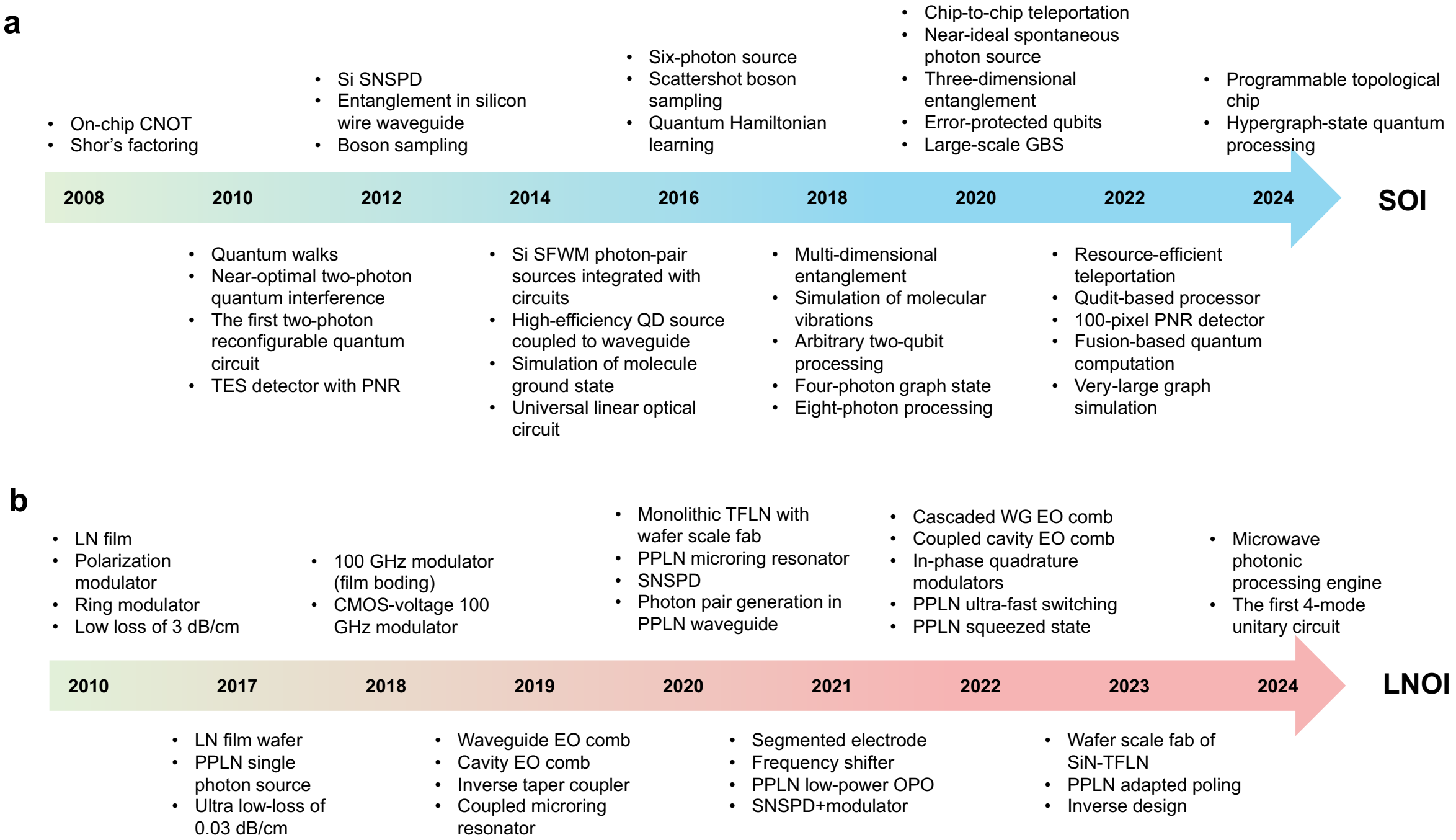

**Fig. 33 | Milestones in photonic quantum computing in SOI and LNOI platforms. a**, key demonstrations of SOI-based photonic quantum computing: 2008-2010: on-chip quantum interference and integrated CNOT gate[138], implementation of Shor's factoring algorithm[139]; 2010-2012: quantum walk of correlated photons involving large-scale quantum interference[173], near optimal two-photon quantum interference[104], the first two-photon reconfigurable quantum circuit for generating, manipulating and measuring entanglement[105], waveguide integrated transition edge sensor (TES) detector with photon-number resolving (PNR) capability[121]; 2012-2014: high-efficiency Si waveguide SNSPD[117], entanglement in silicon wire waveguide[128], boson sampling with multiple photons[190–194]; 2014-2016: Integration of SFWM photon pair sources and quantum circuits[106], high-efficiency quantum dot (QD) coupled to waveguide[87], quantum simulation of molecular vibrations[144], demonstration of a 6-mode universal linear-optical circuit[141]; 2016-2018: on-chip generation of six photons[64], scattershot boson sampling[198], implementation of quantum Hamiltonian learning (QHL) algorithm[130]; 2018-2020: chip-to-chip teleportation[76], near-ideal spontaneous photon source[66], three-dimensional (3D) entanglement[145], error-protected qubits[163], large-scale gaussian bosons sampling (GBS) using many photons[189]; 2020-2022: resource-efficient teleportation of qutrits using quantum autoencoder[143], qudit-based quantum processor[147], 100-pixel PNR detectors[123], fusion-based quantum computation architecture[167], very-large graph simulator[200]; 2022-2024: programmable topological chip[446], hypergraph-state quantum processing[164]. **b**, key demonstrations of LNOI-based photonic quantum computing. 2010-2017: LN film[19,20], Polarization modulator[447], ring modulator[32], low loss waveguide of 3 dB/cm[448]; 2017: LN film wafer using smart cut[449], PPLN single photon source[450], ultra-low loss of 0.03 dB/cm[29]; 2018: 100 GHz electro-optic modulator using film bonding of Si and TFLN[37], CMOS-voltage 100 GHz electro-optic modulator[35]; 2019: waveguide EO comb[324], cavity EO comb[44], inverse taper coupler[301], coupled microring resonator[322]; 2020: Monolithic TFLN with wafer scale fab[451], PPLN microring resonator for single photon anharmonicity[313], LN integrated SNSPD[50], photon pair generation in PPLN waveguide[49]; 2021: segmented electrode[452], frequency shifter[323], PPLN low-power optical parametric oscillation[295,453]; integration of SNSPD and modulators[51]; 2022: cascaded waveguide EO comb[454], coupled cavity EO-Kerr comb[455], in-phase quadrature modulators[444], PPLN ultra-fast switching[422], PPLN squeezed state[294]; 2023: wafer scale fab of film boding of SiN and TFLN[456], PPLN adapted poling[457], inverse design of TFLN devices[458]; 2024: Microwave photon processing engine[34], the first 4-mode universal linear optical circuit[277].

**Table 1 | Fundamental Principles of Discrete and Continuous-Variable MBQC: A Unified Comparison.**

| Dimension | Discrete-Variable MBQC | Continuous-Variable MBQC |
|---|---|---|
| Information carrier | Qubits | Qumodes |
| Physical system | Discrete variables (e.g., photon polarization, electron spin) | Continuous variables (e.g., optical field quadratures $\hat{x}$, $\hat{p}$) |
| Resource state | Cluster state | Gaussian cluster state |
| Gate implementation | Single-qubit measurements + adaptive rotations | Homodyne detection + displacement operations |
| Operation type | Non-Gaussian operations (universal quantum computation) | Gaussian operations (require non-Gaussian resources for universality) |
| Entanglement structure | Graph state | CV graph state |
| Measurement scheme | Discrete projective measurements (e.g., Pauli basis) | Continuous-variable measurements ($\hat{x}$, $\hat{p}$ basis) |
| Error correction difficulty | Qubit-based encoding (e.g., surface codes) | Highly sensitive to Gaussian noise; requires non-Gaussian corrections (e.g., GKP codes) |
| Experimental platform | Superconducting circuits, ion traps, photonic chips | Optical quantum optics (squeezed states, beamsplitter networks) |
| Computational universality | Directly universal | Universality dependent on non-Gaussian operations (e.g., cubic phase gate) |

**Table 2 | Relevant details of some photonic boson sampling experiments reported in the literature**. Note: $n$ is the maximum number of detected photons in the boson sampling experiment after unitary evolution; $m$ is the number of available optical modes. SBS, scattershot boson sampling; GBS, gaussian boson sampling.

| Experiment | $n$ | $m$ | Source | Unitary transformation | Detector |
|---|---|---|---|---|---|
| Spring[192] | 3 | 6 | SPDC | Integrated optics (UV laser written) | SPAD |
| Tillmann[191] | 3 | 5 | SPDC | Integrated optics (fs laser written) | SPAD |
| Crespi[190] | 3 | 5 | SPDC | Integrated optics (fs laser written) | SPAD |
| Spagnolo[197] | 3 | 9 | SPDC | Integrated optics (fs laser written) | SPAD |
| Carolan[193] | 3 | 9 | SPDC | Integrated optics ($SiO_n$) | SPAD |
| Carolan[193] | 4 | 21 | SPDC | Integrated optics (continuous coupling, SiN) | SPAD |
| Bentivegna[198] | 3 | 13 | SPDC and SBS | Integrated optics (fs laser written) | SPAD |
| Carolan[141] | 3 | 6 | SPDC | Integrated optics (reconfigurable, $SiO_2$/Si) | SPAD |
| Zhong[459] | 5 | 12 | SPDC and SBS | Integrated optics (six modes) and polarization | SNSPD |
| Paesani[137] | 4 | 12 | SFWM, SBS, and GBS | Integrated optics | SNSPD |
| Arrazola[189] | 8 | 8 | SFWM, GBS | Integrated optics (reconfigurable, Si) | PNR SNSPD |
| Zhu[460] | 8 | 16 | SFWM, GBS | Integrated optics (reconfigurable, Si) | SNSPD |
| Madsen[199] | 216, time-multiplexed | 16 | SPDC, GBS | Assembled optics | PNR SNSPD |

**Table 3 | A Comparative Analysis of Silicon and Thin-Film Lithium Niobate for Quantum Photonics.**

| Material | $LiNbO_3$ | Si | $Si_3N_4$ |
|---|---|---|---|
| $n_0$ ($n_e$) | 2.21 (2.14) | 3.48 | 2 |
| Propagation loss, dB/cm | 0.027[29,30] | 0.5-2 | 0.005[170] |
| $\chi^{(2)}$ coefficient, pm/V | $d_{33}$: -25.2 (1064 nm) | N/A | N/A |
| $\chi^{(3)}$ coefficient, $m^2/W$ | $1.8 \times 10^{-19}$ (1550 nm) | $5 \times 10^{-18}$ (1550 nm) | $2.5 \times 10^{-19}$ (1550 nm) |
| EO coefficient | $r_{33}$: 30.9 (633 nm) | N/A | N/A |
| Modulation efficiency MZM-conventional $V_\pi L$ typical | 2.3 V·cm[461] | 1.6 V·cm (carrier depletion)[462]<br>30 mW (thermo-optic)[463] | 42.7 mW (thermo-optic)[464]<br>2.7 V·cm (SiN-TFLN)[465] |
| Modulation bandwidth | 170 GHz[461] | 21.7 GHz (carrier depletion)[462]<br>~ kHz (thermo-optic)[463] | ~ kHz (thermo-optic)[464]<br>110 GHz (SiN-TFLN)[465] |
| Extinction ratio MZM | 53 dB[368] | 60.5 dB[103] | 61.2 dB[465] |
| Photon pair source $g_H^{(2)}(0)$ min. | 0.008[265] | 0.005[266] | 0.004[466] |
| Photon pair source PGR @ 1μ$W$ | 2.53 MHz[265] | 292 Hz[266] | 4.2 Hz[466] |
| Squeezing min. dB | -11[294] | -6[467] | -8[283] |
| Squeezing bandwidth max. | 25 THz[294] | 10 MHz[467] | 1 GHz[283] |

**Table 4 | Comparison of common phase matching techniques**

| Dimension | BPM | MPM | QPM |
|---|---|---|---|
| Principle | Compensates phase mismatch using natural crystal birefringence | Matches phase via effective refractive index difference between optical modes in waveguides | Resets phase mismatch by periodically reversing the sign of $\chi^{(2)}$ nonlinear coefficient |
| Advantages | 1. High efficiency at specific wavelengths<br>2. No additional nanofabrication required | 1. Overcomes material birefringence limits<br>2. High power density (waveguide confinement enhances efficiency) | 1. High wavelength flexibility (designable across full transparency window)<br>2. Flexible polarization design (enables strongest $d_{33}$ coefficient)<br>3. No walk-off effect |
| Disadvantages | 1. Angle/temperature sensitivity (requires precise tuning)<br>2. Walk-off effect (limits power/length)<br>3. Material constraints (requires strong birefringent crystals) | 1. High modal loss (especially higher-order modes)<br>2. Low coupling efficiency (mode matching challenge)<br>3. Complex design (nanometer-scale waveguide tolerances) | 1. Requires poling process (challenging for short periods/large sizes)<br>2. Material constraints (only ferroelectric crystals)<br>3. Theoretical efficiency limit |
| Suitable materials | KTP, BBO, LBO birefringent crystals | Si, $Si_3N_4$, LNOI waveguide materials | PPLN, PPKTP poled ferroelectric crystals |
| Typical applications | Low-power visible SHG | Integrated photonic chips (frequency combs, quantum light sources) | High-power conversion (green lasers, mid-IR OPOs, THz sources) |

**Table 5 | Summary of typical experimental results of photon pair generation in $\chi^{(2)}$ and $\chi^{(3)}$ waveguides and microcavities. (PGR- pair generation rate)**

| Reference | Material structure | On-chip pump power | PGR | PGR@1 μW | CAR | $g_H^{(2)}(0)$ |
|---|---|---|---|---|---|---|
| Silverstone[106] | Si waveguide (SFWM, degenerate) | 6.5 mW | 100 kHz | 15.4 Hz | 290 | - |
| Silverstone[106] | Si waveguide (SFWM, non-degenerate) | 7 mW | 100 kHz | 14.3 Hz | 45 | - |
| Paesani[66] | Si waveguide (SFWM) | 500 μW | 15 kHz | 30 Hz | - | 0.053 |
| Lu[63] | Si microdisk (SFWM) | 12 μW | 1.2 kHz | 8.3 Hz | 2610 | 0.003 |
| Liu[70] | Si microring (SFWM) | 61 μW | 1.1 kHz | 18 Hz | 81 | 0.05 |
| Llewellyn[76] | Si microring (SFWM) | 800 μW | 20 kHz | 25 Hz | 50 | 0.05 |
| Engin[61] | Si microring (SFWM) | 19 μW | 827 kHz | 45 Hz | 602 | - |
| Jiang[62] | Si microdisk (SFWM) | 79 μW | 855 kHz | 137 Hz | 274 | - |
| Ma[266] | Si microring (SFWM) | 7.4 μW | 16 kHz | 292 Hz | 12105 | 0.005 |
| Ma[266] | Si microring (SFWM) | 59 μW | 1.1 MHz | 316 Hz | 532 | 0.098 |
| Lu[65] | $Si_3N_4$ microring (SFWM) | 500 μW | 1 MHz | 4 Hz | 50 | - |
| Lu[65] | $Si_3N_4$ microring (SFWM) | 46 μW | 4.8 kHz | 2.27 Hz | 2280 | - |
| Fan[468] | $Si_3N_4$ microring (SFWM) | 1.2 mW | 62 kHz | - | 1243 | 0.014 |
| Chen[466] | $Si_3N_4$ microring (SFWM) | 40 μW | 13.3 kHz | 4.2 Hz | 1438 | 0.004 |
| Guo[58] | AlN microdisk (SPDC) | 1.9 mW | 11 MHz | 5.8 kHz | - | 0.088 |
| Frank[469] | LN microdisk (SPDC) | 16.7 μW | 450 kHz | 26.95 kHz | 6 | - |
| Luo[262] | LN microdisk (SPDC) | 115 μW | 0.5 Hz | 0.004 Hz | 43 | - |
| Ma[265] | PPLN microring (SPDC) | 3.4 μW | 8.5 MHz | 2.53 MHz | 451 | 0.008 |
| Ma[265] | PPLN microring (SPDC) | 13.4 μW | 36.3 MHz | 2.7 MHz | >100 | 0.097 |
| Elkus[48] | PPLN waveguide (SPDC) | 9 μW | 337 kHz | 37 kHz | 6900 | - |
| Zhao[49] | PPLN waveguide (SPDC) | 1.6 μW | 76 kHz | 47.5 kHz | 67224 | - |
| Zhao[49] | PPLN waveguide (SPDC) | 15 μW | 672 kHz | 44.8 kHz | 8361 | 0.022 |
| Zhao[49] | PPLN waveguide (SPDC) | 250 μW | 11.4 MHz | 45.6 kHz | 668 | - |
| Chen[47] | PPLN waveguide (SPDC) | 100 uW | 0.8 MHz | 8 kHz | 631 | - |
| Chen[47] | PPLN waveguide (SPDC) | 1 mW | 7.2 MHz | 7.2 kHz | 23 | - |

**Table 6 | Summary of typical experimental results of squeezed states in $\chi^{(2)}$ and $\chi^{(3)}$ waveguides and microcavities.**

| Reference | Material structure | Measured squeezing | Inferred squeezing | Bandwidth |
|---|---|---|---|---|
| Zhao[284] | Si microring | -1.34 dB | -3.09 dB | - |
| Safavi-Naeini[467] | Si optomechanical resonator | - | -6 dB | <10 MHz |
| Dutt[282] | $Si_3N_4$ microring | -2 dB | -3.9 dB | - |
| Dutt[89] | $Si_3N_4$ microring | -1.7 dB | -5 dB | 5 MHz |
| Zhang[283] | Two coupled $Si_3N_4$ microring | -1.65 dB | -8 dB | 1 GHz |
| Eto[470] | Periodically poled MgO: LN waveguide | -5 dB | -9.7 dB | 76 MHz |
| Domeneguetti[288] | Ti: PPLN waveguide | -3.17 dB | -9 dB | - |
| Lenzini[289] | Ti: PPLN waveguide | -1.38 dB | -7 dB | 35 MHz |
| Kashiwazaki[290] | ZnO: PPLN waveguide on $LiTaO_3$ | -6.3 dB | - | 2.5 THz |
| Kashiwazaki[291] | ZnO: PPLN waveguide on $LiTaO_3$ | -8 dB | <-10 dB | 11 MHz |
| Kashiwazaki[292] | ZnO: PPLN waveguide on $LiTaO_3$ | -6.3 dB | - | 6.0 THz |
| Chen[293] | PPLN waveguide | -0.56 dB | 2.6 dB | 7 THz |
| Nehra[294] | PPLN waveguide | -4.9 dB | -11 dB | 25 THz |
| Park[296] | PPLN microring | -0.55 dB | - | 140 MHz |
| Arge[297] | TFLN microring | -0.46 dB | -2.2 dB | - |

**Table 7 | Comparison of approaches for microwave-to-optical conversion.**

| Dimension | Electro-optomechanical (EOM) | Acousto-optic (AO) | Electro-optic (EO) |
|---|---|---|---|
| Core Mediator | Low-frequency mechanical resonator (phonons) | Propagating acoustic waves/phonons | No mediator particles (material Pockels effect) |
| Mechanism | Piezoelectric effect/Capacitive force + Optomechanical effect | Piezoelectric/driving effect + Photoelastic effect (Brillouin scattering) | Pockels effect |
| Conversion pathway | Microwave -> Mechanical vibration -> Optical modulation | Microwave -> Acoustic wave -> Moving grating -> Optical frequency shift | Microwave field -> Optical modulation/sidebands (direct) |
| Inherent freq. shift | No | Yes (shift magnitude = acoustic frequency) | No (modulation via sideband generation) |
| Speed/Bandwidth | Moderate (limited by mechanical resonator freq. and Q-factor) | Moderate (limited by acoustic excitation and interaction length) | Fastest (bandwidth up to THz) |
| Quantum noise | High (thermal photons difficult to cool to ground state) | High (thermal phonons in acoustic waves) | Lowest (no additional thermal photon channel) |
| Ground-state operation | Challenging (requires mK temperatures) | Challenging (similar constraints) | Relatively feasible (achievable at RT/4K) |
| Material dependence | Piezoelectric/driving materials + Optomechanical materials | Piezoelectric/driving materials + Photoelastic materials | Requires high Pockels-coefficient materials (LN, etc.) |
| Integration complexity | High (MEMS/NEMS processes + photonic integration) | Moderate (planar waveguides + SAW/BAW) | High (demands high-Q microcavities + microwave resonators/electrodes) |
| Maturity level | High (classical)/Moderate (quantum demonstrations) | High (classical modulation)/Low (integrated quantum implementations) | Rapidly evolving (especially on TFLN platforms) |
| Optimal use cases | Low frequency quantum demonstrations, mature technical platforms | Classical optical modulation, frequency conversion | High-efficiency/bandwidth quantum conversion, low-noise, high-speed optical communication |
| Poor suitability for | High-speed applications, ultra-low-noise/room-temperature operation | High-efficiency, low-noise quantum conversion | Low-cost, material-agnostic platforms |